\documentclass[10pt,twocolumn,american,floatfix,longbibliography,aps,prb]{revtex4-1}
\usepackage{mathptmx}

\usepackage[T1]{fontenc}
\usepackage[latin9]{inputenc}
\usepackage[letterpaper]{geometry}
\usepackage{xcolor}
\usepackage{textcomp}
\usepackage{amsmath}
\usepackage{amssymb}
\usepackage{cancel}
\usepackage{graphicx}
\usepackage{esint}
\PassOptionsToPackage{normalem}{ulem}
\usepackage{ulem}
\usepackage{natbib}

\usepackage[american]{babel}
\begin{document}
\title{Thermal activation barriers
for creation and annihilation of
magnetic droplet solitons in the presence of spin transfer torque}
\begin{abstract}
We study noise-induced creation and annihilation of magnetic droplet solitons in experimental parameter regions in which they are linearly stable against drift. Exploiting the rotational symmetry of the problem, we transform to the reference frame rotating with the droplet soliton and introduce an effective potential energy that accounts for the work done against spin-transfer torque to rotate the magnetization between two different orientations. We use this function to compute the activation barrier in both directions between the uniform magnetization state and the droplet soliton state for a variety of nanocontact radii and currents. We investigate droplet soliton structures with both zero and nonzero spin-torque asymmetry parameter. Our approach can be applied to estimate activation barriers for dynamical systems where non-gradient terms can be absorbed by changes of reference frames, and suggests a technique applicable to extended systems that may not be uniformly magnetized.  
\end{abstract}
\date{\today}
\author{Gabriel D.~Chaves-O'Flynn}
\affiliation{Institute of Molecular Physics, Polish Academy of Sciences, ul. Mariana Smoluchowskiego 17,
60-179 Pozna\'n, Poland}
\author{D.L.~Stein}
\affiliation{Department of Physics and Courant Institute of Mathematical Sciences,
New York University, New York, NY 10012 USA}
\affiliation{NYU-ECNU Institutes of Physics and Mathematical Sciences at NYU Shanghai,
3663 Zhongshan Road North, Shanghai, 200062, China}
\affiliation{Santa Fe Institute, 1399 Hyde Park Rd., Santa Fe, NM USA 87501}
\maketitle
\section{Introduction}
\label{sec:intro}
Magnetic droplet solitons are localized, dynamical magnetic textures that preserve their shape on timescales long compared to typical magnon relaxation times~\cite{kosevich_magnetic_1990}. They are typically generated in spin torque oscillators (STOs~\cite{chen_spin-torque_2016}) with a free layer and having perpendicular magnetic anisotropy~\cite{MBD14}, or in nanoconstrictions using pure spin currents generated by the spin Hall effect~\cite{divinskiy_magnetic_2017}. Unlike certain other textures, such as dynamical skyrmions~\cite{zhou_dynamically_2015}, droplet solitons are not topologically protected and are sustained in driven systems via a balance between the competing effects of ordinary dissipation and an energy input provided by an external current acting through spin transfer torque (STT). Droplet solitons, dynamical skyrmions~\cite{zhou_dynamically_2015}, and related textures, both topological and nontopological, have attracted substantial interest due both to their fundamental physical interest and also from the possibility of their applications in information storage, transfer, and manipulation.

Droplet solitons were first predicted by Ivanov and Kosevich~\cite{ivanov_bound_1977}
as ``magnon droplets'' in thin films formed by condensation of spin
waves into a circular, uniformly precessing structure whose magnetization
at the center points in the (nearly) opposite direction to that of the spins far away. In this formulation the deviation of the magnetization from
the ``up'' direction (i.e., perpendicular to the plane) at infinity
is proportional to the number of quantized spin waves excited. Ivanov
and Kosevich restricted their analysis to conservative systems, i.e., no dissipation or spin torque; the resulting droplet structures were argued to be dynamically stable~\cite{Makhankov78}. However, once dissipation --- unavoidable in any real system --- is added, these structures quickly decay. For a review, see Kosevich
\textit{et al.\/}~\cite{kosevich_magnetic_1990}.

After the first experimental evidence~\cite{mohseni_spin_2013} and direct observation~\cite{Backes15} of droplet solitons, multiple experiments have generated
and studied magnetic droplet solitons and related structures over the past several years~\cite{Hang2018,Chung18,Sul18,iacocca_confined_2014,xiao_merging_2016,xiao_parametric_2017,mohseni_magnetic_2018,jiang_using_2018,jiang_impact_2018,
statuto_multiple_2019,mohseni_propagating_2020,mohseni_chiral_2020}. The usual experimental
situation leading to droplet soliton generation and decay occurs in a nanocontact
spin torque oscillator (NC-STO), where two magnetic layers are separated
by a nonmagnetic spacer; the top layer is known as the free layer
and the bottom layer as the fixed (or polarizing) layer. To generate
droplet solitons, a nanocontact (typically with radius of order 100~nm, although
it can vary) through which a current can be sent is placed on the free layer. This type of setup generates a spin transfer torque~(STT)~\cite{Slonczewski96,Berger96} to balance the effects of dissipation within the nanocontact region, thereby stabilizing the droplet soliton.  An external magnetic field,
usually perpendicular to the layers, is also applied, leading to the
usual Larmor precession of spins and additional stabilizing
effects~\cite{BH13}.  

Hoefer, Silva and Keller~\cite{hoefer_theory_2010} were the first to show theoretically that magnetic droplet solitons can be formed in such nanocontacts.  Using both analytical and numerical micromagnetic techniques, they studied droplet soliton shape
profiles, small perturbations of the droplet soliton, nucleation processes,
and related properties. One of their important conclusions was that
the externally applied sustaining current singled out a specific droplet soliton
precession frequency. 

In a subsequent paper, Wills, Iacocca and Hoefer~\cite{WIH16} (see also~\cite{MH19}) 
found that droplet solitons are linearly unstable at large bias currents, being
subject to a drift instability in which the droplet soliton center drifts outside of the nanocontact region. (This had been noticed a year earlier in micromagnetic simulations by Lendinez~{\it et al.\/}~\cite{lendinez_observation_2015}.)  Outside this region, dissipation is uncompensated because of the absence of STT, and the droplet soliton quickly decays. However, they also found that the droplet soliton \textit{is}
linearly stable in a narrow but experimentally accessible region
of parameter space; see Fig.~2 of~\cite{WIH16}. This figure shows three regions in the external field/applied current phase diagram for two nanocontact radii; the central region is where the droplet soliton is both sustained by the applied current and is linearly stable against drift. 

This raises the question of lifetimes of linearly stable droplet solitons in this central region at low temperature: for a droplet soliton in this part of parameter space and centered in the nanocontact region, what are the extreme value statistics of a rare but large thermal (or other) fluctuation spontaneously arising and destroying the droplet soliton? This is important both for experimental studies of droplet solitons and possible future device applications, and this is the question we address in this paper. 

In recent work, Moore and Hoefer~\cite{MH19} computed decay of magnetic droplet solitons in the linearly stable region through a different mechanism, namely {\it thermally activated\/} ejection of the intact droplet soliton from the nanocontact region. In the region of parameter space in which the droplet soliton is linearly stable against drift, the decay mechanism studied here competes with the ejection mechanism, and which one dominates droplet soliton decay will depend on the experimental parameters. We will return to this question in the Discussion section.

The standard framework for analyzing activation over an energy (or
more generally, action) barrier is provided by Kramers' theory~\cite{HTB90}
in which the transition rate $\tau^{-1}$ between two states separated
by an energy barrier $U$ follows to leading order an Arrhenius law
$\tau^{-1}\sim\exp[-U/k_{B}T]$ when $k_{B}T\ll U$. In driven or
otherwise nonequilibrium systems, such a barrier may be more difficult to properly define. In these cases, it is useful to employ a general path-integral approach to large deviations due to Wentzell and Freidlin~\cite{FW12}, which rests on finding the most probable path in state space between two locally stable configurations. This approach is especially useful in dealing with nonequilibrium situations~\cite{MS83,MS93} and has been
applied to thermally induced magnetic reversal under a variety of
circumstances~\cite{Brown63,Braun93,ERvE03,MSK06,PKS16}. We will also utilize this approach in this
paper.

The strategy employed in this paper consists of exploiting the rotational
symmetries in the problem so that Kramers' theory of reversal rates
can be applied in the presence of spin-transfer torques. This is done in two
steps. First, we introduce a pseudopotential~\cite{GT86}
to account for the energy required to reorient the magnetization along
a specific direction against the spin torque. The functional derivative
of this term introduces an additional field, leading to an extra term 
in the Landau-Lifshitz-Gilbert equation that can be cancelled by including
an extra term with nonzero curl. In this way a new dynamic equation
is obtained that is phenomenologically equivalent to the Landau-Lifshitz
equation.

The second step requires transforming to a rotating frame where the polar
axis is oriented along the fixed layer polarization. In this new frame
the nonzero-curl term vanishes and the resulting simplified Landau-Lifshitz-Gilbert equation
can be studied using Kramers' theory. We can then show that some important
features are satisfied: in the rotating frame the effective energy
decreases over time, and most importantly, the new set of equations
leads to an equilibrium distribution for the Fokker-Planck equation.
Satisfying these two requirements provides physical justification for the activation barriers calculated here.  

The plan of the paper is as follows. In Sects.~\ref{sec:setup}-\ref{sec:conservative} we describe the setup of the problem, introduce the magnetization dynamics, and discuss key aspects of the theory of conservative droplet solitons. We then describe in Sect.~\ref{sec:stt} the effects of spin-transfer torque: we introduce the spin-torque pseudopotential, transform to a rotating reference frame, investigate the time evolution of the energy in this frame, and find the magnetization configurations corresponding to energy minima and saddle states. In Sect.~\ref{sec:rotsols} we describe in detail the profiles of stationary configurations used to measure activation barriers. In Sect.~\ref{sec:overdampeddynamics}, we examine the dynamical evolution of overdamped droplet solitons and find the droplet soliton profiles that are saddles, constituting transition states for droplet soliton creation and annihilation. The consequences of spatial dependence of the rate of rotation of the droplet soliton are discussed in Sect.~\ref{sec:smallnu}, where we show that the transition states can still be found for a nonzero spin-torque asymmetry parameter. In Sect.~\ref{sec:FW-connection} we calculate the Freidlin-Wentzell action for a chosen fluctuational trajectory that is traced in the class of shape-preserving precessional configurations and show that the action along that specific path equals double the pseudo-potential energy introduced here, as would occur if the deterministic component of the dynamics were a simple gradient flow. Finally, in Sect.~\ref{sec:discussion} we summarize our results, discuss the regions of parameter space where the theory breaks down, and briefly consider extensions of the theory and future work.

For completeness, the Appendix provides additional derivations of many of the results used in the paper. Some of these can be found elsewhere in the literature, but they have been rederived here using notation consistent with the main text of this paper. 

\section{Setup of the problem}

\label{sec:setup}

We will study the situation of a spin torque oscillator, as described
in the Introduction, in which the free layer is a thin circular slab
of radius~$\rho_{\mathrm{max}}$ and thickness $d$, with $d\ll\rho_{{\rm max}}$,
and the uniform fixed layer magnetization is given by ${\bf m}_{f}=(\sin\theta_{f},0,\cos\theta_{f})$.

In the absence of currents and the resulting spin torques, the magnetization
dynamics is described by the Landau-Lifshitz-Gilbert equation which can be
presented in the following form. We introduce the linear operator
$\mathbb{L}$ acting on normalized fields and torques 
\begin{equation}
\mathbb{L}\equiv-\gamma'M_s\left[\mathbf{m}\times+\alpha\mathbf{m}\times\mathbf{m}\times\right]\,,\label{eq:LLop}
\end{equation}
where $\mathbf{m}$ is the unit magnetization vector $\left(\mathbf{m}=\mathbf{M}/M_{s}\right)$
in the free layer, $M_{s}$ is the saturation magnetization, $\alpha$
is the damping coefficient, $\gamma'=\frac{\gamma_{0}}{1+\alpha^{2}}$
and $\gamma_{0}=2.211\times10^{5}\mathrm{\frac{m}{A\cdot s}}$ \cite{m._j._donahue_oommf_1999}. We hereafter confine ourselves to the constraint $|{\bf m}|\equiv1$.

In this context, a ``field'' is understood as a variational derivative
of an energy density $\mathcal{E}$ and is normalized by $M_s$ (i.e., $\mathbf{h}_{\mathrm{eff}}=-\frac{1}{\mu_{0}M_{s}^{2}}\frac{\delta\mathcal{E}}{\delta\mathbf{m}}$
for large samples, or $\mathbf{h}_{\mathrm{eff}}=-\frac{\nabla_{\mathbf{m}}\mathcal{E}}{\mu_{0}M_{s}^{2}}$
for a macrospin); and a ``torque'' is any term that contributes
to the time rate of change of magnetic moment per unit volume. The
effective field~${\mathbf{h}}_{{\rm eff}}$ is derived in Appendix~\ref{app:bcs} (cf.~Eq.~(\ref{eq:heff2})) and is separately given by~\cite{hoefer_theory_2010}
\begin{equation}
{\mathbf{h}}_{{\rm eff}}=\nabla^{2}{\bf m}+{\bf h}_{Z}+{\bf h}_{{\rm oe}}+(Q-1)\left(\mathbf{m}\cdot\mathbf{n_{\perp}}\right){\bf {n_{\perp}}}\label{eq:heff}\, .
\end{equation}

\begin{figure}
\centering \includegraphics[width=3in]{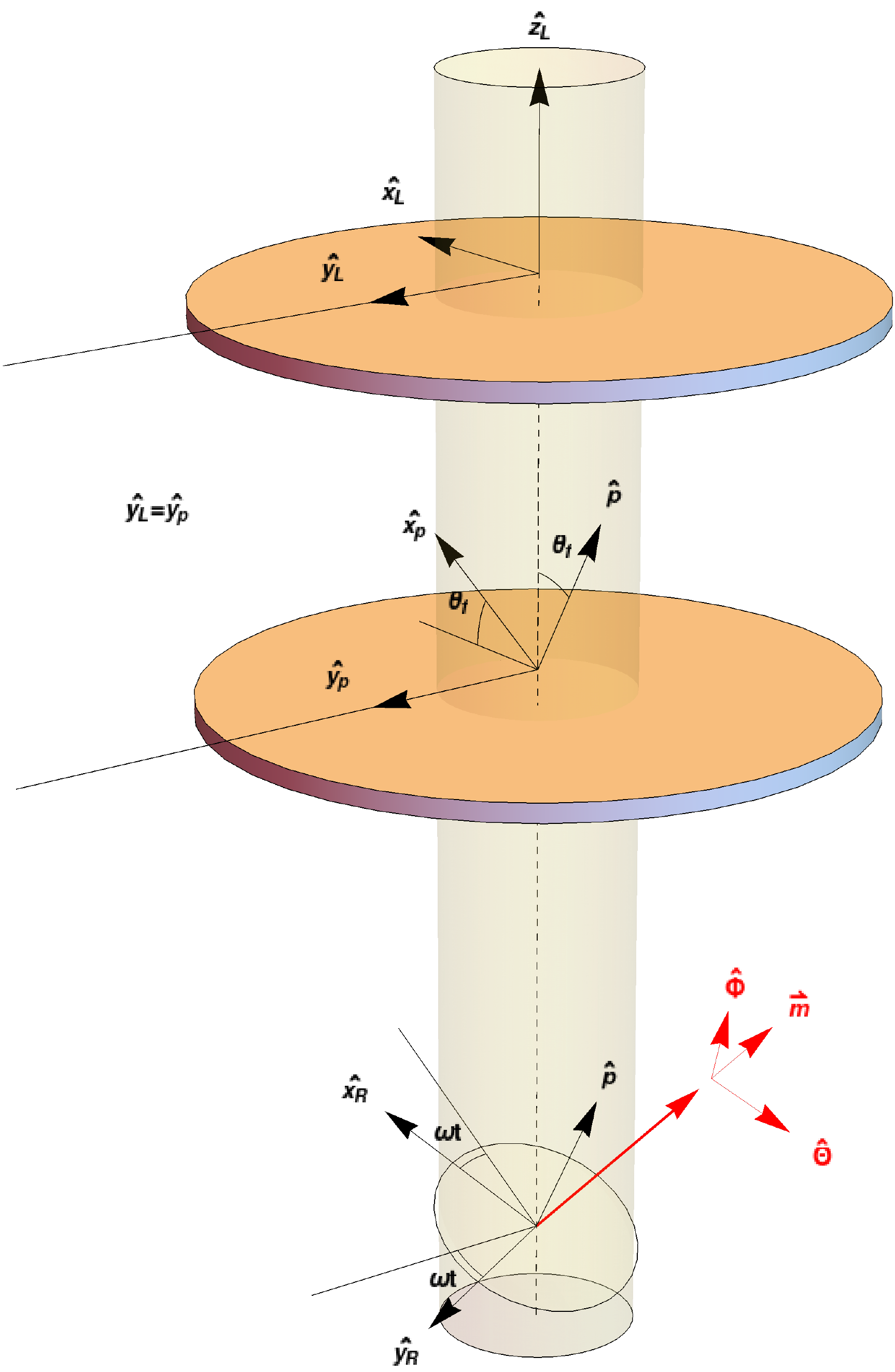} \caption{Geometry of setup and coordinate systems. Current flows upward from
layer. The three coordinate systems of relevance are: (I) the laboratory
frame $\{\mathbf{\hat{x_{L}}},\mathbf{\hat{y_{L}}},\mathbf{\hat{z}_{L}}\}$
where $\mathbf{\hat{z}_{L}=\mathbf{n_{\perp}}}$ is the normal to
the film plane, (II) the fixed layer frame $\{\mathbf{\hat{x_{P}}},\mathbf{\hat{y_{P}}},\mathbf{\hat m}_\mathrm{p}\}$
where the polarization vector $\mathbf{\hat m}_\mathrm{p}$ makes an angle $\theta_{f}$
with $\mathbf{n}$, and (III) the rotating frame $\{\mathbf{\hat{x_{R}}},\mathbf{\hat{y_{R}}},\mathbf{\hat m}_\mathrm{p}\}$,
which rotates about $\mathbf{\hat m}_\mathrm{p}$ with frequency $\omega$. Additionally,
the set of eigenvectors of the operator $\mathbb{L}^{T}\mathbb{L}$
are shown in red; these correspond to the set $\{\text{\ensuremath{\mathbf{m}}},\hat{\Theta},\hat{\Phi}\}$.}
\label{fig:setup}
\end{figure}

The first term on the Right Hand Side (RHS) of~(\ref{eq:heff}) corresponds to the 
exchange energy, $\mathcal{E}_{ex}$, associated with spatially varying
magnetization; the second term is an externally applied static field
${\bf h}_{Z}$ associated
with the Zeeman energy $\mathcal{E}_{Z}=-\mathbf{m}\cdot\mathbf{h}_{Z}$;
the third term corresponds to the Oersted field induced by a current (if present),
and the final term is derived from the perpendicular magnetic anisotropy
energy $\mathcal{E}_{K}$, consisting of two parts: ${\bf Q}=Q{\bf {\mathbf{n_{\perp}}}}$
is the dimensionless crystalline anisotropy field (with $Q>1$, cf.~Appendix~A),
and $-\left(\mathbf{m}\cdot\mathbf{\mathbf{n_{\perp}}}\right)\mathbf{n_{\perp}}$
is the shape anisotropy (equivalently, demagnetizing field) for a
thin film. In this equation all fields are normalized by the saturation
magnetization magnitude $M_{s}$ and space is normalized by the exchange
length~$l_{{\rm ex}}=\sqrt{2 A/\mu_{0}M^2_{s}}$ where
$A$ is the exchange constant and $\mu_{0}$ is the permeability of
free space. The geometry of the setup appears in Fig.~\ref{fig:setup}.

Using this operator the zero-temperature magnetization dynamics can
be written in the compact form~\citep{garcia-palacios_langevin-dynamics_1998,berkov_magnetization_2007}:
\begin{align}
\mathbf{\dot{m}} & =\mathbb{L}\mathbf{h}_{\mathrm{eff}}\,.\label{eq:LLG}
\end{align}
To account for thermal effects in a region of volume $d\mathcal{V}=l^2_\mathrm{ex}d$
an additional noise term $\sqrt{2\eta}\mathbf{\dot{W}}$ can be added
to the effective field~\cite{garcia-palacios_langevin-dynamics_1998,berkov_magnetization_2007}. Here, $\mathbf{\dot{W}}$ is a three-dimensional
normally distributed stochastic white noise process and the noise
strength \emph{$\eta$} is given by $\eta=\frac{\alpha k_{B}T}{2 A d\gamma_0 M_{S}}$.

If the system above is now driven by an STT-inducing current~$I$
with contact radius $\rho^{*}$, an additional nonconservative term
must be added to~(\ref{eq:LLG}). This results in the Landau-Lifshitz-Gilbert-Slonczewksi
(LLGS) equation~\citep{bertotti_nonlinear_2006}
\begin{equation}
\mathbf{\dot{m}}=\mathbb{L}\mathbf{h_{\mathrm{eff}}}-\frac{\sigma \gamma_0 M_s V(\mathbf{r})}{1+\alpha^{2}}\left[\frac{\alpha\mathbf{m}\times\mathbf{\hat m}_\mathrm{p}-\mathbf{m}\times\mathbf{m}\times\mathbf{\hat m}_\mathrm{p}}{1+\nu\mathbf{m}\cdot\mathbf{\hat m}_\mathrm{p}}\right]
\label{eq:LLGS}\, ,
\end{equation}
where $\mathbf{\hat m}_\mathrm{p}$ is the magnetization direction of the fixed layer,
$\nu$ is the spin-torque asymmetry parameter (with $0\le\nu<1$),
$V(\mathbf{r})$ describes the spatial distribution of the (cylindrically
symmetric) current, and 
$\sigma=J/J_{0}$, with
$J_{0}=\frac{\mu_0 e d M_s^2}{\hbar}$
the magnitude of the reduced current density~\cite{hoefer_theory_2010,WIH16}. Eq.~(\ref{eq:LLGS})
is the starting point of the analysis in this paper. 

\section{Magnetization equations of motion}

\label{sec:eom}

Following~\cite{hoefer_theory_2010}, we begin by studying the high-symmetry case
where the free layer lies in the $xy$~plane, ${\bf h}_{{\rm oe}}=0$
and ${\bf h}_{Z}=h_{Z}\mathbf{\hat{z}_{L}}$.  
(This and the following section are
mostly a review of the procedures followed and results obtained in~\cite{hoefer_theory_2010}.)

Given that $|m|=1$, it is useful to parametrize ${\bf m}$ in terms of fixed spherical coordinates 
\begin{equation}
{\bf m}=(\cos\Phi\sin\Theta,\sin\Phi\sin\Theta,\cos\Theta)\label{eq:spherical}
\end{equation}
and to write the field as $\mathbf{h}_{\mathrm{tot}}=h_{\Phi}\hat{\Phi}+h_{\Theta}\hat{\Theta}+h_{m}\mathbf{\mathbf{m}}$,
where the unit vectors are shown in Fig.~\ref{fig:setup}.

Inserting~(\ref{eq:spherical}) into the LLGS equation~(\ref{eq:LLGS})
yields Partial Differential Equations (PDE's) for $\Theta$ and $\Phi$, which are given in~\cite{hoefer_theory_2010} for $\alpha\rightarrow 0$.
Because they will be used below, we reproduce them here for arbitrary $\alpha$ (for easy
comparison, we use the same notation as~\cite{hoefer_theory_2010}): 
\begin{equation}
\frac{\dot{\Theta}}{\gamma' M_s}=F-\alpha\left(G-\frac{\sigma}{\alpha}VP_{\Theta}\right)+\alpha \sigma V P_\Phi\,,\label{eq:Theta}
\end{equation}
and 
\begin{equation}
\sin\Theta\frac{\dot{\Phi}}{\gamma' M_s }=G+\alpha\left(F+\frac{\sigma}{\alpha}VP_{\Phi}\right)+\alpha \sigma V P_\Theta\,,\label{eq:Phi}
\end{equation}
where $V=V(\rho)$ and $F$, $G$, $P_\Phi$, and $P_\Theta$ are functions of $\Phi$ and $\Theta$ and are given by
\begin{equation}
F[\Theta,\Phi]=\sin\Theta\nabla^{2}\Phi+2\cos\Theta{\mathbf{\nabla}}\Phi\cdot{\mathbf{\nabla}}\Theta\,,\label{eq:F}
\end{equation}
\begin{equation}
G[\Theta,\Phi]=-\nabla^{2}\Theta+\frac{1}{2}\sin(2\Theta)\Bigl(|\nabla\Phi|^{2}+Q-1\Bigr)+h_{Z}\sin\Theta\,,\label{eq:G}
\end{equation}

\begin{equation}
P_{\Theta}(\Theta,\Phi)=\frac{-\cos\Theta\cos\Phi\sin\theta_{f}+\sin\Theta\cos\theta_{f}}{1+\nu(\cos\Phi\sin\Theta\sin\theta_{f}+\cos\Theta\cos\theta_{f})}\,,\label{eq:ptheta}
\end{equation}

\begin{equation}
P_{\Phi}(\Theta,\Phi)=\frac{\sin\Phi\sin\theta_{f}}{1+\nu(\cos\Phi\sin\Theta\sin\theta_{f}+\cos\Theta\cos\theta_{f})}\,.\label{eq:pphi}
\end{equation}

Following Hoefer~\textit{et al.\/}~\cite{hoefer_theory_2010}, we rescale length,
time, current, and field parameter to remove the explicit dependence
on $\gamma'$ and $Q$ (recall $Q>1$): 
\begin{equation}
\rho=\rho'/\sqrt{Q-1}\,\,\,\,\,\,\,\,\,\,\,\,\,\,\,\,\,\,\,\,\,\gamma' M_s t=t'/(Q-1)\label{eq:rescale}
\end{equation}
\vskip -.2in 
\[
\sigma=(Q-1)\sigma'\,\,\,\,\,\,\,\,\,\,\,\,\,\,\,h_{Z}=(Q-1)h_{Z}'\,\,\,\,\,\,\,\,\,\,\,\,\,\,\,\,\,\,\,\rho_{{\rm max}}=\rho_{{\rm max}}'/\sqrt{Q-1}\,.
\] Finally, our boundary conditions stipulate that 
\begin{equation}
\Theta(\rho',\phi,t')=0\,\,\,{\rm at}\,\,\,\rho'=\rho'_{{\rm max}}\,\,\,\,\,;\,\,\,\,\frac{\partial\Theta}{\partial\rho'}=0\,\,\,{\rm at}\,\,\,\rho'=0\,.\label{eq:bcs}
\end{equation}
That is, the magnetization is uniformly in the $+{\bf {\mathbf{\hat{z}_{L}}}}$-direction
outside the nanocontact. (Our boundary condition differs from that
of~\cite{hoefer_theory_2010} only in that the latter take $\rho'_{{\rm max}}\to\infty$.)
The boundary condition at $\rho'=0$ ensures regularity of solutions.
A general treatment of boundary conditions is given in Appendix~\ref{app:bcs}.

\section{Conservative droplet soliton profiles}
\label{sec:conservative}

We consider first the case of a conservative system, i.e., $\alpha=\sigma'=0$.
The droplet soliton solution in this case is given by Hoefer~\textit{et
al.\/}~\cite{hoefer_theory_2010}; we briefly review it here. Although this situation
is unphysical, the solution serves as a starting point for consideration
of realistic situations.

The simplest (and most relevant) solution is one where the magnetization
precesses uniformly, i.e., $\Phi(\rho',\phi,t')=\Phi(t')$. In this
case, Eq.~(\ref{eq:F}) gives $F[\theta,\Phi]=0$, and consequently
Eq.~(\ref{eq:Theta}) indicates that $\Theta$ is independent of
time. Introducing $\omega_{0}$ as a precessional frequency parameter,
we can write 
\begin{equation}
\Phi(t')=(\omega_{0}+h_{Z}')t'\label{eq:precession}
\end{equation}
which, using Eq.~(\ref{eq:Phi}), results in a differential equation
for $\Theta=\Theta_{0}(\rho';\omega_{0})$: 
\begin{equation}
\Bigl(\frac{d^{2}}{d\rho'^{2}}+\frac{1}{\rho'}\frac{d}{d\rho'}\Bigr)\Theta_{0}-\frac{1}{2}\sin(2\Theta_{0})+\omega_{0}\sin\Theta_{0}=0\,.\label{eq:profile}
\end{equation}
This equation plus the boundary conditions~(\ref{eq:bcs}) determine
the conservative droplet soliton profile~$\Theta(\rho';\omega_{0})$.

Before proceeding, we note two features of these equations. First,
because $0<\omega_{0}<1$~\cite{hoefer_theory_2010}, the droplet soliton precession frequency~$\omega_{0}+h_{Z}'$
varies between the Zeeman frequency~$h_{Z}'$ and the FMR frequency~$1+h_{Z}'$.
This fact plus Eqs.~(\ref{eq:bcs}) and~(\ref{eq:profile}) show
that $m_{z}(\rho'=0;\omega_{0})>-1$ independently of $\omega_{0}$.

Second, the frequency~$\omega_{0}$ appears as a free parameter in
the solution; it can take any value between~0 and~1 in a conservative
system (if it existed). This is an important difference with realistic
situations in which STT counterbalances dissipation; there the droplet soliton
precessional frequency is determined by the strength of the applied
current~\cite{hoefer_theory_2010}. We return to nonconservative systems later.
For now, we simply note that droplet soliton profiles $\Theta(\rho';\omega_{0})$
are parametrized by $\omega_{0}$.

\begin{figure}
\includegraphics[angle=-90,width=3.5in]{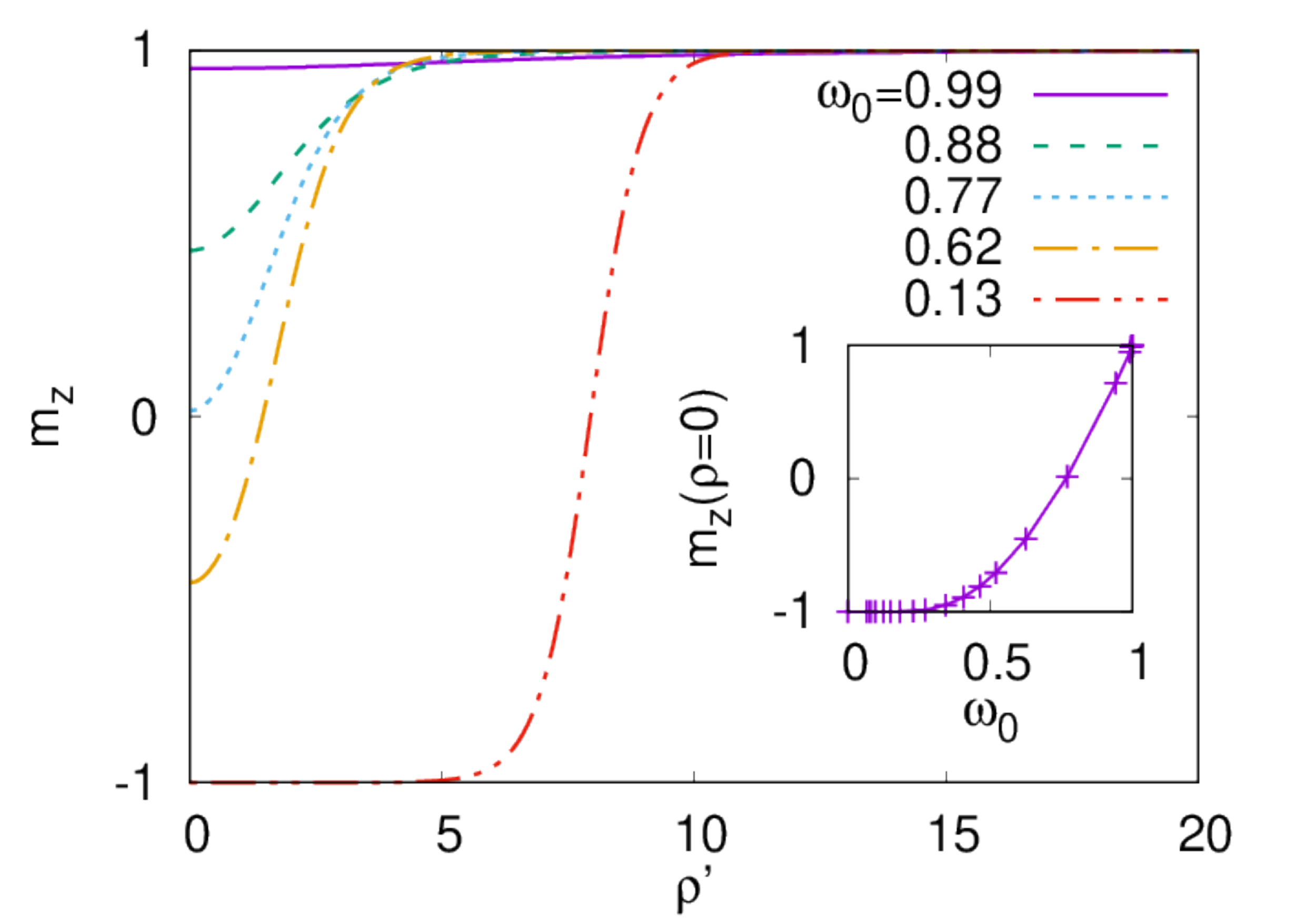}
\caption{\label{fig:figcp} Conservative droplet soliton profiles for different $\omega_{0}$. The amplitude
of each droplet soliton at the origin $m_{z}(\rho=0)$ is given next to the
frequency. The inset shows the values of the amplitude of the droplet soliton
for different shape parameters $\omega_{0}$.
}
\end{figure}
Eq.~(\ref{eq:profile}) must be solved using numerical means~\cite{chaves-oflynn_sup}. Fig.~\ref{fig:figcp}
shows several solutions for different~$\omega_{0}$ (conservative
droplet soliton profiles are also given in Fig.~2 of~\cite{hoefer_theory_2010}). 

We remark on several features of the conservative droplet soliton profiles
shown in Fig.~\ref{fig:figcp}. We note first that
$m_{z}(0;\omega_{0})$ is a monotonically increasing function of~$\omega_{0}$;
in particular, $m_{z}(0;\omega_{0})\to-1$ as $\omega_{0}\to0$ and
$m_{z}(0;\omega_{0})\to+1$ as $\omega_{0}\to1$. (The latter is unsurprising;
as $\omega_{0}\to1$ the droplet soliton becomes the uniform state precessing
at the FMR frequency.) We also note that, except for $\omega_{0}$
close to~1, the region in which $\Theta_{0}$ varies significantly
with~$\rho'$ is relatively narrow.

Even though these soliton droplet profiles correspond to an unphysical
situation (i.e., absence of dissipation), they will become important
in the subsequent discussion where $\sigma\ne0$ and $\alpha\ne 0$. In particular we will be focusing on solutions of Eqs.~(\ref{eq:Theta}) and (\ref{eq:Phi}) that preserve shape (i.e. $\dot \Theta=0$) and have uniform precessional frequency ($\frac{ d\Phi}{dt'}=\omega$). It can be easily verified that for the highly symmetric case $P_\Phi=0$, these are equivalent conditions: $(1+\alpha^2) G[\Theta,\Phi]=\omega\sin \Theta$ and $F[\Theta,\Phi]=\alpha G[\Theta,\Phi]-\sigma V P_\Theta$.  

\section{Effects of Spin Transfer Torque}

\label{sec:stt}

In order to account for the effects of spin  transfer torque,
we introduce a term in the action corresponding to an energy density
of a pseudopotential: 
\begin{equation}
\mathcal{E}_{\mathrm{ST}}=\begin{cases}
\mu_{0}M_{s}^{2}\frac{\sigma V(\rho)}{\alpha}\mathbf{m}\cdot\mathbf{\hat m}_\mathrm{p} & \nu=0\\
\mu_{0}M_{s}^{2}\frac{\sigma V(\rho)}{\alpha}\frac{\ln\left[1+\nu\mathbf{m}\cdot\mathbf{\hat m}_\mathrm{p}\right]}{\nu} & \nu\ne0\,.
\end{cases}\label{eq:spintorquepseudopotential}
\end{equation}

This can be used to derive an effective field on the system arising
from the spin torque: 
\begin{equation}
\mathbf{h}_{\mathrm{ST}}=-\frac{1}{\mu_{0}M_{s}^{2}}\frac{\delta\mathcal{E}_{\mathrm{ST}}}{\delta\mathbf{m}}=-\frac{\sigma V(\rho)}{\alpha}\frac{\mathbf{\hat m}_\mathrm{p}}{1+\nu\mathbf{m}\cdot\mathbf{\hat m}_\mathrm{p}}.\label{eq:spintorquefield}
\end{equation}

The total field is now given by $\mathbf{h}_{\mathrm{tot}}=\mathbf{h}_{\mathrm{eff}}+\mathbf{h}_{\mathrm{th}}+\mathbf{h}_{\mathrm{ST}}$,
and the equations for magnetization dynamics, now including spin polarized
currents and thermal fluctuations, become: 
\begin{align}
\mathbf{\dot{m}} & =\mathbb{L}\mathbf{h_{\mathrm{tot}}}-\gamma_0 M_s\frac{\sigma V(\rho)}{\alpha}\left[\frac{\mathbf{m}\times\mathbf{\hat m}_\mathrm{p}}{1+\nu\mathbf{m}\cdot\mathbf{\hat m}_\mathrm{p}}\right]\label{eq:crossLLGS} \\
 & =\mathbb{L}\mathbf{h_{\mathrm{tot}}}-\frac{\gamma_0 }{\mu_0 M_s}\nabla_{\mathbf{m}}\times\left(\mathbf{m}\mathcal{E}_{\mathrm{ST}}\right)\,.\label{eq:curlLLGS}
\end{align}

Eq.~(\ref{eq:crossLLGS}) is a compact version of the stochastic~Landau-Lifshitz-Gilbert-Slonczewski
equation, and will be our starting point for further analysis. We
note in particular that the curl term is inversely proportional to
$\alpha$, and so becomes important for small~$\alpha$. Notice that, in contrast with the factor of $\gamma'$ used in the definition of $\mathbb{L}$, the curl term uses $\gamma_0$ instead. 

With the addition of $\mathcal{E}_{ST}$, the pseudoenergy 
becomes $\mathcal{E}_{\mathrm{tot}}=\mathcal{E}_{ex}+\mathcal{E}_{K}+\mathcal{E}_{Z}+\mathcal{E}_{\mathrm{ST}}$;
where the terms correspond, respectively, to: exchange, effective
anisotropy, Zeeman, and spin-torque pseudopotential. When convenient, we will use the reduced version of these energies $\mathcal{E'}=\frac{\mathcal{E}}{\mu_0M_s^2}$.

\subsection{Transformation to a rotating reference frame}

\label{subsec:rotframe}
The spin torque term on the RHS of~(\ref{eq:curlLLGS}) cannot
be algebraically manipulated into a form that is the gradient of a
smooth potential. However, in the limit of low asymmetry parameter~$\nu\rightarrow0$
and uniform current density, it provides useful information on how
the time derivatives of the various vectors transform between a fixed
reference frame and a rotating reference frame. We now provide this
expression explicitly.

Consider a reference frame uniformly rotating with frequency $\omega$
about a fixed axis $\hat{\Omega}$. The transformation of the time
derivatives of a vector $\mathbf{u}$ between the fixed and the rotating
frames is given by: 
\begin{equation}
\mathbf{\dot{u}}=\mathbf{\dot{\tilde{u}}}+\omega\hat{\Omega}\times\mathbf{u},\label{eq:timederivativeofvectorsintherotatingframe}
\end{equation} Where $\mathbf{\dot{u}}$ and $\mathbf{\dot{\tilde{u}}}$ denote the vector in the fixed and rotating frame, respectively.

Eq.~(\ref{eq:crossLLGS}) describes the dynamics of the magnetization
in a fixed reference frame. It will be useful to transform to a reference
frame that rotates about $\mathbf{\hat m}_\mathrm{p}$ with frequency $\omega_{\mathrm ST}=\frac{\sigma\gamma_0 M_s V(\rho)}{\alpha}$.
In this frame the magnetization evolves as: 
\begin{equation}
\mathbf{\dot{\tilde{m}}}_{\mathrm{rotating\ frame}}=\mathbb{L}\mathbf{\tilde{h}_{\mathrm{tot}}}.\label{eq:rotatingframe}
\end{equation}

If the energy functional is rotationally symmetric with respect to
$\mathbf{\hat m}_\mathrm{p}$, the fields in the rotating frame are time independent, and~(\ref{eq:rotatingframe}) describes an autonomous dynamical system.
The approach of Serpico~\textit{et al.\/}~\citep{serpico_nonlinear_2007},
can be used to show that the magnetization in this frame evolves toward
the nearest energy minimum. It will further become evident that in
the rotating frame the magnetization of a macrospin reaches thermal
equilibrium even when $\nu\ne0$.

To analyze droplet soliton behavior in the presence of STT, we insert~(\ref{eq:spherical})
into the LLGS equation~(\ref{eq:crossLLGS}) and rescale with~(\ref{eq:rescale}) to obtain equations
for $\Theta$ and $\Phi$: 
\begin{equation}
\dot{\Theta}=h_{\Phi}+\alpha h_{\Theta}-\frac{\sigma' V(\rho')}{\alpha'}P_{\Phi}\label{eq:Theta-1-1}
\end{equation}
and 
\begin{equation}
\sin\Theta\ \dot{\Phi}=-h_{\Theta}+\alpha h_{\Phi}+\frac{\sigma' V(\rho')}{\alpha'}P_{\Theta}\,\,,\label{eq:Phi-1-1}
\end{equation}
where $P_{\Phi}=-\frac{\mathbf{\hat m}_\mathrm{p}\cdot\hat{\Phi}}{1+\nu\mathbf{m}\cdot\mathbf{\hat m}_\mathrm{p}}$,
$P_{\Theta}=-\frac{\mathbf{\hat m}_\mathrm{p}\cdot\hat{\Theta}}{1+\nu\mathbf{m}\cdot\mathbf{\hat m}_\mathrm{p}}$,
$\ h_{\Phi}=\mathbf{h}_{\mathrm{tot}}\cdot\hat{\Phi}$, $h_{\Theta}=\mathbf{h}_{\mathrm{tot}}\cdot\hat{\Theta}$ and $\alpha'=\frac{\alpha}{1+\alpha^2}$.

Each component of the field can be separated into terms corresponding
to each type of energy density (e.g. $h_{\Theta}=h_{ex,\Theta}+h_{Z,\Theta}+h_{K,\Theta}+h_{ST,\Theta}$).
We provide explicit expressions for all field terms in Appendix \ref{sec:Explicity-expresions-for}.

As they are written above, the time evolution equations~(\ref{eq:Theta-1-1})
and~(\ref{eq:Phi-1-1}) are valid for any fixed coordinate system
regardless of the direction of the polar axis. If $\mathbf{\hat m}_\mathrm{p}$ is
chosen as the polar axis, the last term in~(\ref{eq:Theta-1-1})
vanishes, and $P_{\Theta}$ becomes independent of $\Phi$. Furthermore,
when passing to a reference frame rotating about $\mathbf{\hat m}_\mathrm{p}$ with
frequency $\frac{\sigma'}{\alpha'}$, an additional
term $\frac{\sigma'}{\alpha'}\sin\Theta$ must be subtracted from (\ref{eq:Phi-1-1}). In that frame $\Theta=\tilde{\Theta}$. Dropping the explicit references
to the dependencies on $\Theta$ and $\Phi$ we have, in the rotating frame: 
\begin{equation}
\dot{\Theta}=h_{\Phi}+\alpha h_{\Theta}\label{eq:thetamovingframe}
\end{equation}
and 
\begin{equation}
\sin\Theta\ \dot{\tilde{\Phi}}=-h_{\Theta}+\alpha h_{\Phi}+\frac{\sigma'}{\alpha'}\left[V P_{\Theta}-\sin\Theta\right]\,.\label{eq:phimovingframe}
\end{equation}
The bracket vanishes when $\nu=0$ and the current strength is uniform in the region occupied by the droplet soliton.

Eqs.~(\ref{eq:Theta-1-1}) and~(\ref{eq:Phi-1-1}) for the static
reference frame are closely related to those given in~\cite{hoefer_theory_2010}
(Eqs.~(3) and (4) of that paper) where $\mathbf{\hat m}_\mathrm{p}$ and $\mathbf{h}$ both
point perpendicular to the plane. In that specific case the following
relations hold: 
\begin{align}
F[\Theta,\Phi] & =h_{\Phi}-\frac{\sigma' V(\rho')}{\alpha}P_{\Phi}\label{eq:F-1}
\end{align}
and 
\begin{align}
G[\Theta,\Phi] & =\frac{\sigma' V(\rho')}{\alpha}P_{\Theta}-h_{\Theta}\,.\label{eq:G-1}
\end{align}

For magnetization configurations where the azimuthal angle is not uniform, the exchange interaction introduces
complications that will be examined perturbatively for small $\nu$~(Sec.~\ref{sec:smallnu}).
To understand the origin of these complications, we use (cf.~Eq.~(\ref{eq:spherical}))
spherical coordinates for the magnetization $\mathbf{m}$, where now
$\Theta$ is the polar angle that $\mathbf{m}$ makes with $\mathbf{\hat m}_\mathrm{p}$
and $\Phi$ is the azimuthal angle in the plane perpendicular to $\mathbf{\hat m}_\mathrm{p}$.
The rotational symmetry requirement indicates that for constant $\Phi$ 
the azimuthal component of the exchange field is zero: $\left(\mathrm{i.e.,\ }h_{\mathrm{ex},\Phi}=-\frac{1}{\sin\Theta}\frac{\delta\mathcal{E}_{ex}}{\delta\Phi}=0\right)$.
However, in a non-uniform configuration this no longer need be the
case: the azimuthal field depends on the spatial profile of $\Phi$.

Using the scaling Eq.~(\ref{eq:rescale}) the exchange induced azimuthal field is given by: 
\begin{align}
h_{\mathrm{ex},\Phi} & =-\frac{1}{\sin\Theta}\left[\cancelto{0}{\frac{\partial\mathcal{E}_{ex}}{\partial\Phi}}-\nabla\left(\frac{\partial\mathcal{E}_{ex}}{\partial\nabla\Phi}\right)\right]\\
 & =\left[\nabla^{2}\Phi\sin\Theta+\nabla\Phi\cdot\nabla\Theta\cos\Theta\right].
\end{align}

Because this expression is generally nonzero, the energy will change
as the magnetization rotates about $\mathbf{\hat m}_\mathrm{p}$. The case
of constant $\Phi$ is not valid when $\nu\ne0$ or $\rho^{*}\ne\infty$, as it changes with a $\Theta$-dependent rate obtained from the second term in Eq.~(\ref{eq:crossLLGS}):
\begin{equation}
-\omega_{\mathrm ST} \left[\frac{\mathbf{m}\times\mathbf{\hat m}_\mathrm{p}}{1+\nu\mathbf{m}\cdot\mathbf{\hat m}_\mathrm{p}}\right] \cdot \left[\frac{\hat{\Phi}}{\sin\Theta}\right]=\frac{\omega_{\mathrm ST}}{1+\nu\cos\Theta}.
\end{equation}

This would imply that in an extended magnet the value of $\Phi$ will lag in some regions compared to others. In general, the curl term in~(\ref{eq:curlLLGS})
will not be absorbed into a common rotating frame for the full magnet
and, unless certain conditions are satisfied \citep{wedemann_curl_2016,berry_curl_2016,berry_m._v._hamiltonian_2015}
the Gibbs distribution is no longer a stationary solution of the Fokker-Planck
equation for the micromagnetic system. 

In appendix \ref{sec:uniaxial-macrospin}, we discuss the $\Theta$-dependence of this precessional rate in the case of uniaxial macrospins. The rest of this paper is concerned with shape-preserving precessional states which may have, in addition to $\omega_{\mathrm ST}$, additional frequency shifts ${\omega^{(1)}}$ which must have the same value in any reference frame. 

A quantity that will appear repeatedly in the subsequent discussion is 
\begin{equation}
\Xi\equiv\mathbf{m}\cdot\left(\frac{\delta\mathcal{E}'_{\mathrm{tot}}}{\delta\mathbf{m}}\times\nabla_{\mathbf{m}}\mathcal{E}'_{\mathrm{ST}}\right)\,,
\end{equation}
which measures the degree of misalignment between fields obtained
from the total energy functional $\mathcal{E}_{\mathrm{tot}}$ and
from the spin torque pseudo-potential $\mathcal{E}_{\mathrm{ST}}$.
More generally, it is a measure of the time rate of energy change
for a magnetization configuration that appears stationary in the rotating
frame. We highlight the fact that for the rotationally symmetric systems
discussed here this reduces to 
\begin{equation}
\Xi\equiv h_{\mathrm{ex},\Phi}\cdot h_{\mathrm{ST},\Theta},\label{eq:xisymmetric}
\end{equation}
and becomes zero if $h_{\mathrm{ex},\Phi}=0$.

\subsection{Magnetization Dynamics in the Rotating Frame}

\label{subsec:rotdyn}

Moving to the rotating frame when $\nu=0$ has the advantage that
configurations which are critical points of the energy landscape ($h_{\Theta}=h_{\Phi}=0$)
also become stationary dynamical points $\left(\dot{\Theta}=\dot{\tilde{\Phi}}=0\right)$.
This allows us to identify the saddle states through which the magnetization
passes in switching from the basin of attraction of one local energy
minimum to that of another. In this section we examine the effects
on the various relevant physical quantities of moving to the rotating
frame.

To proceed it is useful to define a set of Euler rotations. We start
by selecting $\mathbf{\hat{y}}=\mathbf{\hat{p}}\times\mathbf{\hat{z_{L}}}$
as the axis of nodes, and define $\mathbf{\hat{x}}$ to complete a
right handed set of coordinates: $\mathbf{\hat{x}}=\mathbf{\hat{y}_{L}}\times\mathbf{\hat{z}}$.
Transformation of vectors from the laboratory frame to the stationary
frame aligned with $\mathbf{\hat m}_\mathrm{p}$ can be effected using the rotation
matrix 
\begin{equation}
\mathcal{R}_{P\leftarrow L}=\left(\begin{array}{ccc}
\cos\theta_{f} & 0 & \sin\theta_{f}\\
0 & 1 & 0\\
-\sin\theta_{f} & 0 & \cos\theta_{f}
\end{array}\right)\,.
\end{equation}

Similarly, transforming from the polarized frame to the rotating frame
can be done using the matrix 
\begin{equation}
\mathcal{R}_{R\leftarrow P}=\left(\begin{array}{ccc}
\cos\omega t & \sin\omega t & 0\\
-\sin\omega t & \cos\omega t & 0\\
0 & 0 & 1
\end{array}\right)\,.
\end{equation}

The combined transformation matrix, from the laboratory frame to the
rotating frame, is therefore 
\begin{align}
\mathcal{R}_{\omega t} & =\mathcal{R}_{R\leftarrow P}\mathcal{R}_{P\leftarrow L}\nonumber \\
 & =\left(\begin{array}{ccc}
\cos\omega t\cos\theta_{f} & \sin\omega t & \cos\omega t\sin\theta_{f}\\
-\sin\omega t\cos\theta_{f} & \cos\omega t & -\sin\omega t\sin\theta_{f}\\
-\sin\theta_{f} & 0 & \cos\theta_{f}
\end{array}\right)\,.
\end{align}
Notice that in this convention the angle $\theta_{f}$ is measured
from $\mathbf{\hat m}_\mathrm{p}$ to $\mathbf{\hat{z_{L}}}$. The reader is reminded
that, because the transformation is unitary, the inverse matrix $\mathcal{R}_{\omega t}^{-1}$
is simply the transpose $\mathcal{R}_{\omega t}^{T}$.

The exchange energy term $\nabla^{2}{\bf m}$ is invariant under rotation
of the coordinate system. For the remaining terms in the energy, the
corresponding fields transform and become time-dependent: 
\begin{equation}
\mathbf{h}_{\mathrm{rotating\ frame}}=\mathrm{\mathcal{R}_{\omega t}}\mathbf{h}_{\mathrm{lab\ frame}}\,.\label{eq:transformationofvectors}
\end{equation}

The Landau-Lifshitz operator $\mathbb{L}$ transforms as well: 
\begin{equation}
\mathbb{L}_{\mathrm{rotating\ frame}}=\mathrm{\mathcal{R}_{\omega t}}\mathbb{L}_{\mathrm{lab\ frame}}\mathrm{\mathcal{R}_{\omega t}^{-1}}.
\end{equation}

Combining Eqs.~(\ref{eq:timederivativeofvectorsintherotatingframe})
and~(\ref{eq:transformationofvectors}) allows us to determine how
the time derivatives of vectors transform on switching from the laboratory
to the rotating frame: 
\begin{equation}
\mathbf{\dot{u}}_{\mathrm{rotating\ frame}}=\mathrm{\dot{\mathcal{R}}_{\omega t}}\mathbf{u}_{\mathrm{lab\ frame}}+\mathcal{R}_{\omega t}\dot{\mathbf{u}}_{\mathrm{lab\ frame}}.
\end{equation}

We choose $t=0$ to be the time when the coordinate axes of the rotational
frame coincide with the polarizer frame. Then $\mathrm{\mathcal{R}}_{\omega t=0}=\mathcal{R}_{P\leftarrow L}$
and $\mathrm{\dot{\mathcal{R}}_{\omega t}}=\dot{\mathcal{R}}_{R\leftarrow P,t=0}\mathcal{R}_{P\leftarrow L}$,
and the time derivative matrix at $t=0$ can be written succintly
as the cross product 
\begin{equation}
\dot{\mathcal{R}}_{R\leftarrow P}=\omega\left(\begin{array}{ccc}
0 & 1 & 0\\
-1 & 0 & 0\\
0 & 0 & 0
\end{array}\right)=-\omega\mathbf{\hat m}_\mathrm{p}\times.
\end{equation}

Finally, the full transformation of time derivatives from the laboratory
frame into the rotating frame is: 
\begin{equation}
\mathbf{\dot{u}}_{\mathrm{rotating\ frame}}=-\omega\mathbf{\hat m}_\mathrm{p}\times\mathcal{R}_{P\leftarrow L}\mathbf{u}_{\mathrm{lab\ frame}}+\mathcal{R}_{P\leftarrow L}\dot{\mathbf{u}}_{\mathrm{lab\ frame}}.
\end{equation}

\medskip{}

We next examine how vectors that are constant in the laboratory frame
behave in the rotating frame. For small misalignments between $\mathbf{\hat m}_\mathrm{p}$
and $\mathbf{n}$, $\mathcal{R}_{P\leftarrow L}\approx\left(\mathbf{1}+\theta_{f}\mathbf{\hat{y}}\times\right)$
and so 
\begin{align}
\mathbf{\dot{u}}_{\mathrm{rotating\ frame}} & =-\omega\mathbf{\hat m}_\mathrm{p}\times\mathcal{R}_{P\leftarrow L}\mathbf{u}_{\mathrm{lab\ frame}}\label{eq:hrotatingframe}\\
 & \approx-\omega\mathbf{\hat m}_\mathrm{p}\times\left(\mathbf{1}+\theta_{f}\mathbf{\hat{y}_{L}}\times\right)\mathbf{u}_{\mathrm{lab\ frame}}\,.
\end{align}
Fields that are constant in the laboratory frame will
become time-dependent in the rotating frame if the RHS of~(\ref{eq:hrotatingframe})
is nonzero. However, the RHS does become zero in the
symmetric scenario in which $\theta_{f}=0$ and the external fields
are aligned with $\mathbf{\hat m}_\mathrm{p}$. (This is the case considered by Hoefer
\textit{et al.\/}~\cite{hoefer_theory_2010}.)

We emphasize that the symmetry condition discussed above is somewhat
more restrictive than simple cylindrical symmetry. The rotational
symmetry must be satisfied point-by-point in the full sample; it is
not sufficient that the full problem alone is invariant under a global
rotation about a common axis. In particular, the Oersted field and
the edge-anisotropy fields do not satisfy this more restrictive symmetry
requirement. In the rotational frame centered at an arbitrary point
away from the axis of the disk, both fields appear to be circularly
polarized. As a result, this will introduce self-oscillations in the
rotating frame that will be reflected as quasi-periodicities in the
laboratory frame. The proper treatment of this quasiperiodic scenario
is beyond the scope of this work; we restrict ourselves here to the
study of reversal in the high symmetry case. However, we anticipate
that the concepts discussed below for cases in which $\theta_{f}\ne0$
will be of importance for the proper treatment of the Oersted field.

The pseudoenergy defined in Appendix~A is, after rescaling (Eq.~(\ref{eq:rescale}):
\begin{equation}
\mathcal{E'}_{\mathrm{tot}}=\left|\nabla'\mathbf{m}\right|^{2}-\left(\mathbf{m}\cdot\hat{\mathbf{z}}_{L}\right)^{2}-\mathbf{m}\cdot\mathbf{h'}_{0}+\mathcal{E'}_{\mathrm{ST}}\,.
\end{equation}
This quantity is a scalar and will therefore be invariant under rotational
transformations. The first and last terms on the RHS are unaffected
by changes in reference frame, but the second (magnetostatic) and
third (Zeeman) terms are specified for fields that are constant in
the laboratory frame. As we move to the rotating frame their energy
densities become harmonically dependent on time. The rate of change
of those quantities is given by: 
\begin{equation}
\frac{d}{dt}\left[\left(\mathbf{m}\cdot\hat{\mathbf{z}}_{L}\right)^{2}\right]=2\left(\mathbf{m}\cdot\hat{\mathbf{z}}_{L}\right)\left[\dot{\mathbf{m}}\cdot\hat{\mathbf{z}}_{L}+\mathbf{m}\cdot\dot{\mathbf{z}_{L}}\right]
\end{equation}
and 
\begin{equation}
\frac{d}{dt}\left[\mathbf{m}\cdot\mathbf{h}_{0}\right]=\dot{\mathbf{m}}\cdot\mathbf{h}_{0}+\mathbf{m}\cdot\dot{\mathbf{h}}_{0}.
\end{equation}
More generally, for anisotropy energies of polynomial type with respect
to specific directions $\hat{\mathbf{u}}$, i.e. $\mathcal{E'}_{K}=\sum_{n}K_{n}\left(\mathbf{m}\cdot\hat{\mathbf{u_{n}}}\right)^{n},$we
can decompose the
field $\mathbf{h}=\sum_{n}\mathbf{h}_{\mathrm{n}}$ into terms $\mathbf{h}_{\mathrm{n}}=-nK_{n}(\mathbf{m}\cdot\hat{\mathbf{u_{n}}})^{n-1}\hat{\mathbf{u_{n}}}$
and rewrite the energy as $\mathcal{E'}_{K}=-\sum_{n}\frac{\mathbf{m}\cdot\mathbf{h}_{n}}{n}$ so that
\begin{align}
\frac{d\mathcal{E}'_{K}}{dt} & =\sum_{n}nK_{n}\left(\mathbf{m}\cdot\hat{\mathbf{u_{n}}}\right)^{n-1}\left[\hat{\mathbf{u_{n}}}\cdot\mathbf{\dot{m}}+\dot{\mathbf{u_{n}}}\cdot\mathbf{m}\,\right]\\
 & =\mathbf{h}\cdot\mathbf{\dot{m}}+\sum_{n}nK_{n}\left(\mathbf{m}\cdot\hat{\mathbf{u_{n}}}\right)^{n-1}\dot{\mathbf{u_{n}}}\cdot\mathbf{m}\,.
\end{align}
The scalar product is unaffected when switching to the moving frame.
Since in the rotating frame $\dot{\mathbf{m}}=\mathbb{L}\mathbf{h}$,
and $\mathbf{\dot{u}}$ is given by (\ref{eq:hrotatingframe}), we
arrive finally at 
\begin{equation}
\frac{d\mathcal{E}'}{dt}\approx-\alpha\left|\mathbf{h}_{\mathrm{tot}}\right|^{2}-\omega\left[\mathbf{\hat m}_\mathrm{p}\times\left(\mathbf{1}+\theta_{f}\mathbf{\hat{y}_{L}}\times\right)\mathbf{h}_{\mathrm{lab\ frame}}\right]\cdot\mathbf{m}\,.
\end{equation}

It is convenient to keep in mind that $\mathbf{\hat m}_\mathrm{p}\times\mathbf{h}_{\mathrm{ST}}=0$,
so the cross product in the parenthesis will  require only the terms
for $\mathbf{h}_{\mathrm{eff}}$ given in~(\ref{eq:heff}). In the
symmetric case ($\theta_{f}=0$) we find: 
\begin{eqnarray}
\frac{d\mathcal{E}'_{\mathrm{tot}}}{dt} & = & -\alpha\left|\mathbf{h}_{\mathrm{tot}}\right|^{2}-\left(\omega\mathbf{\hat m}_\mathrm{p}\times\frac{\delta\mathcal{E'}}{\delta\mathbf{m}}_{\mathrm{lab}}\right)\cdot\mathbf{m}\nonumber \\
 & = & -\alpha\left|\mathbf{h}_{\mathrm{tot}}\right|^{2}-\left(\nabla_{\mathbf{m}}\mathcal{E'}_{\mathrm{ST}}\times\frac{\delta\mathcal{E'}}{\delta\mathbf{m}}_{\mathrm{lab}}\right)\cdot\mathbf{m}\nonumber \\
 & = & -\alpha\left|\mathbf{h}_{\mathrm{tot}}\right|^{2}+\Xi\,.
\end{eqnarray}

From this we see that, in the absence of $\Xi$, the system will evolve
toward lower energies. The term $\Xi$ is zero when the fields are
aligned with $\mathbf{\hat m}_\mathrm{p}$; if not, then its magnitude oscillates
as the system rotates about $\mathbf{\hat m}_\mathrm{p}$. Self-oscillations occur
if the total energy change is zero after a full periodic orbit: 
\begin{equation}
\Delta\mathcal{E'}=\ointclockwiseop\left(-\alpha\left|\mathbf{h}_{\mathrm{tot}}\right|^{2}+\Xi\right)dt=0.\label{eq:melnikov}
\end{equation}

A similar quantity, known as the Melnikov function~\cite{bertotti_nonlinear_2006},
quantifies the energy change as the magnetization performs one cycle
in the conservative dynamics $\left(\alpha=0\right)$. The calculation
of that quantity requires precise knowledge of the equal energy orbits
and is the basis of calculations based on the method of averaging~\cite{newhall_averaged_2013,pinna_thermally_2013}.
Eq.~(\ref{eq:melnikov}) can be used from either the stationary or
the rotating reference frames. At this point, the meaning of $\Xi$
becomes apparent: it is a spin torque-induced power density influx.
For a given magnetization orientation energy is dissipated by damping
at a rate $\alpha\left|\mathbf{h}_{\mathrm{tot}}\right|^{2}$ while
the spin torque term injects energy into the system; the curl term
in Eq.~(\ref{eq:curlLLGS}) is the cause of this power influx. Because
it is perpendicular to $\mathbf{h}_{\mathrm{ST}}$, the effective
spin torque energy cannot increase; rather, the energy will flow into the
other energy terms ($\mathcal{E}_{ex},\mathcal{E}_{K},\mathrm{\ or\ }\mathcal{E}_{Z}$).

Because the external fields considered here are aligned with $\mathbf{\hat m}_\mathrm{p}$,
they do not contribute to a change in the energy as the reference
frame rotates. However, the same cannot be said for the exchange field
since its direction depends on the instantaneous spatial distribution
of the magnetization. Nevertheless, $\Xi=0$ for the set of solutions
of the equation $F\left[\Theta,\Phi\right]=0$. If $\Phi$ is small,
then contributions of its non-uniformities can be neglected when calculating
$G\left[\Theta,\Phi\right]$.

Summarizing the above discussion: by moving to the rotating frame,
it is possible to find stationary solutions if external fields are
parallel to the fixed layer polarization; if not, then the solutions
display small self-oscillations. In the vicinity of these configurations,
the energy is guaranteed to decrease at every cycle. We postulate
(and will justify below) that these configurations either play the
role of transition barriers or else are metastable configurations.

\section{Stationary solutions in the rotating frame}

\label{sec:rotsols}

We are now in a position to discuss the stationary solutions of Eqs.~(\ref{eq:thetamovingframe})
and~(\ref{eq:phimovingframe}); as described above, we restrict ourselves
to the case where $\Phi$ is spatially uniform, so that $h_{\Phi}=0$.
In the rotating frame the bracket vanishes when $\nu=0$ and as
a consequence stationary solutions require that $h_{\Theta}=0$ as
well.

We will refer to this highly symmetric scenario in which $\theta_{Z}=\theta_{f}=\nu=0$
as the unperturbed case, with solutions $\Phi_{0}=0$ and $\Theta_{0}(\mathbf{r})$.
The equations then become
\begin{equation}
\begin{array}{cc}
h_{\Theta}^{(0)} & =h_{ex,\Theta}^{(0)}+h_{Z,\Theta}^{(0)}+h_{K,\Theta}^{(0)}+h_{ST,\Theta}^{(0)}\\
0 & =\nabla'^{2}\Theta_{0}-h_{Z}\sin\Theta_{0}-\frac{1}{2}\sin2\Theta_{0}+\frac{\sigma' V}{\alpha}\sin\Theta_{0}
\end{array}\label{eq:unperturbedthetafieldequation}
\end{equation}
and 
\begin{equation}
\begin{array}{cc}
h_{\Phi}^{(0)} & =h_{ex,\Phi}^{(0)}+h_{Z,\Phi}^{(0)}+h_{K,\Phi}^{(0)}+h_{ST,\Phi}^{(0)}\end{array}
\end{equation}
is trivially zero due to the uniformity of $\Phi_{0}$ and the rotational
symmetry of the setup. In cylindrical coordinates this becomes
\begin{equation}
\dot{\Theta_{0}}=\frac{\partial^{2}\Theta_{0}}{\partial \rho'^{2}}+\frac{1}{\rho'}\frac{\partial\Theta_{0}}{\partial \rho'}-\frac{1}{2}\sin\left(2\Theta_{0}\right)+\omega_{h}\sin\Theta_{0}=0\label{eq:stationarysolutions}
\end{equation}
and 
\begin{equation}
\omega_{h}=\left[\frac{\sigma' V(\rho')}{\alpha}-h'_{Z}\right]\,.\label{eq:shapeparameter}
\end{equation}
Eqs.~(\ref{eq:stationarysolutions}) and~(\ref{eq:shapeparameter})
are identical to Eqs.~(13)-(17) in~\cite{hoefer_theory_2010}. As in~\cite{hoefer_theory_2010},
we search for stationary solutions in which the spin-torque term counteracts
the damping. However, Hoefer~\textit{et al.\/}~\cite{hoefer_theory_2010} analyzed the case where
the dissipative terms were a small perturbation; we do not make this
assumption here. 

Moving to the rotating frame allows us to ignore the precession
of the system and gives us direct access to the slow dynamics in which
the solutions to~(\ref{eq:stationarysolutions}) correspond to
dynamic equilibrium configurations which could be stable or unstable
depending on the values of $\omega_{h}$ and current profile $\sigma' V$.  As in~\cite{hoefer_theory_2010}, we define the current profile using
the Heaviside step function $\mathcal{H}$, with a step at the nanocontact
radius $\rho^{*}$: $V(\rho)=\mathcal{H}(\rho^{*}-\rho)$. The nontrivial
solutions are identical to those found in~\cite{hoefer_theory_2010} for the case
$h_{Z}=0$, $\nu=0$ and $\rho^{*}=\rho_{\mathrm{max}}\rightarrow\infty$.
In~\cite{hoefer_theory_2010} these are found to be unstable to small
perturbations (see in particular Fig.~4 of that paper). In fact, our analysis
indicates that these unstable solutions play an important physical role in the current context: they are the saddle configurations
separating the two stable constant solutions $\Theta=0$ and $\Theta=\pi$.

For intermediate nanocontact radii and current strength we obtained equilibrium
configurations by performing a time integration of $\dot{\Theta}$
using conservative droplet soliton profiles as the initital condition~\cite{chaves-oflynn_sup}. Final
results are shown  in 
\begin{figure}
\includegraphics[angle=-90,width=3in]{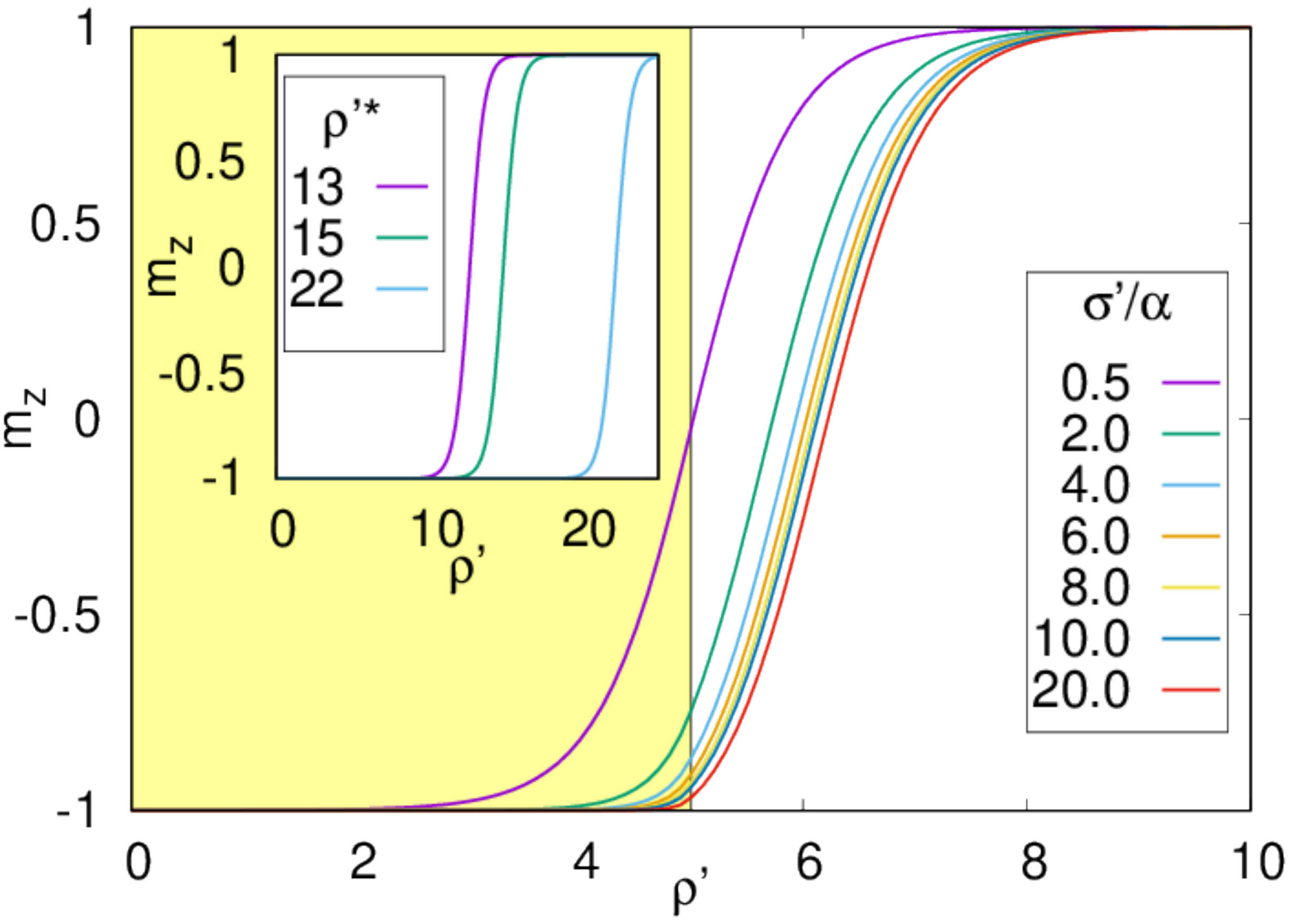}\caption{\label{fig:stableconfigurations}(Main figure) Stable droplet solitons for a variety of
currents $\sigma'/\alpha$ (with $Q=2$) for $\rho^{'*}=5$. In general, the transition
region between $m_{z}=-1$ and $m_{z}=+1$ is approximately aligned
with the nano-contact's edge. The asymmetry parameter $\nu=0$ in
these cases. The yellow region indicates the range of $\rho'$ within the nanoconact region. (Inset) $m_z$ vs.~$\rho'$ at $\sigma'/\alpha=0.13$ for several nanocontact radii.}
\end{figure}
Fig.~\ref{fig:stableconfigurations}. For low currents the magnetization
relaxes towards $\Theta=0$, but if the current magnitude passes a
certain threshold $\sigma_{\mathrm{crit}}$, then inside the nanocontact region the magnetization switches to $\Theta(\rho=0)\rightarrow\pi$. Once
the magnetization stops evolving, the domain wall separating the $\Theta=0$
orientation from the $\Theta=\pi$ orientation is in the vicinity
of the nanocontact's edge. The overall configuration inside the nanocontact corresponds to the stable droplet soliton for
a given radius and current strength. These stable states are related to the stationary low frequency droplet solitons of~\cite{BH13} which are described to be stable against small displacements from the center of the nanocontact and are described as circular domain walls. 

Notice that there will be an abrupt change in the precessional frequency at the nanocontact edge (since $\left. \omega_{ST}\right|_{\rho\rightarrow\rho*^-}=\sigma'/\alpha'$, but $\left. \omega_{ST}\right|_{\rho\rightarrow\rho*^+}=0$). However, since $\Theta(\rho<\rho*)\rightarrow\pi$ for very large currents, $\Phi$ becomes undefined. Additionally the $\Theta$ profile inside will merge nicely with the outside $\Theta$ profile of low-frequency stationary droplet solitons \cite{BH13} which are of the form $\cos(\Theta)=\tanh(\rho-1/\omega)$. The droplet soliton radii for the stable droplet solitons considered here will be larger than the radius of the nanocontact.

As we've already seen, given a spin-polarized current density, there
is a reference frame rotating at frequency $\omega_{ST}=\frac{\sigma'}{\alpha'}$
for which the unstable solutions of~(\ref{eq:stationarysolutions})
appear stationary. These are the saddle states between the configurations
parallel and antiparallel to $\mathbf{\hat m}_\mathrm{p}$. In the laboratory frame
they precess with frequency $\omega_{ST}$. This interpretation of Eq.~(\ref{eq:stationarysolutions}) allows us to define a critical current
above which reversal is guaranteed. That is, if we set $\rho^{*}=\infty$,
and consider a small, global fluctuation $\delta\Theta$ away from
$\Theta=0$, we find 
\begin{equation}
\dot{\Theta}\approx\left(-1+\left[\frac{\sigma'}{\alpha}\left(\frac{1}{1+\nu}\right)-h'_{Z}\right]\right)\text{\ensuremath{\delta\Theta}}>0,
\end{equation}
whenever 
\begin{equation}
\frac{\sigma'}{\alpha}>\left(\frac{\sigma'}{\alpha}\right)_{\mathrm{crit}}=\left(1+\nu\right)\left(h'_{Z}+1\right).
\end{equation}
For a finite nanocontact radius, the critical current must be larger
and its meaning needs to be be re-evaluated. For finite $\rho^{*}$ there
will be a current magnitude for which the solution $\Theta=0$ becomes
unstable, but the reversed region does not expand much further than
the nanocontact radius: doing so would severely increase
the exchange energy, by increasing the size of the reversed region
and consequently the size of the domain wall that surrounds such fluctuations.
As long as the domain wall is enclosed inside the nanocontact region
any exchange energy cost will be compensated by a corresponding decrease
in the value of the pseudo-potential energy $\mathcal{E}_{ST}$. This
stops when the droplet soliton radius becomes comparable to $\rho^{*}$: further
increases in the current density only serve to pin the magnetization
closer to $\Theta=\pi$ but do not produce further growth of
the droplet soliton radius.

In summary, if $\frac{\sigma'}{\alpha}>\left(\frac{\sigma'}{\alpha}\right)_\mathrm{crit}$, the solution $\Theta=0$ is no longer stable, and a droplet
soliton will be formed \textit{necessarily} as the energy-minimizing configuration. When $\frac{\sigma'}{\alpha}<\left(\frac{\sigma'}{\alpha}\right)_\mathrm{crit}$, the solution $\Theta=0$ is
stable, and two nonuniform solutions of~(\ref{eq:stationarysolutions})
exist: one is an energy minimum (stable droplet soliton), and the other is the saddle state (unstable droplet soliton) (these correspond to the stable and unstable branches shown in~Fig.~4 of~\cite{hoefer_theory_2010}). In the limit of infinite nanocontact
radius all droplet solitons are saddles and the uniform $\Theta=\pi$ configuration is the stable state. (It has been suggested that these saddle states appear similar to experimentally realized spin torque oscillator states observed
in~\cite{rippard_spin-transfer_2010} and~\cite{Mohseni11}.)

\subsection{\label{subsec:Numerical-solutions-for}Numerical solutions for the
stationarity condition.}

To find solutions of Eq.~(\ref{eq:stationarysolutions}) we use the
BVP4C method from Matlab, as was done in \cite{hoefer_theory_2010}.
The key difference is that we solve the equation for finite nanocontact
radii, so the values of $\sigma'/\alpha$ need to be larger. We set
$\rho'_{\mathrm{max}}=100$, fix the value of $\Theta$ at $\rho'=0$,
and let $\sigma'/\alpha$ be an eigenvalue parameter to be found numerically. We impose the following boundary conditions: 
\begin{equation}
\left.\frac{\partial\Theta}{\partial\rho}\right|_{\rho=0}=0\qquad\Theta\left(\rho=0\right)=\Theta_{\mathrm{max}}\qquad\Theta(\rho=\rho_{\mathrm{max}})=0.
\end{equation}
By performing this ``poor's man continuation'' we can feed the solutions
as initial guesses and solve for a wide range of parameters ~\cite{chaves-oflynn_sup}. Every
set of values $\nu$, $\rho^{*}$ and $\Theta_{\mathrm{max}}$ is
associated with a value of $\sigma'/\alpha$.

For comparison to the limit $\rho^{*}\rightarrow\infty$, we will
identify $\omega_{0}$ with the eigenvalue corresponding to the parameter set $\nu=0$, $\rho^{*}=\rho_{\mathrm{max}}$, and $\Theta=\Theta_{\mathrm{max}}$.
Keeping the same $\Theta_{\mathrm{max}}$ but varying $\nu$ and $\rho^{*}$
we find new values of $\sigma'/\alpha$ that have the same amplitude
at the origin. In this way, we can compare the frequency that a finite
nanocontact droplet soliton would have if it had the same amplitude $\Theta_{\mathrm{max}}$
of the conservative droplet soliton.

\begin{figure}
\includegraphics[angle=-90,width=3in]{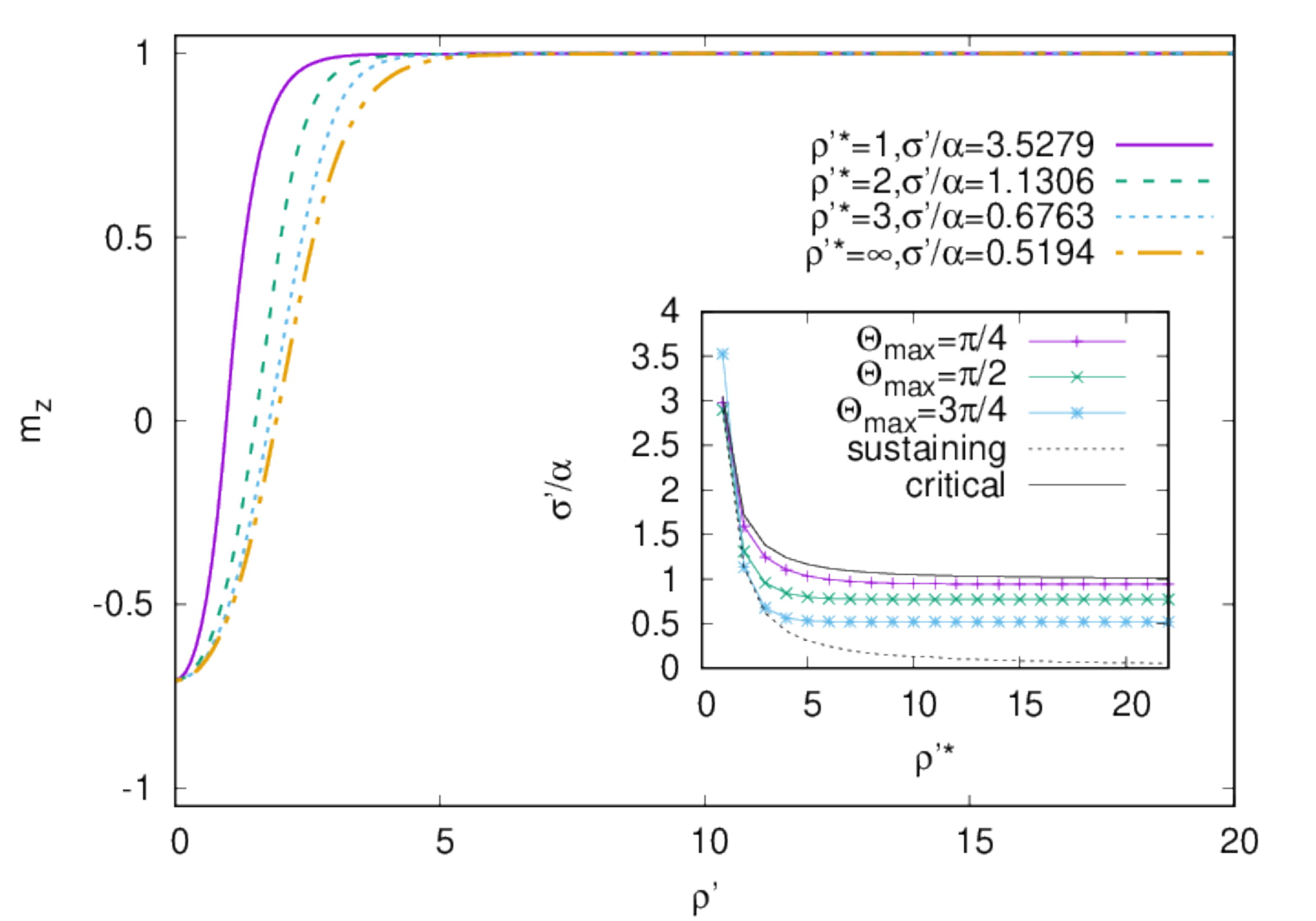}\caption{\label{fig:radialdependence} Profiles of stationary fluctuations
of amplitude $\Theta_{\mathrm{max}}=3\pi/4$ at different nanocontact
radii $\rho^{*}$ (Q=2). As indicated in the legend a reduction in $\rho^{*}$
must be accompanied by an increase on the current density. Inset,
currents required to maintain the fluctuation stationary. In these
curves, $\nu=0$. The area between the critical (solid black line) and the sustaining current (dotted black  line) provides the range of currents in the region of bi-stability. These curves for critical and sustaining currents are obtained from the data of Fig. \ref{fig:omegavssoa}.}
\end{figure}

In Fig.~\ref{fig:radialdependence} we show results of this procedure
for different values of $\rho^{*}$ and fixed $\Theta_{\mathrm{max}}=\frac{3}{4}\pi$.
As the nanocontact area is reduced, a larger current is required to keep
the fluctuations stationary. The profile is narrower in the radial
direction. Because very strong currents are necessary for small nanocontact
radii, the uniform $\Theta=0$ state ceases to be metastable and the
droplet soliton is sustained. For large radii, fluctuations of this amplitude
require smaller currents but they represent energy saddles.

\section{Overdamped dynamics and activation barriers}
\label{sec:overdampeddynamics}

In this section we perform two types of simulations to confirm that solutions of Eq.~(\ref{eq:stationarysolutions}) are indeed saddle configurations in the energy landscape. The first is a one-dimensional relaxation calculation which drops the precessional terms in Eqs.~(\ref{eq:Theta-1-1}) and (\ref{eq:Phi-1-1}); the second is a full micromagnetic simulation using OOMMF~\cite{m._j._donahue_oommf_1999} in which the spin torque term ${\mathcal E}_{\mathrm{ST}}$ is added to the energy ~\cite{chaves-oflynn_sup}. 

Both simulations approximate the same physical system, modeling a magnetic material with exchange constant $A=13\mathrm{pJ/m}$ and saturation magnetization $M_s=8\times10^5\mathrm{A/m}$. To simplify the algebra, we selected an unphysically large $K$ (so that $Q=2$), but our approach will remain valid for more realistic values. The nanocontact diameter used is $r^{*}=50$ nm ($\rho^{*'}=4.4$) and the thickness is $d=1$ nm. The critical current, $I_c=\pi J_c r*^2$, for these parameters is $2.40$ mA.

\subsection{\label{subsec:overdamped1D}1D relaxation}

In the rotating reference frame, the long-term evolution can be captured using only the damping terms in Eqs.~(\ref{eq:Theta-1-1}) and (\ref{eq:Phi-1-1}):
\begin{equation}
\dot{\Theta}=\alpha h_\Theta\qquad\dot{\Phi}=\alpha h_{\Phi}
\end{equation}
Here we assume uniform precessional states, i.e., $\Phi=\omega t$ and $\nabla\Phi$=0, which guarantee that $h_{\Phi}=0$ always. The time evolution of the droplet soliton profiles, $\Theta(\rho,t)$, contains enough information to verify whether a given configuration corresponds to a thermally activated transition state.

\begin{figure}
\includegraphics[angle=0,width=3.1in]{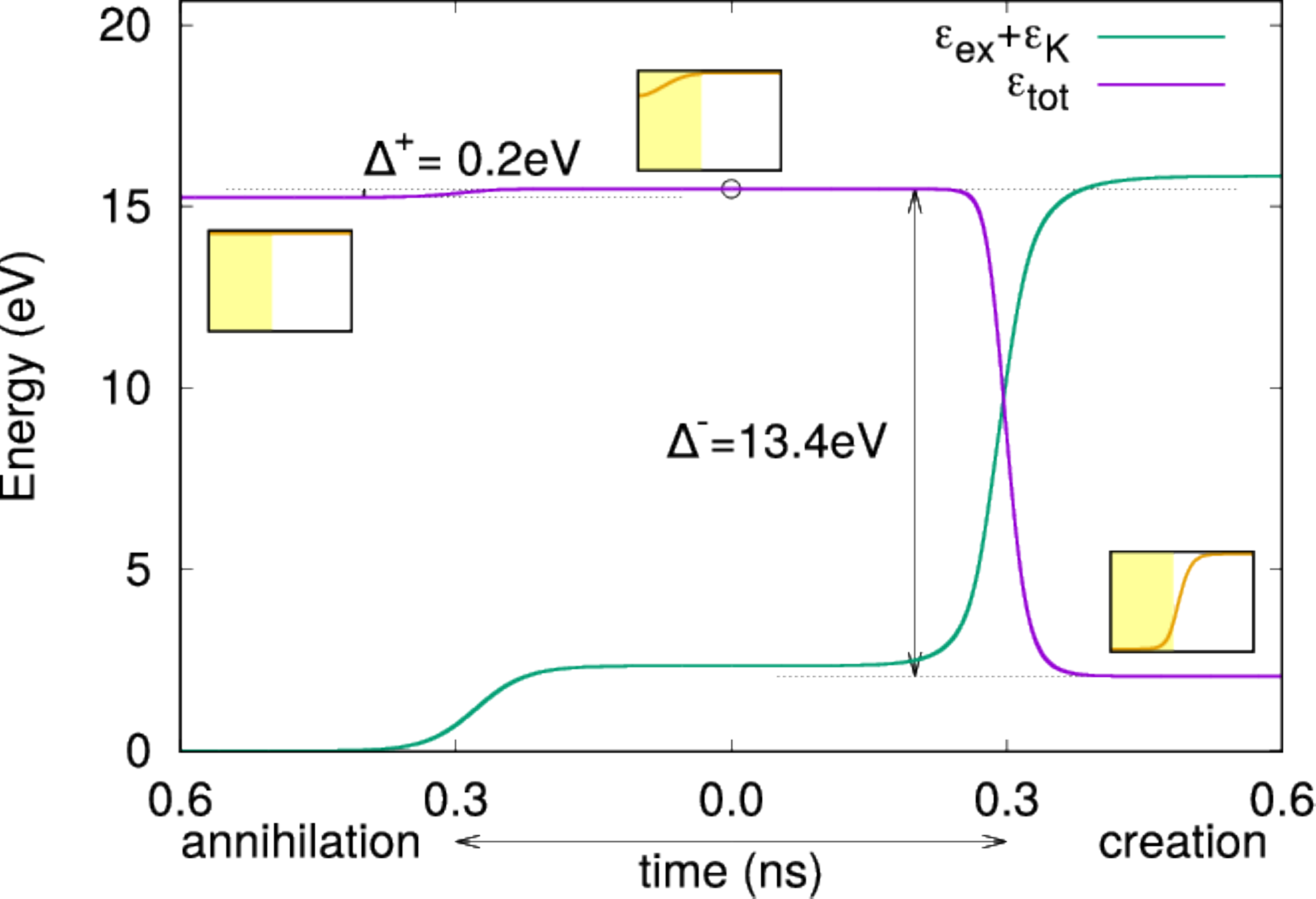}\caption{
\label{fig:overdamped1D}
Time evolution of the 1D overdamped dynamics. The initial state at time~0 is the droplet soliton saddle. Time increases from the center to the left for $\sigma^-$, corresponding to droplet annihilation, and from the center to the right for $\sigma^+$ corresponding to droplet creation. The spatial profiles, $m_z(\rho)$, of the final states are shown as in opposite margins and for the initial state in the center above the curve. The energy barriers for creation, $\Delta^+$, and annihilation, $\Delta^-$, are shown in the figure with vertical arrows. The light yellow shade represents the area covered by the nanocontact. In this figure the initial profile is defined with $\omega_h=0.9911$, $m_z(\rho=0)=0.52$. The two currents used for the time integration shown are $\sigma^-=0.992218$ and $\sigma^+=0.992249$. 
}
\end{figure}

Starting with solutions of Eq.~(\ref{eq:stationarysolutions}), we set $\alpha=1$ and integrate $\dot\Theta=h_\Theta$ forward in time. We vary $\sigma$ to find two values $\sigma^+$ and $\sigma^-$ that satisfy two conditions: first, $0<\sigma^+-\sigma^-<5\times10^{-5}$; and second, the initial profile relaxes to the uniform state for $\sigma^-$ and to the stable droplet soliton for $\sigma^+$. If the slope of the curve $\mathcal{E}_\mathrm{tot}(t)$ is close to zero at $t=0$, we take this as confirmation that the initial state constitutes the transition configuration for thermal activation~\cite{chaves-oflynn_sup}. 

Fig.~\ref{fig:overdamped1D} illustrates this procedure. The difference of $\mathcal{E}_{tot}$ values between the saddle droplet soliton and the uniform state~($\Delta^+$) constitutes the barrier for droplet soliton creation; the difference between the saddle droplet soliton and the stable droplet soliton ($\Delta^-$) constitutes the barrier for droplet soliton annihilation. Besides $\mathcal{E}_\mathrm{tot}$, Fig.~\ref{fig:overdamped1D} also shows the energy terms excluding the spin torque contribution ($\mathcal{E}_\mathrm{K}-\mathcal{E}_\mathrm{ex}=\mathcal{E}_\mathrm{tot}-\mathcal{E}_\mathrm{ST}$). It is clear from this figure that $\mathcal{E}_\mathrm{tot}$ is a better descriptor of thermal stability than the value without $\mathcal{E}_\mathrm{ST}$.

In subsection \ref{subsec:energybarriers1D}, we will repeat this procedure systematically to find the energy barriers for different applied currents and nanocontact radii. 

\subsection{\label{subsec:OOMMF_simulations}Overdamped micromagnetic simulations.}

We reproduce the qualitative features of the overdamped 1D dynamics with an equivalent system using OOMMF. To reduce computational time we drop the demagnetizing field term and set the crystalline anisotropy to $K_{\mathrm{eff}}=\frac{K}{2}=\frac{\mu_0 M_s^2}{2}$, which in turn sets $Q=2$. Because the precessional term cannot be dropped independently of the STT term, we allow for precession and set $\alpha=1$. The disk has a diameter of 500nm and we discretize the problem with 1nm cubic cells.

The extension for spin torque dynamics by Xiao~et~al.~\cite{xiao_boltzmann_2004} was modified so that $\mathcal{E}_\mathrm{ST}$ was included as an energy term~\cite{chaves-oflynn_sup}. The dynamical part of the code was left unchanged and is based on Eq.~(\ref{eq:LLS}). The secondary spin torque term, an additional curl-like term in the OOMMF extension, was set to zero. We set OOMMF's polarization and asymmetry to $P=2$ and  $\Lambda=1.0$, repectively. This corresponds to $\nu=0.0$ and fully efficient polarization.

We emphasize that our modification of the OOMMF code does not alter the dynamics; it is merely a bookkeeping change to incorporate the spin-torque pseudopotential into the energy. However, because the new term is relatively large we were forced to relax the standard restriction that energy changes remain below a predetermined threshold. 
As a consequence, traces of energy evolution in time shown below appear noisy.

\begin{figure}
\includegraphics[angle=0,width=3.1in]{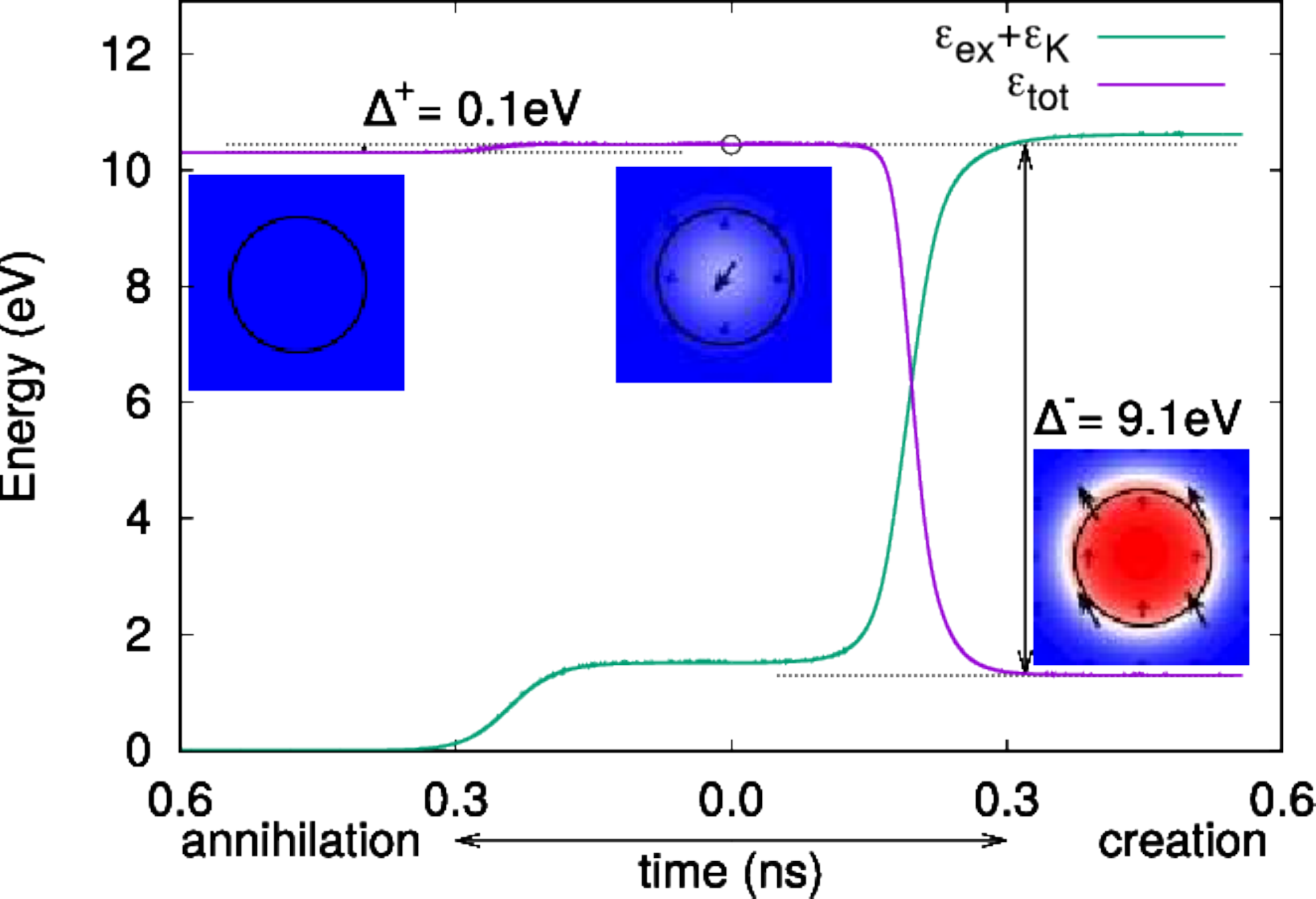}\caption{
\label{fig:OOMMFoverdamped2D}
Overdamped ($\alpha=1$) OOMMF dynamics. The initial state is the droplet soliton saddle. The time evolution for creation and annihilation is the same as in Fig.~\ref{fig:overdamped1D}. The spatial profiles, $m_z(x,y)$, of the final states are shown as in opposite margins and for the initial state in the center above the curve. The energy barriers for creation ($\Delta^+$) and annihilation ($\Delta^-$) are again depicted with vertical arrows. The area covered by the nanocontact is indicated by a white circle. In this figure the initial profile is defined with $\omega_h=0.9911$, $m_z(\rho=0)=0.52$. The two currents used for the time integration shown are $I^-=2.491$ mA and $I^+=2.492$ mA, so that $\sigma=1.04$. 
}
\end{figure}

\begin{figure*}
\includegraphics[angle=0,width=7in]{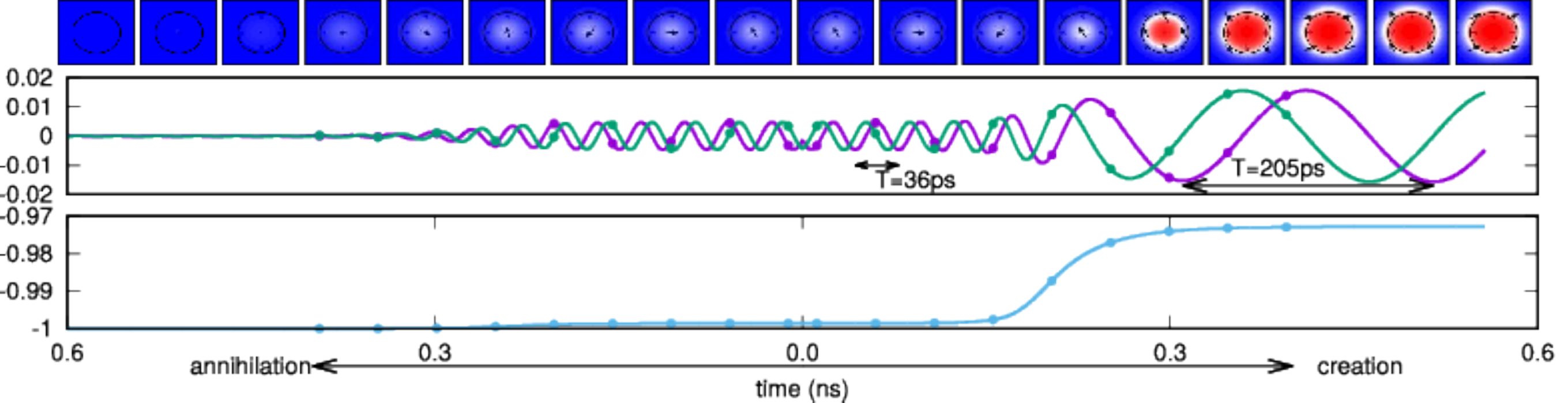}\caption{
\label{fig:oommfdynamics}
Magnetization configurations of the overdamped ($\alpha=1$) OOMMF dynamics, and spatially averaged magnetization components vs time. Both simulations start with the same initial configuration. As in the previous two figures, time evolution occurs to the left for $I^-=2.491$ mA and to the right for $I^+=2.492$ mA. Points in the curve are associated to the figures in the top row. Consistent with expectations, the low amplitude droplet soliton (saddle state) has a higher frequency than the large amplitude droplet soliton (stable droplet soliton). The precessional frequency of each configuration can be estimated visually from this plot after rescaling time $\omega=\frac{2\pi}{\gamma_0 M_s T}$ giving $\omega_{\mathrm{saddle}}=0.9964$ and $\omega_{\mathrm{stable}}=0.1735$.
}
\end{figure*}

The initial magnetization is a 2D realization of the saddle configuration used in Sect.~\ref{subsec:energybarriers1D}. Successive simulations find two currents, $I^+$ and $I^-$ with difference $\delta I=I^+-I^-=10^{-4}$; the system evolves towards the uniformly magnetized state $m_z=1$ for $I^-$ and to the stable droplet soliton for $I^+$. Results are shown in Figs.~\ref{fig:OOMMFoverdamped2D} and~\ref{fig:oommfdynamics}. These figures show qualitative agreement between the 1D model and the OOMMF micromagnetic simulations. In the latter  the initial state is a saddle for a slightly larger current, $\sigma=1.04$. 

\begin{figure}
\includegraphics[angle=-90,width=3in]{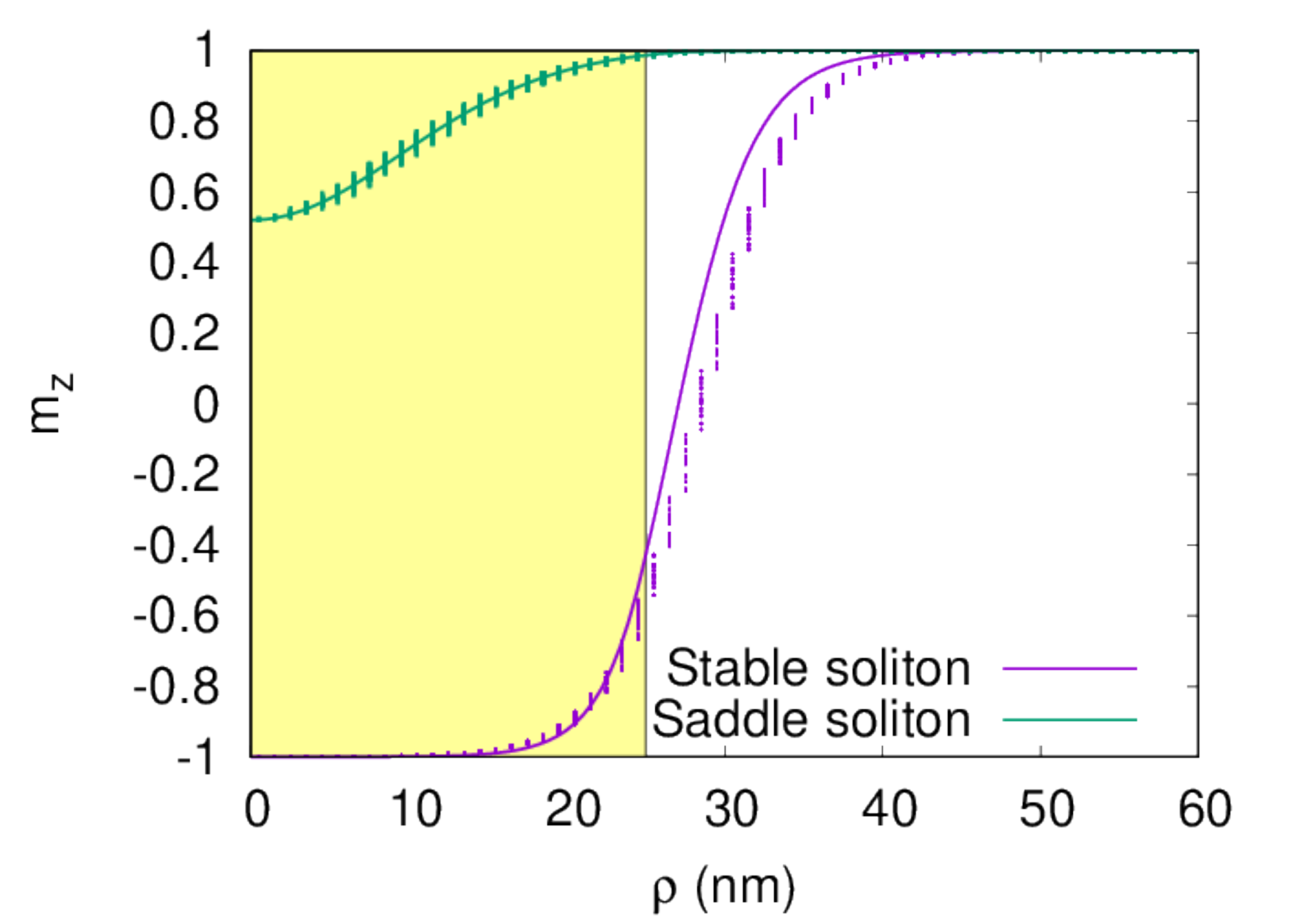}\caption{
\label{fig:overdampedprofiles}
Profiles of $\Theta(\rho)$ for overdamped micromagnetic simulations. Solid lines represent the droplet soliton profiles obtained from the overdamped 1D dynamics; the superimposed dots are obtained from OOMMF simulations. We attribute the small quantitative discrepancies partly to discretization effects and partly to the fact that exchange energy contributions from variations in the azimuthal are neglected in the 1D model.
}
\end{figure}

There are quantitative discrepancies between the two simulations that can be attributed to the fact that in the full micromagnetic simulations $\nabla \Phi$ may be nonzero. Furthermore, the slightly larger current of the OOMMF simulation widens the stable droplet soliton profile (see Fig.~\ref{fig:overdampedprofiles}), changing all energy terms.
\subsection{\label{subsec:energybarriers1D}Energy barriers in the 1D approximation.}
Here we use the 1D overdamped approach to obtain the energy barriers for the droplet soliton at different scaled currents $\sigma/\alpha$ and nanocontact radii $\rho^*$.

\begin{figure}
\includegraphics[angle=-90,width=3.5in]{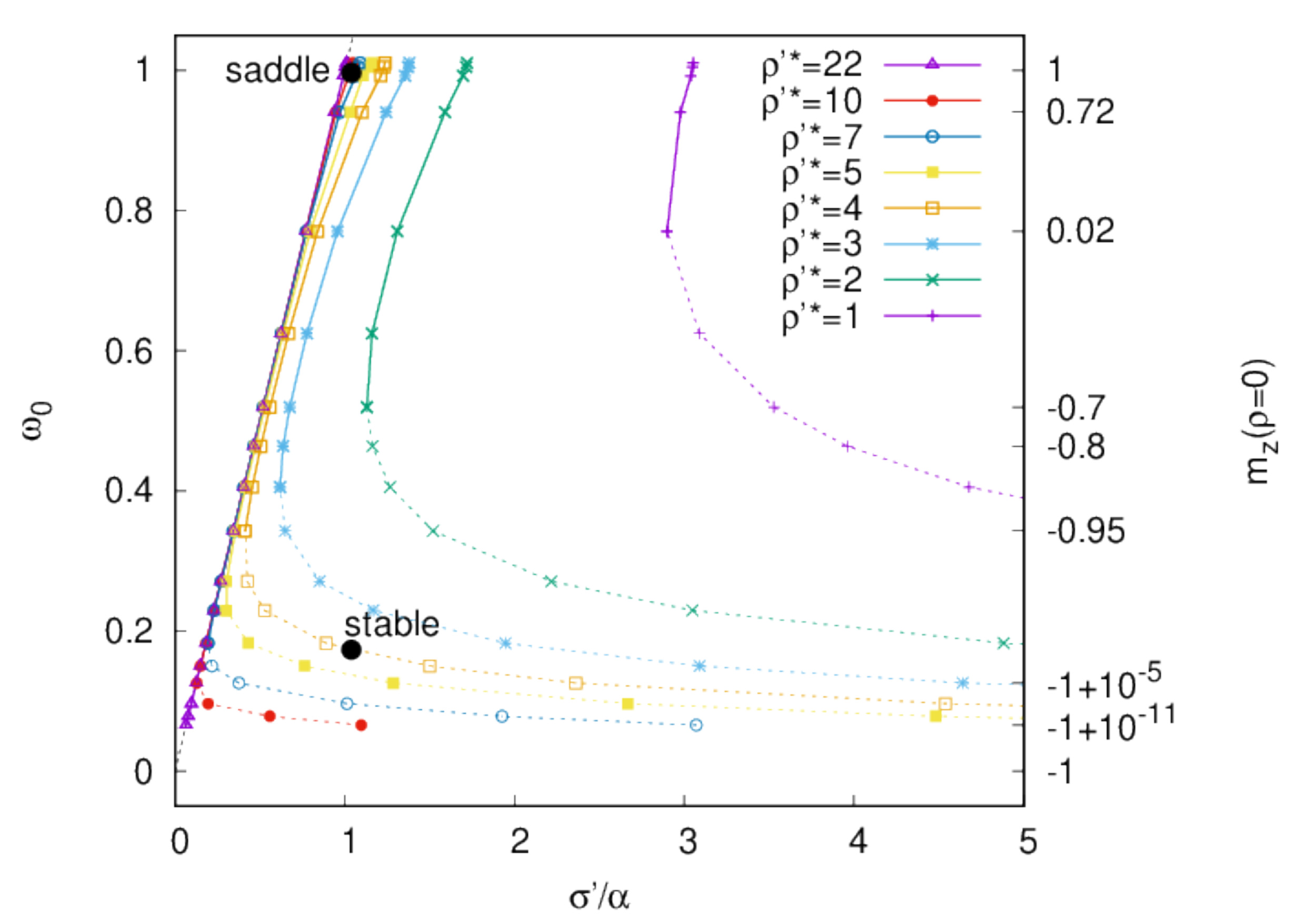}\caption{\label{fig:omegavssoa} Solutions of Eq.~(\ref{eq:stationarysolutions}) for various $m_z$ amplitudes at the origin: given $m_z(\rho=0)$, the algorithm finds $\sigma/\alpha$ as an eigenvalue of~(\ref{eq:stationarysolutions}). Along the left axis the frequency for the infinite nanocontact is shown; a dashed line of slope~one depicts the infinite nanocontact limit. The different finite-nanocontact curves superimpose over the infinite-nanocontact line already at moderate sizes. All curves are shown with two branches: the solid lines, where the slope is positive, represent the saddle states, and the dashed lines indicate the corresponding stable states. The critical current for a given nanocontact radius is the point at which the corresponding curve crosses the uniform state, i.e., where $m_z(\rho=0)=1$. The sustaining current for each radius corresponds to the vertex where the two branches meet. The filled black circles show the frequencies obtained from visual inspection of the OOMMF simulation; the points are expected to lie in the curve corresponding to $\rho^*=4.4$.}
\end{figure}

The first step is to find the values of $\sigma/\alpha$ for different droplet soliton amplitudes and nanocontact radii $\rho*$, which is done by solving Eq.~(\ref{eq:stationarysolutions}). The resulting values are shown in Fig.~\ref{fig:omegavssoa}. It can immediately be seen that as $\rho^*$ increases, the infinite nanocontact limit is rapidly approached. For the thermal stability calculations, the relevant segments are the solid parts of the curve; for those values the solutions represent saddle configurations which are then used as starting configurations for the overdamped simulations, as explained in Sect.~\ref{subsec:overdamped1D}.
 
\begin{figure*}
\includegraphics[angle=0,width=7.0in]{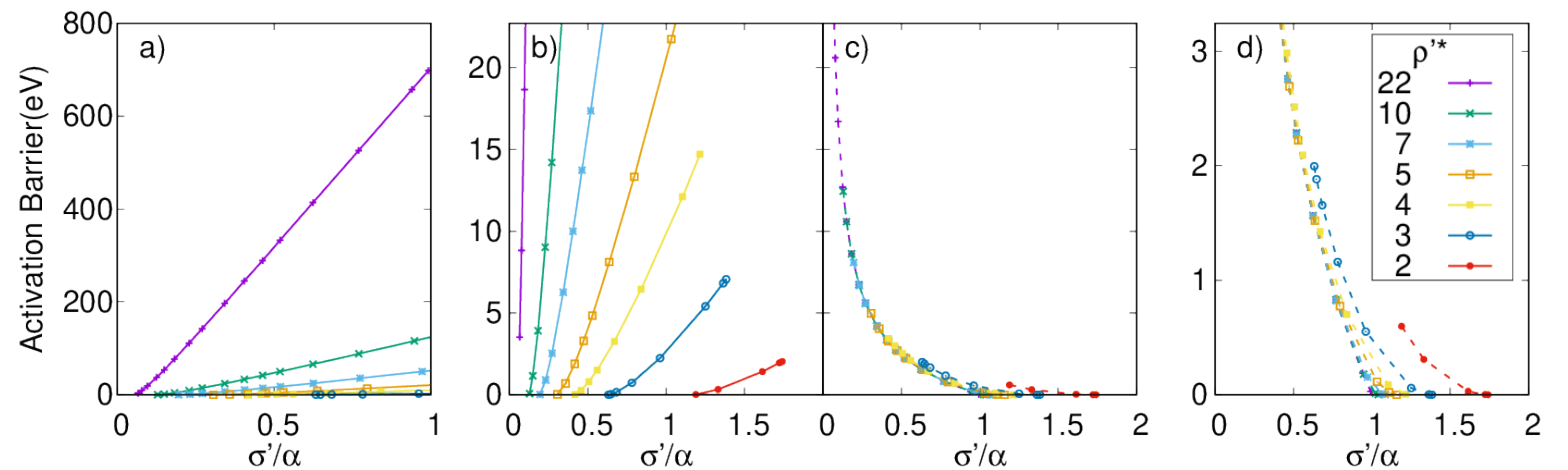}\caption{\label{fig:summaryofbarriers} Energy barrier dependence on current when $h=0$, $\sigma/\alpha$, calculated from the overdamped 1D dynamics with $h_Z=\nu=0$ and different nanocontact radii $\rho*$. (a,b) $\Delta^-$, barrier for droplet soliton annihilation. (c,d) $\Delta^+$, barrier for droplet soliton creation. As shown, the values of $\Delta^+$ are generally much smaller than $\Delta^-$. To better illustrate the range of currents for which the barriers $\Delta^+$ and $\Delta^-$, (b) and (c) are juxtaposed with the same axis scale.
}
\end{figure*}

After simulating all saddle points the energy barriers were measured. Results are shown in Fig.~\ref{fig:summaryofbarriers}. For large values of $\sigma/\alpha$, the annihilation barrier $\Delta^-$ is close to linear, reflecting the fact that once the nanocontact region is saturated the profile remains unchanged, and $\mathcal{E}_\mathrm{ST}$ is roughly $\omega_\mathrm{ST}$ multiplied by the nanocontact area. As $\sigma/\alpha$ is reduced (Fig.~\ref{fig:summaryofbarriers}b), this quasilinearity is lost for smaller radii. 

The creation barrier $\Delta^+$  (Figs.~\ref{fig:summaryofbarriers}c and \ref{fig:summaryofbarriers}d) has a singularity at $\sigma/\alpha=0$ and decreases rapidly toward the critical current, consistent with expectations. As the nanocontact radius is reduced, the region of bistability is reduced and the barriers for droplet soliton creation grow.

\section{Perturbation expansions near stationary dynamical points in the rotating
frame}
\label{sec:smallnu}

\subsection{Small asymmetry parameter $\nu$, $\theta_{f}=0$, $\theta_{Z}=0$}
\label{subsec:smallasymmetry}

In this case, the fields are time-independent, and we expect stationary states in the rotating frame. More generally, can
expect a frequency shift $\omega^{(1)}$ that will be observable both
in the static and the rotating frames. The first order perturbations
of the equations of motion obey:
\begin{equation}
\dot{\Theta_{1}}=h_{\Phi}^{(1)}+\alpha h_{\Theta}^{(1)}
\end{equation}
\begin{equation}
\dot{\Phi_{1}}=-h_{\Theta}^{(1)}+\alpha h_{\Phi}^{(1)}-\frac{\sigma' V\nu}{2\alpha'}\sin2\Theta_{0}.
\end{equation}

As they are written above, both equations contain differential operators
acting on $\Phi_{1}$ and $\Theta_{1}$, but they can be decoupled
at points of uniform precession about $\mathbf{\hat m}_\mathrm{p}$, i.e., where $\dot{\Theta_{1}}=0$ and $\dot{\Phi}_{1}=\omega^{(1)}$. This decoupling results in 
\begin{equation}
h_{\Phi}^{(1)}=\left(\frac{\alpha}{1+\alpha^{2}}\right)\left[\omega^{(1)}+\frac{\sigma' V\nu}{2\alpha'}\sin2\Theta_{0}\right]\label{eq:hphifirstorder}
\end{equation}
and
\begin{equation}
h_{\Theta}^{(1)}=-\left(\frac{1}{1+\alpha^{2}}\right)\left[\omega^{(1)}+\frac{\sigma' V\nu}{2\alpha'}\sin2\Theta_{0}\right].
\end{equation}
Thanks to our choice of coordinate system, a similar procedure can
be performed to obtain decoupled equations to higher orders
by substituting $h_{\Phi}^{(n)}=-\alpha h_{\Theta}^{(n)}$, from $\dot{\Theta}_{n}=0$,
into the equation for $\dot{\Phi}_{n}=\omega^{(n)}$. However, for now we  restrict our discussion to the first-order case.

Setting $\theta_{f}$ and $\theta_{Z}$ to zero (cf. appendix
\ref{sec:Explicity-expresions-for}) we obtain the first order perturbation
equation for $\Theta_{1}$:
\begin{equation}
\begin{array}{ccc}
h_{\Theta}^{(1)} & = & h_{ex,\Theta}^{(1)}+h_{Z,\Theta}^{(1)}+h_{K,\Theta}^{(1)}+h_{ST,\Theta}^{(1)}\end{array}
\end{equation}
which after some manipulation simplifies to
\begin{equation}
L_{\Theta}\Theta_{1}=-\frac{\omega^{(1)}}{\left(1+\alpha^{2}\right)},\label{eq:schrodingertheta}
\end{equation}
with the Schr\"{o}dinger operator $L_{\Theta}$ defined as
\begin{equation}
\begin{array}{ccc}
L_{\Theta} & \equiv & \left(-h'_{Z}+\frac{\sigma' V}{\alpha}\right)\cos\Theta_{0}-\cos2\Theta_{0}+\nabla'^{2}\end{array}\, .
\end{equation}
Similarly for $\Phi_{1}$:
\begin{equation}
\begin{array}{ccc}
h_{\Phi}^{(1)} & = & h_{ex,\Phi}^{(1)}\\
L_{\Phi}\Phi_{1} & = & \alpha'\left[\omega^{(1)}+\frac{\sigma' V\nu}{2\alpha'}\sin2\Theta_{0}\right]
\end{array}\label{eq:schrogingerphi}
\end{equation}
with the corresponding Schr\"{o}dinger operator
\begin{equation}
\begin{array}{ccc}
L_{\Phi} & \equiv & \left[\left(\nabla'\Theta_{0}\right)^{2}-\cos\Theta_{0}\left(\cos\Theta_{0}-\omega_{h}\right)\right]+\nabla'^{2}\end{array}.
\end{equation}

Since the two operators $L_{\Phi}$ and $L_{\Theta}$ are linear
in $\Theta_{1}$ and $\Phi_{1}$, the solution of Eqs.~(\ref{eq:schrodingertheta})
and~(\ref{eq:schrogingerphi}) must be linear combinations of their
eigenfunctions:
\begin{equation}
\Theta_{1}=\sum_{n}c_{n}\Theta_{n}^{(1)}\qquad\Phi_{1}=\sum_{n}d_{n}\Phi_{n}^{(1)}
\end{equation}
where $\Theta_{n}^{(1)}$ and $\Phi_{n}^{(1)}$ are the basis vectors
which satisfy
\begin{equation}
L_{\Theta}\Theta_{n}^{(1)}=\vartheta_{n}\Theta_{n}^{(1)}\qquad L_{\Phi}\Phi_{n}^{(1)}=\varphi_{n}\Phi_{n}^{(1)}\label{eq:eigenfunctionandeigenvalus}
\end{equation}
and appropriate boundary conditions. With the solutions to this eigenproblem
the stationary configuration is described by the set of coefficients
$\left\{ c_{n},d_{n}\right\} $ which satisfy
\begin{equation}
\sum_{n}c_{n}\vartheta_{n}\Theta_{n}^{(1)}=-\frac{\omega^{(1)}}{\left(1+\alpha^{2}\right)}
\end{equation}
and
\begin{equation}
\sum_{n}d_{n}\varphi_{n}\Phi_{n}^{(1)}=\frac{\alpha\left[\omega^{(1)}+\frac{\sigma' V\nu}{2\alpha'}\sin2\Theta_{0}\right]}{1+\alpha^{2}}\ .
\end{equation}
In each of these equations, the right-hand-side should be orthogonal
to the kernel of the corresponding Schrodinger operator. It can be
verified that the function $\varphi_{0}=\sin\Theta_{0}$ belongs to
the kernel of $L_{\Phi}$, which provides a necessary condition for
the existence of a coherent perturbation:
\begin{equation}
\omega^{(1)}=-\frac{\sigma'\nu}{\alpha'}\frac{\int_{0}^{\rho^{*}}\cos\Theta_{0}\sin^{2}\Theta_{0}\rho d\rho}{\int_{0}^{\infty}\sin\Theta_{0}\rho d\rho}.\label{eq:nufrequencyshift}
\end{equation}
For this condition to be sufficient, we need to complement equation
Eq.~(\ref{eq:schrogingerphi}) with appropriate boundary conditions satisfied
by $\varphi_{0}$ so that, based on the uniqueness and completeness
theorem, it becomes the unique element of the kernel of $L_{\Phi}$.

From the condition of regularity at the origin for $\Theta_{0}$ we
get:
\begin{equation}
\left.\frac{\partial\varphi_{0}}{\partial\rho}\right|_{\rho=0}=\cos\Theta_{0}\left.\frac{\partial\Theta_{0}}{\partial\rho}\right|_{\rho=0}=0.
\end{equation}
For the second boundary condition, we can postulate that the azimuthal
angle far from the nanocontact is unperturbed:
\begin{equation}
\varphi(\rho\rightarrow\infty)=\left.\sin\Theta_{0}\right|_{\rho\rightarrow\infty}=0\ .
\end{equation}

\begin{figure}
\includegraphics[angle=-90,width=3.5in]{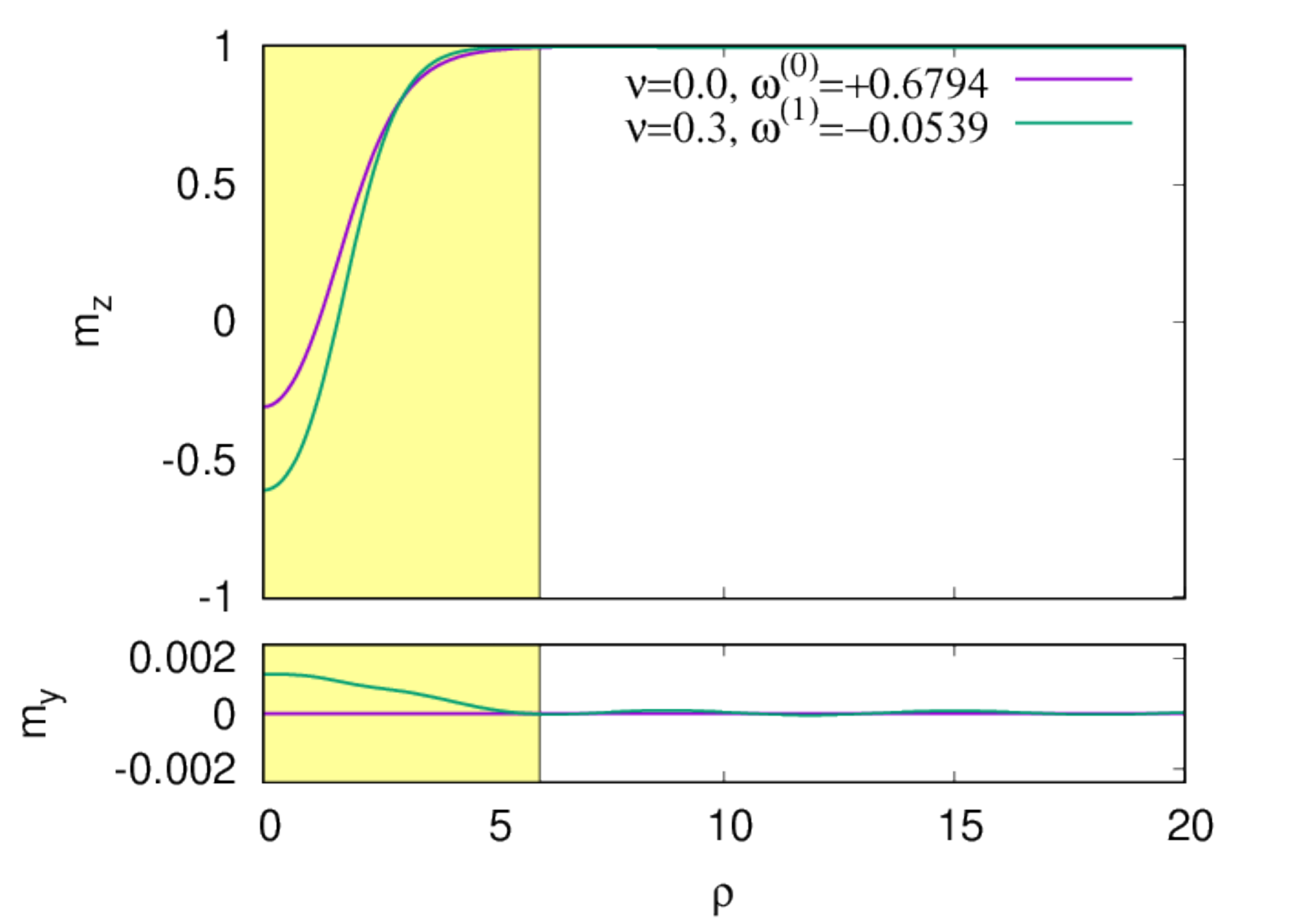}\caption{\label{fig:nuperturbation}The effect of finite asymmetry parameter $\nu$ in droplet soliton profiles: the amplitude, $\Theta(\rho)$, is increased and the azimuthal angle, $\Phi(\rho)$,
is no longer constant.   Here $m_y=\sin\Phi^{(1)}\sin\left(\Theta^{(0)}+\Theta^{(1)}\right)$  and $\rho^*=6.0,\alpha=0.01$.}
\end{figure}

Using a similar procedure to that described in  Sect.~\ref{subsec:Numerical-solutions-for},
we find the numerical solutions to the perturbation equations. The
procedure is summarized as follows: for a given $\nu$ a frequency
shift is obtained using Eq.~(\ref{eq:nufrequencyshift}). We guess the initial
values of the perturbation function at the origin (i.e., $\Phi_{1}(\rho=0)$
and $\Theta_{1}(\rho=0)$) and let the BVP4C method from Matlab find
values for $\omega^{(1)}$ corresponding to those boundary conditions~\cite{chaves-oflynn_sup}.
We iteratively tune $\Phi_{1}(\rho=0)$ and $\Theta_{1}(\rho=0)$
so that the output from Matlab approaches the value obtained from
Eq.~(\ref{eq:nufrequencyshift}). This results in functions $\Phi_{1}(\rho)$
and $\Theta_{1}(\rho)$ with frequency shifts given by $\omega^{(1)}$.
An example of such a solution is shown in Fig.~\ref{fig:nuperturbation}.

These results show three effects that can be expected when the spin-torque induced precessional frequency is rotationally symmetric but no longer radially uniform: a frequency shift ($\omega^{(1)}\ne0$), an amplitude change ($\Theta^{(1)}\ne0$), and a radially dependent buckling ($\Phi^{(1)}\ne 0$).

From the uniqueness argument we conclude that $\rho^{*}, \nu, \sigma$, and $\alpha$ determine uniquely the shape of the perturbed droplet soliton and the new
frequency of precession. It is clear that in a frame rotating with
frequency $\omega_{0}+\omega^{(1)}$ the droplet soliton appears stationary.
Unfortunately, this situation occurs for a configuration
for which $h_{\Phi}^{(1)}\ne0$ and $h_{\Theta}^{(1)}\ne0$. As a
consequence, an estimate of the escape rates based solely on potential energy
differences is no longer applicable. In this case, is is necessary to use the
more general approach of Freidlin and Wentzell. We will return to this issue in Sec. \ref{sec:FW-connection} and calculate the action for a specific class of fluctuational trajectories.

Nevertheless, we point out that the droplet soliton profiles found by solving
Eqs.~(\ref{eq:schrodingertheta}) and (\ref{eq:schrogingerphi}) have
a well defined energy to first order in $\nu$. This can be 
seen from Eq.~(\ref{eq:xisymmetric}), using the first-level expansion
for $h_{ST,\Theta}$:
\begin{equation}
h_{ST,\Theta}^{(0)}\approx\frac{\sigma'  V}{\alpha}\left[\sin\Theta_{0}+\epsilon\Theta_{1}\cos\Theta_{0}+\frac{\nu\sin2\Theta_{0}}{2}\right].
\end{equation}

Using $h_{ex,\Phi}^{(1)}$ from Eq.~(\ref{eq:hphifirstorder}) we obtain the
spin-torque power density:
\begin{align}
\Xi^{(1)} & =\left(h_{ST,\Theta}^{0}+h_{ST,\Theta}^{1}\right)\times h_{ex,\Phi}^{(1)}\\
 & =h_{ST,\Theta}^{(0}\times h_{ex,\Phi}^{(1)}\\
 & =\frac{\sigma V}{\alpha}\sin\Theta_{0}\times\frac{\alpha\left[\omega^{(1)}+\frac{\sigma' V\nu}{2\alpha'}\sin2\Theta_{0}\right]}{1+\alpha^{2}}\, .
\end{align}

Finally, integrating over the full disk we find
\begin{equation}
2\pi\int\Xi^{(1)}\rho d\rho=0\ 
\end{equation}
from the orthogonality condition Eq.~(\ref{eq:nufrequencyshift}). This
means that for the droplet soliton profiles discussed above, the spin-torque
power inflow in some regions of the disk is balanced by a spin-torque
power outflow in another region. Since the fields are calculated only to first order, the total power dissipation, given by Eq.~(\ref{eq:melnikov}), is also zero.

\section{Connection with Freidlin-Wentzell theory}
\label{sec:FW-connection}
\selectlanguage{american}%

We next establish a connection between the pseudo-energy
and the Freidlin-Wentzell action~\cite{FW12} for randomly perturbed dynamics.
For simplicity, we use polar coordinates in the fixed laboratory
frame, keeping track of the energy functions $\mathcal{E}_{0}=\mathcal{E}_{\mathrm{ex}}+\mathcal{E}_{K}+\mathcal{E}_{Z}$
and $\mathcal{E}_{\mathrm{tot}}=\mathcal{E}_{0}+\mathcal{E}_{ST}$.
Writing the Landau-Lifshitz operator and its inverse as
\begin{equation}
\mathbb{L}_{\Theta,\Phi}=\left(\begin{array}{cc}
\alpha & 1\\
-\frac{1}{\sin\Theta} & \frac{\alpha}{\sin\Theta}
\end{array}\right)\qquad\mathbb{L}_{\Theta,\Phi}^{-1}=\frac{\left(\begin{array}{cc}
\alpha & -\sin\Theta\\
1 & \alpha\sin\Theta
\end{array}\right)}{1+\alpha^{2}}
\end{equation}
the equation of motion in the presence of noise can be written 
\begin{equation}
\label{eq:eomnoise}
\dot{\mathbf{m}}_{\Theta,\Phi}-\mathbb{L}_{\Theta,\Phi}\mathbf{h}_{\mathrm{tot}}-\mathbf{f}=\sqrt{2\eta}\mathbb{L}_{\Theta,\Phi}\dot{\mathbf{W}}\ 
\end{equation}
where three two-component vectors have been introduced:  $\dot{\mathbf{m}}_{\Theta,\Phi}=\{\dot{\Theta},\dot{\Phi}\}$, $\mathbf{f}=\{0,\omega_{ST}\frac{1}{1+\nu\cos\Theta}\}=\{\frac{1}{\sin\Theta}\frac{\delta\mathcal{E}_{ST}}{\delta\Phi},-\frac{1}{\sin\Theta}\frac{\delta\mathcal{E}_{ST}}{\delta\Theta}\}$, 
and $\mathbf{h}_{\mathrm{tot}}=\{-\frac{\delta\mathcal{E}_{\mathrm{tot}}}{\delta\Theta},-\frac{1}{\sin\Theta}\frac{\delta\mathcal{E}_{tot}}{\delta\Phi}\}$.

\medskip

The deterministic part of Eq.~(\ref{eq:eomnoise}) can be simplified using the vector $\mathbf{b}_{\Theta,\Phi}\equiv\mathbb{L}_{\Theta,\Phi}\mathbf{h}_{\mathrm{tot}}+\mathbf{f}$, resulting in the equation
\begin{equation}
\mathbb{L}_{\Theta,\Phi}^{-1}\left[\dot{\mathbf{m}}_{\Theta,\Phi}-\mathbf{b}_{\Theta,\Phi}\right]=\sqrt{2\eta}\mathbf{W}.
\end{equation}

The Freidlin-Wentzell action per unit thickness is
\begin{equation}
S=\iiint\mathcal{L}\rho d\rho d\phi dt
\end{equation}
where 
\begin{equation}
\mathcal{L}=\frac{1}{2}\mathbb{A}_{ii}^{-1}\left[\dot{\mathbf{m}}_{\Theta,\Phi}-\mathbf{b}_{\Theta,\Phi}\right]_{i}^{2}
\end{equation}
is the Freidlin-Wentzell Lagrangian and 
\begin{equation}
\mathbb{A}^{-1}=\left(\mathbb{L}_{\Theta,\Phi}^{-1}\right)^{T}\mathbb{L}_{\Theta,\Phi}^{-1}=\left(\begin{array}{cc}
1 & 0\\
0 & \sin^{2}\Theta
\end{array}\right)
\end{equation}
is the inverse of the diffusion tensor.

Because the white noise modelling thermal effects does not destroy the global rotational symmetry of the magnetization, we can integrate out the azimuthal coordinates, leading to the reduced action
\begin{equation}
S_{\rho}=\iint\mathcal{L}_{\rho}d\rho dt
\end{equation}
with
\begin{equation}
\mathcal{L}_{\rho}\left(h_{\Theta},h_{\Phi},\dot{\Phi},\dot{\Theta};t,\rho\right)=\frac{2\pi\rho}{2}\mathbb{A}_{ii}^{-1}\left[\dot{\mathbf{m}}_{\Theta,\Phi}-\mathbf{b}_{\Theta,\Phi}\right]_{i}^{2}\, .
\end{equation}

Because of the rotational symmetry we can write
\begin{align}
\mathbb{A}_{\rho}^{-1}=\left(\mathbb{L}_{\Theta,\Phi}^{-1}\right)^{T}\mathbb{L}_{\Theta,\Phi}^{-1} & =\left(\begin{array}{cc}
\rho & 0\\
0 & \rho\sin^{2}\Theta
\end{array}\right).\\
 & =\left(\sqrt{2\pi\rho}\right)^{2}\mathbb{A}^{-1}\, .
\end{align}

The extra radial factor accounts for the fact that coherent thermal fluctuations for annuli of radius $\rho$
and width $d\rho$ become rarer as $\rho$ increases. This can be interpreted
as follows: if the sample is discretized in infinitesimal segments
of $d\mathcal{V}$ and the thermal field on each segment has standard deviation
$h_{th}$, then the averaged field fluctuation on the annulus as a whole grows only as
$h_{\rho}=\sqrt{\rho}h_{th}$. At any instant of time, a coherent
fluctuation of unit size observed at radius $\rho$ requires a thermal
energy density proportional to $h_{\rho}^{2}=\rho h_{th}^{2}$, thereby decreasing in likelihood as
$\rho$ increases. The Freidlin-Wentzell Lagrangian
is a measure of such thermal energy and therefore grows proportionally to the volume.

We will calculate the action for two specific translation paths $\mathbf{m}_{\downarrow}$,
$\mathbf{m}_{\uparrow}$ that we now define. First we 
separate the damping from the precessional part of the Landau-Lifshitz operator:
\begin{equation}
\mathbb{L}_{\Theta,\Phi}=\mathbb{L}_{\Theta,\Phi}^{\alpha}+\mathbb{L}_{\Theta,\Phi}^{\gamma}
\end{equation}
\begin{equation}
\mathbb{L}_{\Theta,\Phi}^{\alpha}=\alpha\left(\begin{array}{cc}
1 & 0\\
0 & \frac{1}{\sin\Theta}
\end{array}\right)\qquad\mathbb{L}_{\Theta,\Phi}^{\gamma}=\left(\begin{array}{cc}
0 & 1\\
-\frac{1}{\sin\Theta} & 0
\end{array}\right)
\end{equation}
which allows us to write the (deterministic) drift field as
\begin{equation}
\mathbf{b}_{\Theta,\Phi}=\mathbb{L}_{\Theta,\Phi}^{\alpha}\mathbf{h}_{\mathrm{tot}}+\mathbb{L}_{\Theta,\Phi}^{\gamma}\left(\mathbf{h}_{\mathrm{tot}}+\mathbf{h}_{\mathrm{ST}}\right).
\end{equation}
The paths of interest are then defined as follows:
\begin{equation}
\dot{\mathbf{m}}_{\downarrow}=\mathbf{b}_{\Theta,\Phi}=\mathbb{L}_{\Theta,\Phi}^{\alpha}\mathbf{h}_{\mathrm{tot}}+\mathbb{L}_{\Theta,\Phi}^{\gamma}\left(\mathbf{h}_{\mathrm{tot}}+\mathbf{h}_{\mathrm{ST}}\right)
\end{equation}
and
\begin{align}
\dot{\mathbf{m}}_{\uparrow} & =\mathbf{b}_{\Theta,\Phi}-2\mathbb{L}_{\Theta,\Phi}^{\alpha}\mathbf{h}_{\mathrm{tot}}2\nonumber\\
 & =-\mathbb{L}_{\Theta,\Phi}^{\alpha}\mathbf{h}_{\mathrm{tot}}+\mathbb{L}_{\Theta,\Phi}^{\gamma}\left(\mathbf{h}_{\mathrm{tot}}+\mathbf{h}_{\mathrm{ST}}\right).
\end{align}

Our choice of fluctuational paths is motivated as follows: one of the two pieces of the drift field, $\mathbb{L}_{\Theta,\Phi}^{\alpha}\mathbf{h}_{\mathrm{tot}}$,is curl-free and so can be derived as the gradient of a potential. The other piece, $\mathbb{L}_{\Theta,\Phi}^{\gamma}\left(\mathbf{h}_{\mathrm{tot}}+\mathbf{h}_{\mathrm{ST}}\right)$, is divergenceless and so can be represented as the curl of a (separate) potential function. We choose the  fluctuational trajectory so that the nonzero-curl term does not contribute to the Freidlin-Wentzell action. 

The action for the ``downhill'' path $\mathbf{m}_{\downarrow}$ is 
zero, since $\dot{\mathbf{m}}_{\downarrow}$ runs parallel to the
deterministic field. The Lagrangian for the ``uphill'' path $\dot{\mathbf{m}}_{\uparrow}$ is
\begin{align}
\mathcal{L}= & \frac{1}{2}\mathbb{A}_{ii}^{-1}\left[\dot{\mathbf{m}}_{\uparrow}-\mathbf{b}_{\Theta,\Phi}\right]_{i}^{2}\nonumber\\
= & \frac{1}{2}\mathbb{A}_{ii}^{-1}\left[\dot{\mathbf{m}}_{\uparrow}-\mathbb{L}_{\Theta,\Phi}^{\alpha}\mathbf{h}_{\mathrm{tot}}-\mathbb{L}_{\Theta,\Phi}^{\gamma}\left(\mathbf{h}_{\mathrm{tot}}+\mathbf{h}_{\mathrm{ST}}\right)\right]_{i}^{2}\nonumber\\
= & \frac{1}{2}\mathbb{A}_{ii}^{-1}\left[\dot{\mathbf{m}}_{\uparrow}+\mathbb{L}_{\Theta,\Phi}^{\alpha}\mathbf{h}_{\mathrm{tot}}-\mathbb{L}_{\Theta,\Phi}^{\gamma}\left(\mathbf{h}_{\mathrm{tot}}+\mathbf{h}_{\mathrm{ST}}\right)\right]_{i}^{2}\nonumber\\
 & -2\left[\mathbb{L}_{\Theta,\Phi}^{\alpha}\mathbf{h}_{\mathrm{tot}}\right]^{T}\mathbb{A}_{ii}^{-1}\left[\dot{\mathbf{m}}_{\uparrow}-\mathbb{L}_{\Theta,\Phi}^{\gamma}\left(\mathbf{h}_{\mathrm{tot}}+\mathbf{h}_{\mathrm{ST}}\right)\right]\nonumber\\
= & -2\left[\mathbb{L}_{\Theta,\Phi}^{\alpha}\mathbf{h}_{\mathrm{tot}}\right]^{T}\mathbb{A}_{ii}^{-1}\left[\dot{\mathbf{m}}_{\uparrow}-\mathbb{L}_{\Theta,\Phi}^{\gamma}\left(\mathbf{h}_{\mathrm{tot}}+\mathbf{h}_{\mathrm{ST}}\right)\right]\nonumber\\
= & -2\mathbf{h}_{\mathrm{tot}}^{T}\left(\mathbb{L}_{\Theta,\Phi}^{\alpha}\right)^{T}\mathbb{A}_{ii}^{-1}\dot{\mathbf{m}}_{\uparrow}\nonumber\\
 & +2\mathbf{h}_{\mathrm{tot}}^{T}\left(\mathbb{L}_{\Theta,\Phi}^{\alpha}\right)^{T}\mathbb{A}_{ii}^{-1}\mathbb{L}_{\Theta,\Phi}^{\gamma}\left(\mathbf{h}_{\mathrm{tot}}+\mathbf{h}_{\mathrm{ST}}\right)\nonumber\\
= & \mathcal{L}_{0}+\mathcal{L}_{ST}
\end{align}

The first term is
\begin{align}
\mathcal{L}_{0} & =-2\mathbf{h}_{\mathrm{tot}}^{T}\left(\mathbb{L}_{\Theta,\Phi}^{\alpha}\right)^{T}\mathbb{A}_{ii}^{-1}\dot{\mathbf{m}}_{\uparrow}\nonumber\\
 & =+2\alpha\rho\left[\frac{\delta\mathcal{E}_{\mathrm{tot}}}{\delta\Theta}\dot{\Theta}+\frac{\delta\mathcal{E}_{tot}}{\delta\Phi}\dot{\Phi}\right]
\end{align}
which, after integrating over time and space, gives
\begin{equation}
\label{eq:gradient}
S_{0}=\iint\mathcal{L}_{0}d\rho dt=2\Delta E_{tot}\, .
\end{equation}
The second term, using $\mathbf{h}_{ST}=\{-\frac{\delta\mathcal{E}_{ST}}{\delta\Theta},0\}$
gives
\begin{equation}
\mathcal{L}_{ST}=2\alpha h_{ex,\Phi}\cdot h_{ST,\Theta}=2\alpha\rho\Xi.
\end{equation}

Therefore, integration over time and space gives the spin-torque induced action
\begin{equation}
S_{ST}=\iint\mathcal{L}_{ST}=2\alpha\int\int\Xi\rho d\rho dt\, .
\end{equation}
As described before, the total power input in the droplet soliton configurations
described satisfies 
\begin{equation}
\label{eq:nongradient}
\int\Xi\rho d\rho=0\, .
\end{equation}

So, combining Eqs.~(\ref{eq:gradient})-(\ref{eq:nongradient}) for escape trajectories which include configurations having
zero spin-torque induced energy flow, the Freidlin-Wentzell
action is simply twice the energy difference between the initial
and final states, as is the case in simpler systems with gradient drift fields arising from a potential.

\section{Discussion}
\label{sec:discussion}

In this paper we have studied the thermal stability of magnetic droplet solitons which are linearly stable against the drift instability~\cite{hoefer_theory_2010,MH19}. Taking advantage of the rotational symmetry, we transform the problem to  a reference frame that rotates with the droplet soliton, thereby simplifying the problem. We introduced a pseudo-potential that can incorporate the nongradient spin torque terms, allowing us to analyze the problem using the Kramers approach to computing reversal rates. Numerical simulations of the dynamical LLGS equations demonstrate that the system indeed evolves toward one or another (depending on initial conditions) of the pseudo-potential minima.

Our central result is summarized in Fig.~\ref{fig:summaryofbarriers}, which shows the activation barriers for thermally (or other noise)-induced escape from the droplet soliton to the uniform magnetization state, for a variety of nanocontact radii and spin-torque-inducing currents. As a function of current, barriers grow faster for larger radii. 

We next discuss possible scenarios in which these results can be generalized, and discuss the model's limitations, and in particular situations where it breaks down.

Large Oersted fields are known to induce modulational instabilities that force the droplet soliton out of the nanocontact region at relatively short times~\cite{WIH16,Hang2018}. For this regime, thermal activation over a barrier is irrelevant since the droplet soliton decays before thermal processes can play any role. But even for moderate currents the Oersted field introduces a time-harmonic perturbation of the dynamics that we have omitted. A continuous generalization of the Poincare-Melnikov method is worth pursuing to incorporate fields with in-plane components. This will be the focus of future work.

In our approach, we initially assumed that the magnetization has a common frequency of precession with $\nabla\Phi=0$. This requires $\nu=0$. For $\nu\ne 0$, to first order the spin-torque dynamical term produces exchange-driven shearing of the magnetization, resulting in a nonuniform $\Phi$. As shown in Fig.~\ref{fig:nuperturbation}, three main effects are observed: first, a distortion of the $\Theta(\rho)$ profile of the droplet soliton; second, a frequency shift common to all reference frames; and third, a 'buckling' of the azimuthal angle $\Phi(\rho)\ne0$. 

An extreme situation involved with this shearing occurs for small nanocontact radii. Inside the nanocontact the spin-torque-induced precessional term is proportional to $\sigma'/\alpha$, and outside it is zero. Because the two regions precess at different rates, large exchange torques are created at the nanocontact edge. The interfacial shear at the boundary between the two regions will produce similar effects to those found for $\nu=0$, and can be neglected if $\sin\theta\approx0$ at the nanocontact edge. This is the case for small and large amplitude droplet solitons, or for large $\rho^*$. However, as $\rho^*\rightarrow0$ this shearing becomes important and the spin-torque energy might `leak' beyond the nanocontact region resulting in 
spatial modulation of the magnetization far beyond the nano-contact region (see \cite{Hoefer_2008} for examples of 
such modulation). For this scenario, the total energy of the system will continue to increase as oscillations develop far from the nanocontact. However, as we discussed in the main text of the paper, the limit $\rho^*\rightarrow\infty$ is valid for radii that are experimentally relevant.

As noted in the introduction, even in the region of the current/field phase diagram where the droplet soliton is linearly stable to the drift instability, thermal activation over a barrier can still eject the droplet soliton from the nanocontact region. Numerical estimates of this thermally-induced ejection have been provided in~\cite{MH19}, where two mechanisms for thermally activated droplet soliton annihilation were considered: ``noise-induced damping'', which corresponds to the droplet soliton decay mechanism considered in this paper, and ejection via drift instability. Because our analysis mostly considers different regions of the parameters space from those addressed in~\cite{MH19}, a direct comparison between our results and theirs is not possible at the present time. We expect this to be a topic for future study. Nevertheless, as a first step we summarize below the differences in parameters chosen so that the difference in the regions of parameters space studied in the two papers is clarified. 

The droplet soliton profile configurations used in~\cite{MH19} are restricted to the case of small $\sigma$ and $\alpha$, while this constraint is not present in the current work. Fig.~\ref{fig:omegavssoa} summarizes the result that the conservative droplet soliton profiles correspond to saddle configurations in the $\rho^*\rightarrow\infty$ limit. The procedure we have followed allows us to find droplet soliton profiles for saddle states at arbitrary values of $\rho^*$, and also at $\nu\ne 0$ when $h_Z=0$. 

Ref.~\cite{MH19} examines mostly cases where $h_Z>1.0$ and the applied current is  moderate-to-large. Our discussion focuses instead on zero or small fields (cf.~Fig.~\ref{fig:summaryofbarriers}), given that the low-field regime provides the widest range of droplet soliton stability with respect to applied current (cf.~Fig.~2 of~\cite{WIH16}). Turning on an external field has several effects: first, it produces a shift of the droplet soliton frequency with respect to the current, similar to that observed for conservative droplet solitons~\cite{BH13}. This has the effect of reducing the droplet soliton amplitude. More relevant to the present discussion is that it lowers the activation barrier for both modes of droplet soliton decay. In addition,~\cite{MH19} found that increasing the current in large fields facilitates escape via the drift instability.

It is important to note that the formulation used for STT both in our paper and in the sequence of papers leading up to~\cite{BH13} assume a fixed layer polarization $\mathbf{\hat{m_p}}$ perpendicular to the film plane \cite{hoefer_theory_2010,BH13,WIH16,BH15}. This setup differs from actual experiments that typically use an easy-plane polarization layer~\cite{chung_magnetic_2016,MBD14,lendinez_observation_2015,lendinez_effect_2017}. As a consequence, large external fields are needed to reorient $\mathbf{\hat{p}}$ along $\mathbf{n}_\perp$ in order to approximate  the highly symmetric case. Such large fields reduce the range of currents in which the stationary droplet soliton is stable against drift, as discussed in~\cite{MH19}.

Because our methods exploit the presence of circular symmetry, and used $\mathbf{\hat{p}}$ as a privileged direction for the polar axes to define the rotational reference frame, our predictions are relevant to a so-called ``all-perpendicular'' device (i.e. $\mathbf{\hat p}||\mathbf{n}_\perp$)~\cite{Chung18}. 

We now estimate droplet soliton lifetimes in the presence of thermal noise due to the decay mechanism described in this paper, based on data from Fig.~\ref{fig:summaryofbarriers}. 
We emphasize that these estimates are based on the leading-order asymptotics, i.e., the action barriers which determine the exponential Arrhenius factor, of the activation times for droplet soliton decay. The prefactors, which correspond to subdominant terms, are estimated below.  We caution that these transition times should not be confused with deterministic creation and annihilation times, which are on the nanosecond timescale.

We assume a Co/Ni free layer in zero field with magnetization $\mu_0M_s=0.9$T and $H_K=1.35$T ($Q$=1.5), as in \cite{chung_magnetic_2016}. We assume $A=13$pJ/m and $\alpha=0.01$. With these parameters, we examine the droplet soliton stability in a nanocontact of scaled radius $\rho'^*=5$ ($r^*=22.5$nm). The sustaining current is $I_\mathrm{min}=2.3$mA and the critical current is $I_\mathrm{max}=9.0$mA. If we assume an attempt rate given by $f_0=\omega_{ST}\alpha=\sigma\gamma_0M_s$ we can estimate lifetimes using $f_0^{-1}e^{\frac{\Delta}{k_BT}}$ at room temperature $T=300K$. With a current of 2.8mA the mean lifetime of a droplet soliton in the presence of thermal noise is approximately  12~minutes ($\Delta^-\approx26k_BT$) while the uniform state is very stable ($\Delta^+\approx127k_BT$). For a larger current of 6.5mA, the droplet soliton generation time is slightly under 2~minutes ($\Delta^+\approx25k_BT$). Once nucleated, however, it is very stable ($\Delta^-\approx190k_BT$).  In a highly symmetric experimental setup, the signal from the resonant frequency of the droplet soliton is very weak, and one must rely on magnetoresistance measurements to determine the presence or absence of a droplet soliton. 

\bigskip

\noindent {\bf Acknowledgments.}  The authors are grateful to Andy Kent for useful discussions and comments on the manuscript. We are also grateful to Mark Hoefer for a useful correspondence in the early stages of this work, and an anonymous referee for the suggestion that the saddle states discussed in this paper appear similar to experimentally observed spin torque oscillator states.  This research was supported in part by U.S.~National Science Foundation Grant DMR~1610416 (DLS) and the National Science Centre Poland under OPUS funding Grant No.~2019/33/B/ST5/02013 (GDC).

\selectlanguage{american}%

\appendix

\section{Effective field and boundary conditions}

\label{app:bcs}

Consider a quasi-2D system with nanocontact radius~$\rho^{*}$, free layer radius~$\rho_{\mathrm{max}}$, surface area
$\Omega$ and thickness $d$. Using the Kohn-Slastikov approximation~\cite{KS05},
we write the micromagnetic energy as 
\begin{eqnarray}
\label{eq:energy}
\frac{E}{d} & = & \int_{\Omega}\Bigl\{ A\left|\nabla\mathbf{m}\right|^{2}-\left(K_{S}-\frac{\mu_{0}M_{s}^{2}}{2}\right)m_{z}^{2}-\mu_{0}M_{s}\mathbf{m}\cdot\mathbf{H}_{0}
 \Bigr\} d^{2}\mathbf{r}\nonumber \\
 & + & K_{\mathrm{edge}}\oint_{\partial\Omega}\left(\mathbf{m}\cdot\mathbf{n_{\parallel}}\right)^{2}d\mathbf{r}\, ,
\end{eqnarray}
where $K_{\mathrm{edge}}=\frac{\mu_{0}M_{s}^{2}d}{4\pi}\left|\ln\left(\frac{t}{2R}\right)\right|$, $\mathbf{n}_{\parallel}$ is the in-plane vector perpendicular
to the sample's edge and the field $\mathbf{H}_{0}=\mathbf{H}_\mathrm{oe}+\mathbf{H}_{Z}$ includes external and Oersted contributions.

Varying both sides of~Eq.~(\ref{eq:energy}) yields 
\begin{align}
\frac{\delta E}{d} & =\int_{\Omega}2A\left(\nabla\mathbf{m}\right)\cdot\left(\nabla\mathbf{\delta m}\right)d\mathbf{^{2}r}\\
 & -\int_{\Omega}2\left(K_{S}-\frac{\mu_{0}M_{s}^{2}}{2}\right)\left(\mathbf{\hat{z}}\cdot\mathbf{m}\right)\left(\mathbf{\hat{z}}\cdot\delta\mathbf{m}\right)d\mathbf{^{2}r}\\
 & -\int_{\Omega}\mu_{0}M_{s}\mathbf{H}_{0}\cdot\delta\mathbf{m}d\mathbf{^{2}r}\\
 & +2\oint_{\partial\Omega}\left[K_{\mathrm{edge}}\left(\mathbf{m}\cdot\mathbf{n}_{\parallel}\right)\right]\mathbf{\delta m}\cdot\mathbf{n}d\mathbf{r}
\end{align}

The first term contains derivatives of $\mathbf{\delta\mathbf{m}}$
and corresponds to the effects of exchange. Following Brown and Miltat~\citep{miltat_numerical_2007} we use the relation $\nabla\cdot\left(\nabla\mathbf{m}\cdot\delta\mathbf{m}\right)=\left(\nabla^{2}\mathbf{m}\right)\delta\mathbf{m}+\nabla\mathbf{m}\cdot\nabla\delta\mathbf{m}$
to integrate by parts, yielding
\begin{equation}
\int_{\Omega}\left(\nabla\mathbf{m}\right)\cdot\left(\nabla\mathbf{\delta m}\right)=\int_{\Omega}\nabla\cdot\left(\nabla\mathbf{m}\cdot\delta\mathbf{m}\right)d\mathbf{^{2}r}-\int_{\Omega}\left(\nabla^{2}\mathbf{m}\right)\delta\mathbf{m}d\mathbf{^{2}r}\, .
\end{equation}
Using Gauss' theorem to rewrite the first term on the RHS yields
\begin{equation}
\int_{\Omega}\left(\nabla\mathbf{m}\right)\cdot\left(\nabla\mathbf{\delta m}\right)=\oint_{\partial\Omega}\left(\frac{\partial\mathbf{m}}{\partial\mathbf{n}_{\parallel}}\cdot\delta\mathbf{m}\right)d\mathbf{r}-\int_{\Omega}\left(\nabla^{2}\mathbf{m}\right)\delta\mathbf{m}\,d\mathbf{^{2}r}\, .
\end{equation}

Rearranging this variational expression into bulk and surface terms we get
\begin{eqnarray}
\frac{\delta E}{d} & = & \int_{\Omega}\Bigl[-2A\left(\nabla^{2}\mathbf{m}\right)-2\mathbf{\hat{z}_{L}}\left(K_{S}-\frac{\mu_{0}M_{s}^{2}}{2}\right)\left(\mathbf{\hat{z}_{L}}\cdot\mathbf{m}\right)\nonumber \\
& 
+ & \mu_{0}M_{s}\mathbf{H}_{0}\Bigr]\delta\mathbf{m}d\mathbf{^{2}r}\nonumber \\
 & + & 2\oint_{\partial\Omega}\left[K_{\mathrm{edge}}\left(\mathbf{m}\cdot\mathbf{n}_{\parallel}\right)\left(\mathbf{n}_{\parallel}\cdot\mathbf{\delta m}\right)+A\frac{\partial\mathbf{m}}{\partial\mathbf{n}_{\parallel}}\cdot\delta\mathbf{m}\right]d\mathbf{r}\, .
\end{eqnarray}

From the bulk integral we obtain the effective field
\begin{eqnarray}
\label{eq:heff2}
\mathbf{h}_{{\rm {eff}}} & = & -\frac{1}{\mu_{0}M_{s}^{2}}\frac{\delta\mathcal{E}}{\delta\mathbf{m}}\\ 
&=&l_{\mathrm{ex}}^{2}\left(\nabla^{2}\mathbf{m}\right)+(Q-1)\left(\mathbf{\hat{z}_{L}}\cdot\mathbf{m}\right)\mathbf{\hat{z}_{L}}\nonumber 
+\mathbf{h_{\mathrm{0}}}\,,
\end{eqnarray}
where $Q=2K_{S}/(\mu_{0}M_{s}^{2})>1$ is the (dimensionless) magnitude
of the crystal anisotropy field.

Turning to the surface integral 
\begin{equation}
\oint_{\partial\Omega}\left[K_{\mathrm{edge}}\left(\mathbf{m}\cdot\mathbf{n_{\parallel}}\right)\left(\mathbf{n}_{\parallel}\cdot\mathbf{\delta m}\right)+A\frac{\partial\mathbf{m}}{\partial\mathbf{n}_{\parallel}}\cdot\delta\mathbf{m}\right]d\mathbf{r}\,,
\end{equation}
we rewrite the magnetization variation as $\delta\mathbf{m}=\delta\theta\times\mathbf{m}$
\begin{equation}
\oint_{\partial\Omega}\left[K_{\mathrm{edge}}\left(\mathbf{m}\cdot\mathbf{n}_{\parallel}\right)\mathbf{n}_{\parallel}\cdot\left(\delta\theta\times\mathbf{m}\right)+A\frac{\partial\mathbf{m}}{\partial\mathbf{n}_{\parallel}}\cdot\left(\delta\theta\times\mathbf{m}\right)\right]d\mathbf{r}
\end{equation}
which shows that any variation conserves the norm of $\mathbf{m}$.

Finally, using the cyclic permutation of the triple vector product
gives the surface integral contribution to the energy variation 
\begin{equation}
\oint_{\partial\Omega}\left[K_{\mathrm{edge}}\left(\mathbf{m}\cdot\mathbf{n}_{\parallel}\right)\left(\mathbf{m}\times\mathbf{n}_{\parallel}\right)+A\left(\mathbf{m}\times\frac{\partial\mathbf{m}}{\partial\mathbf{n}_{\parallel}}\right)\right]\delta\theta d\mathbf{r}\,.
\end{equation}

For this contribution to be extremal, it must be zero independently
of $\delta\theta$, so 
\begin{equation}
K_{\mathrm{edge}}\left(\mathbf{m}\cdot\mathbf{n}_{\parallel}\right)\left(\mathbf{m}\times\mathbf{n}_{\parallel}\right)+A\left(\mathbf{m}\times\frac{\partial\mathbf{m}}{\partial\mathbf{n}_{\parallel}}\right)=0\,.
\end{equation}

Cross-multiplying with $\mathbf{m}$ gives the boundary condition
\begin{equation}
\frac{K_{\mathrm{edge}}}{A}\left(\mathbf{m}\cdot\mathbf{n}_{\parallel}\right)\Bigl[\left(\mathbf{m}\cdot\mathbf{n}_{\parallel}\right)\mathbf{m}-\mathbf{n}_{\parallel}\Bigr]=\frac{\partial\mathbf{m}}{\partial\mathbf{n}_{\parallel}}
\end{equation}

Since the vector normal to the surface lies along $\hat{\rho}$ we
get 
\begin{equation}
\frac{K_{\mathrm{edge}}}{A}\left(\mathbf{m}\cdot\hat{\rho}\right)\Bigl[\left(\mathbf{m}\cdot\hat{\rho}\right)\mathbf{m}-\hat{\rho}\Bigr]=\frac{\partial\mathbf{m}}{\partial\mathbf{\rho}}
\end{equation}
so, recalling that $\Theta$ is the angle the magnetization makes
with respect to the normal to the free layer disk, we have 
\begin{equation}
\frac{K_{\mathrm{edge}}}{A}\sin\Theta(\mathbf{m}\sin\Theta-\hat{\rho})=\frac{\partial\mathbf{m}}{\partial\mathbf{\rho}}\,.\label{eq:general}
\end{equation}
For rescaling in a way compatible with Eq.~(\ref{eq:rescale}) we need to use $K_{\mathrm{edge}}=K'_{\mathrm{edge}}\sqrt{Q-1}$.

The boundary condition~(\ref{eq:general}) is compatible with the
cases we have been studying: if we set $\Theta=0$ at $\rho_{\mathrm{max}}$
then $\frac{\partial\mathbf{m}}{\partial\mathbf{\rho}}$ must also
be zero at $\rho_{\mathrm{max}}$. That is, the homogenous Dirichlet
boundary condition enforces the free boundary condition (FBC).

However, an expression for more general scenarios is also possible:
if $\Theta\ne0$ at the boundary, the proper boundary condition would
be: 
\begin{align}
\frac{\partial\Theta}{\partial\rho} & =-\frac{K_{\mathrm{edge}}}{2A}\sin\left(2\Theta\right)\cos^{2}\left(\Phi-\phi\right)\\
\frac{\partial\Phi}{\partial\rho} & =\frac{K_{\mathrm{edge}}}{2A}\sin\left(2(\Phi-\phi)\right)\,.
\end{align}

These boundary conditions are important for studying dynamical skyrmions~\cite{statuto_generation_2018,zhou_dynamically_2015}.


\section{Derivation of the Landau-Lifshitz-Slonczewksi equation}

The magnetization dynamics in the presence of spin-polarized currents
are obtained by adding the Slonczewski spin~torque to the Landau-Lifshitz-Gilbert
equation~\citep{ralph_spin_2008}: 
\begin{equation}
\mathbf{\dot{m}}=
\alpha\mathbf{m}\times\mathbf{\dot{m}}
+\gamma_{0}M_s
\left[\sigma V\frac{\mathbf{m}\times\mathbf{m}\times\mathbf{\hat m}_\mathrm{p}}{1+\nu\mathbf{m}\cdot\mathbf{\hat m}_\mathrm{p}}
-\mathbf{m}\times\mathbf{h_{\mathrm{eff}}}
\right]\, .
\label{eq:LLS}
\end{equation}
Cross-multiplying Eq.~(\ref{eq:LLS})
with $\mathbf{m}$, substituting the result back into the second term
of Eq.~(\ref{eq:LLS}), expanding triple cross products, and exploiting the
fact that $\mathbf{m}\cdot\mathbf{\dot{m}}=0$ yields
\begin{equation}
\mathbf{\dot{m}}=\mathbb{L}\mathbf{h_{\mathrm{eff}}}-\frac{\gamma_{0}M_s\sigma V}{1+\alpha^{2}}
\left[
\frac{\alpha\mathbf{m}\times\mathbf{\hat m}_\mathrm{p}-\mathbf{m}\times\mathbf{m}\times\mathbf{\hat m}_\mathrm{p}}{1+\nu\mathbf{m}\cdot\mathbf{\hat m}_\mathrm{p}}
\right]\, .
\end{equation}
Because $\alpha$ is of order $10^{-2}$, the $\alpha^2$ term in the denominator is usually dropped. In our case, this is not necessary.

Because the first term in brackets is proportional to $\alpha$ it
has been named the 'damping-like' torque, and the second term the 'field-like'
torque. This is unfortunate since the second term acts along the field
lines while first term is curl-like. Following our discussion about thermal
equilibrium it should be clear that the second term permits the definition
of a potential, while the first does not. In our framework, the first
term gets absorbed by the transformation to the rotating frame while the second term is absorbed into the energy functional.

In standard micromagnetic packages \cite{vansteenkiste_design_2014,m._j._donahue_oommf_1999} the third term in~Eq.~(\ref{eq:LLS}) is written as
\begin{equation}
\left|
\gamma_0
\right|\beta
\epsilon 
\left(
\mathbf{m}
\times
\mathbf{m_p}
\times
\mathbf{m}
\right), 
\end{equation} with the parameters
$$
\epsilon=\frac{P\Lambda^2}{(\Lambda^2+1)+(\Lambda^2-1)(\mathbf{m\cdot m_p})}
\mathrm{\quad and\quad } 
\beta=\left|
\frac{\hbar}{\mu_0 e}
\right|
\frac{J}{dM_s}$$
with polarization $P$, asymmetry $\Lambda$, and current density $J$ expressed in units of $\mathrm{A/m^2}$. The parameters $\nu$ and $\Lambda$ are related by
$$\nu=\frac{\Lambda^2-1}{\Lambda^2+1}\qquad\Lambda^2=\frac{1+\nu}{1-\nu}.$$
The normalized spin torque coefficient is $\sigma=\frac{J}{J_c}$, where the critical current density
$J_c=\frac{\mu_0edM_s^2}{\xi \hbar}$ depends on an efficiency factor $\xi=\frac{P(\nu+1)}{2}$ which in this paper we fixed to $\xi=1$.

\section{Equilibrium distribution of the Landau-Lifshitz and Landau-Lifshitz-Gilbert equations}

In this section we summarize the well known result that dynamical
systems of the form Eq.~(\ref{eq:rotatingframe}) evolve toward a thermal equilibrium distribution for the magnetization. Consider a segment $d\mathcal{V}=l^2_\mathrm{ex}d$ of the sample with uniform magnetization; the exchange interaction with nearest-neighbor regions can be assumed to be included in the effective field. In this section we assume that the restrictions described above are satisfied.

To study thermal effects we separate the stochastic field
$\mathbf{h}_{\mathrm{th}}$ from the deterministic field $\mathbf{h}_{\mathrm{det}}=\mathbf{h}_{\mathrm{eff}}+\mathbf{h}_{\mathrm{ST}}$. In this way, Eq.~(\ref{eq:rotatingframe}) can be written in Langevin form as
\cite{garcia-palacios_langevin-dynamics_1998} 
\begin{align}
\mathbf{\dot{m}}_{\mathrm{rotating\ frame}} & =\mathbb{L}\mathbf{h_{\mathrm{det}}}+\mathbb{L}\mathbf{h}_{\mathrm{th}}\label{eq:langevin}
\end{align}
which, due to the constraint of constant magnetization magnitude,
can be interpreted using either the Ito or Stratonovich intepretation~\cite{berkov_magnetization_2007}.

For the following discussion it is convenient to rescale the energy functional
$\mathcal{E}'=-\frac{\mathcal{E_{\mathrm{tot}}}}{\mu_{0}M_{s}^{2}}$
and rewrite the deterministic field as $\mathbf{h}_{\mathrm{det}}=-\frac{\delta\mathcal{E}_{\mathrm{tot}}'}{\delta\mathbf{m}}$.
The standard way of writing the RHS of~Eq.~(\ref{eq:langevin}) 
is as the sum of a ``deterministic
drift'', representing the evolution at zero temperature, $\mathbf{b}=\mathbb{L}\mathbf{h_{\mathrm{det}}}$,
and a stochastic noise process $\mathbf{h}_{\mathrm{th}}=\sqrt{2\eta}\mathbf{\dot{W}}$:
\begin{equation}
\mathbf{\dot{m}}_{\mathrm{rotating\ frame}}=\mathbf{b}+\sqrt{2\eta}\mathbb{L}\mathbf{\dot{W}}.\label{eq:driftandnoise}
\end{equation}
In the jargon of stochastic equations $\mathbb{L}$ is known as the
diffusion matrix, and the product $a=\mathbb{L}\mathbb{L}^{T}$ as
the diffusion tensor.

The diffusion matrix can be separated into a symmetric and antisymmetric
part: 
\begin{equation}
\mathbb{L}=\mathbb{L_{S}}+\mathbb{L_{A}}.
\end{equation}
Conveniently, the two parts satisfy $\mathbb{L_{A}}\mathbb{L_{A}}^{T}=-\frac{\gamma'M_s}{\alpha}\mathbb{L_{S}}$.
This is seen more easily using tensor notation 
\begin{align}
\mathbb{L_{A}}_{i,k} & =-\gamma'M_s\epsilon_{ijk}m_{j}\\
\mathbb{L_{S}}_{i,k} & =\gamma'M_s\alpha\left[\delta_{i,k}-m_{i}m_{k}\right]\, .
\end{align}
Using the property of the Levi-Civita tensor $\epsilon_{ijk}\epsilon_{imn}=\left(\delta_{jm}\delta_{mn}-\delta_{jn}\delta_{km}\right)$, the diffusion tensor and its ``inverse'' can
be shown to be, respectively, 
\begin{equation}
a=\gamma'^2M_s^2\left(1+\alpha^{2}\right)\left[\delta_{i,k}-m_{i}m_{k}\right]
=\frac{\gamma_0}{\alpha}M_s\mathbb{L_S}
\end{equation}
\begin{equation}
a^{-1}=\frac{\left[\delta_{i,k}-m_{i}m_{k}\right]}{\gamma'^2M_s^2\left(1+\alpha^{2}\right)}.
\end{equation}

Because the diffusion tensor has a zero eigenvalue, it
is not invertible. As a consequence, the component of
the thermal field parallel to $\mathbf{m}$ has no effect on the time
evolution, and its inverse is defined only for vectors perpendicular
to $\mathbf{m}$, which is the case for both $\mathbf{\dot{m}}$ and
$\mathbf{b}$. For this reason, the non-invertibility of $a$ 
in three dimensions presents no obstacle for analysis.

The analysis presented in~\cite{garcia-palacios_langevin-dynamics_1998}
shows that the Langevin-type equation~(\ref{eq:driftandnoise}) is associated
with a Fokker-Planck equation of the form 
\begin{equation}
\dot{P}=\nabla_{\mathbf{m}}\cdot\left(\left[\mathbf{b}-\eta\mathbb{L}\mathbb{L}^{T}\nabla_{\mathbf{m}}\right]P\right).
\end{equation}

For the noise strength $\eta$ to be directly related to temperature,
the Boltzmann distribution $P\left(\mathbf{m}\right)=\frac{1}{\mathcal{Z}}e^{-\frac{\mathcal{E}_{\mathrm{tot}}d\mathcal{V}}{k_{B}T}}$
must be a stationary solution of the Fokker-Planck equation, i.e.,

\begin{equation}
\dot{P}=\nabla_{\mathbf{m}}\cdot\left(\left[\mathbf{b}-\eta\mathbb{L}\mathbb{L}^{T}\nabla_{\mathbf{m}}\right]\frac{1}{\mathcal{Z}}e^{-\frac{\mathcal{E}_{\mathrm{tot}}d\mathcal{V}}{k_{B}T}}\right)
\end{equation}
must vanish.

To proceed, it is useful to write ${\bf b}$ as
\begin{equation}
\mathbf{b}=\left(\mathbb{L_{S}}+\mathbb{L_{A}}\right)\left(-\frac{\delta\mathcal{E}_{\mathrm{tot}}'}{\delta\mathbf{m}}\right).
\end{equation}
The antisymmetric parts result in a curl-like term

\begin{align}
\mathbb{L_{A}}\left(-\frac{\delta\mathcal{E}'_{\mathrm{tot}}}{\delta\mathbf{m}}\right)P & =\gamma'M_s\mathbf{m}\times\frac{\delta\mathcal{E'}_{\mathrm{tot}}}{\delta\mathbf{m}}\frac{1}{\mathcal{Z}}e^{-\frac{\mathcal{E}_{\mathrm{tot}}}{k_{B}T}}\nonumber\\
 & =\frac{\gamma'}{\mu_0M_s}\nabla_{\mathbf{m}}\times(\mathbf{m}\mathcal{E}_{\mathrm{tot}})\frac{1}{\mathcal{Z}}e^{-\frac{\mathcal{E}_{\mathrm{tot}}}{k_{B}T}}\nonumber\\
 & =-\frac{\gamma'k_{B}T}{\mu_0M_sd\mathcal{V}}\nabla_{\mathbf{m}}\times\left(\mathbf{m}P\right)
\end{align}
which is divergenceless. In a similar fashion, the symmetric piece becomes gradient-like:
\begin{align}
P\mathbb{L_{S}}\left(-\frac{\delta\mathcal{E}'_{\mathrm{tot}}}{\delta\mathbf{m}}\right) & =-\mathbb{L_{S}}\frac{1}{\mathcal{Z}}e^{-\frac{\mathcal{E}_{\mathrm{tot}}}{k_{B}T}}\nabla_{\mathbf{m}}\mathcal{E}'_{\mathrm{tot}}\nonumber\\
 & =\frac{k_{B}T}{\mu_0 M_s^2d\mathcal{V}}\mathbb{L_{S}}\nabla_{\mathbf{m}}P\, .
\end{align}

The stationarity condition, $\dot{P}=0$, can be simplified as
\begin{align}
0 & =\nabla_{\mathbf{m}}\cdot\left(
\frac{k_{B}T\mathbb{L_{S}}\nabla_{\mathbf{m}}P}{\mu_0 M_s^2d\mathcal{V}}
-\frac{\eta\gamma_0M_s\mathbb{L_{S}}\nabla_{\mathbf{m}}P}{\alpha}\right)\nonumber\\
 & =\nabla_{\mathbf{m}}\cdot\left(\left[ 
 \frac{k_{B}T}{\mu_0 M_s^2d\mathcal{V}}
-\frac{\eta\gamma_0 M_s}{\alpha}\right]\mathbb{L_{S}}\nabla_{\mathbf{m}}P\right).
\end{align}
This establishes the relation between the noise strength and temperature as $\eta=\frac{\alpha k_{B}T}{2 A d\gamma_0 M_s}$.
We emphasize that the second term in Eq.~(\ref{eq:curlLLGS})
is the reason that most spin-torque driven systems are not in thermal
equilibrium.

In the stationary reference
frame, the Fokker-Planck equation leads to a differential equation for the distribution $P$ of the form 
\begin{equation}
\dot{P}=\nabla_{\mathbf{m}}\cdot\left(-\frac{\gamma_0}{\mu_0M_s}\nabla_{\mathbf{m}}\times\left(\mathbf{m}\mathcal{E}_{\mathrm{ST}}\right)P\right)
\end{equation}
since all other terms on the RHS still cancel. This can be manipulated
into the more compact form 
\begin{equation}
\dot{P}=-P\frac{2Ad\gamma_0M_s}{k_BT}\Xi=-P\frac{\alpha}{\eta}\Xi\, ,
\end{equation}
which implies that fluctuations near equilibrium evolve at a rate $\frac{\alpha}{\eta}\Xi$.

\section{\label{sec:uniaxial-macrospin}Uniaxial macrospin}

Previous work (see, for example~\cite{butler_switching_2012,taniguchi_thermally_2011,he_switching_2007,pinna_large_2016}) has studied the problem of reversal rates of a uniaxial macrospin in the presence of STT. Here, we briefly discuss how our approach compares to previous work on these systems.

Refs.~\cite{butler_switching_2012,taniguchi_thermally_2011,he_switching_2007} introduce an effective energy term as an ansatz for the stationary solution of the Fokker-Planck equation; this term is consistent with our definition of the pseudo-energy. The equations of motion are phenomenologically equivalent, but our version has an explicit curl-like term. As explained in the main text, this term can be absorbed by transforming to an appropiate reference frame when $\nu=0$.

However, when $\nu\ne 0$, one cannot drop the second term of Eq.~(\ref{eq:crossLLGS}). Nevertheless, in the highly symmetric case such a term does not result in an increase of the total energy $\mathcal{E}_{tot}$. Its inclusion in the dynamics affects the {\it deterministic\/} trajectories but not the energy distribution of the ensemble. For this reason, the Boltzmann distribution for $\nu=0$ is also a solution of the Fokker-Planck equation when $\nu\ne0$ {\it as long as\/} the fields are axially symmetric.

A simple geometrical analogue helps clarify this point.
\begin{figure}
a)\includegraphics[width=1in]{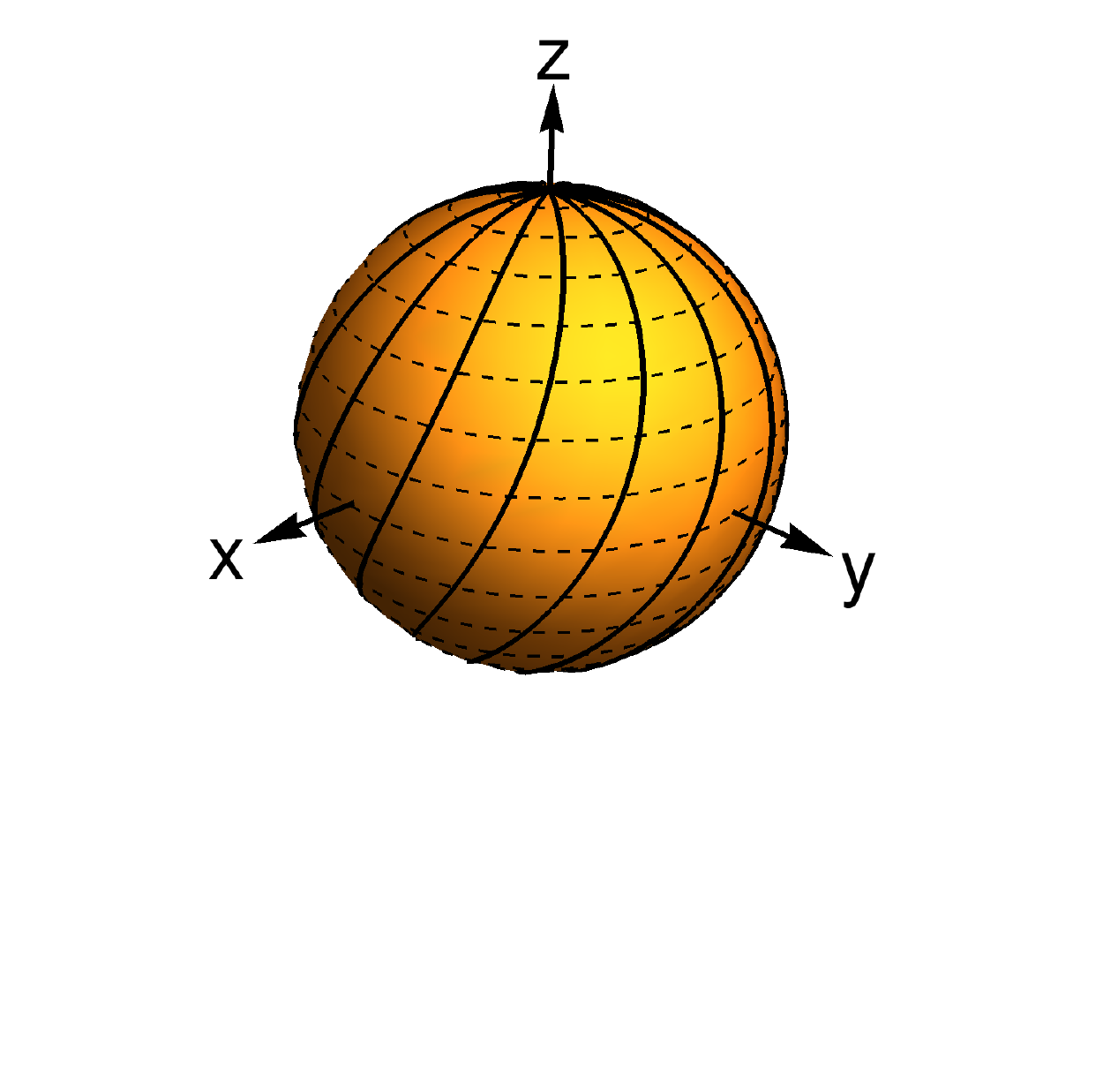}b)\includegraphics[width=1in]{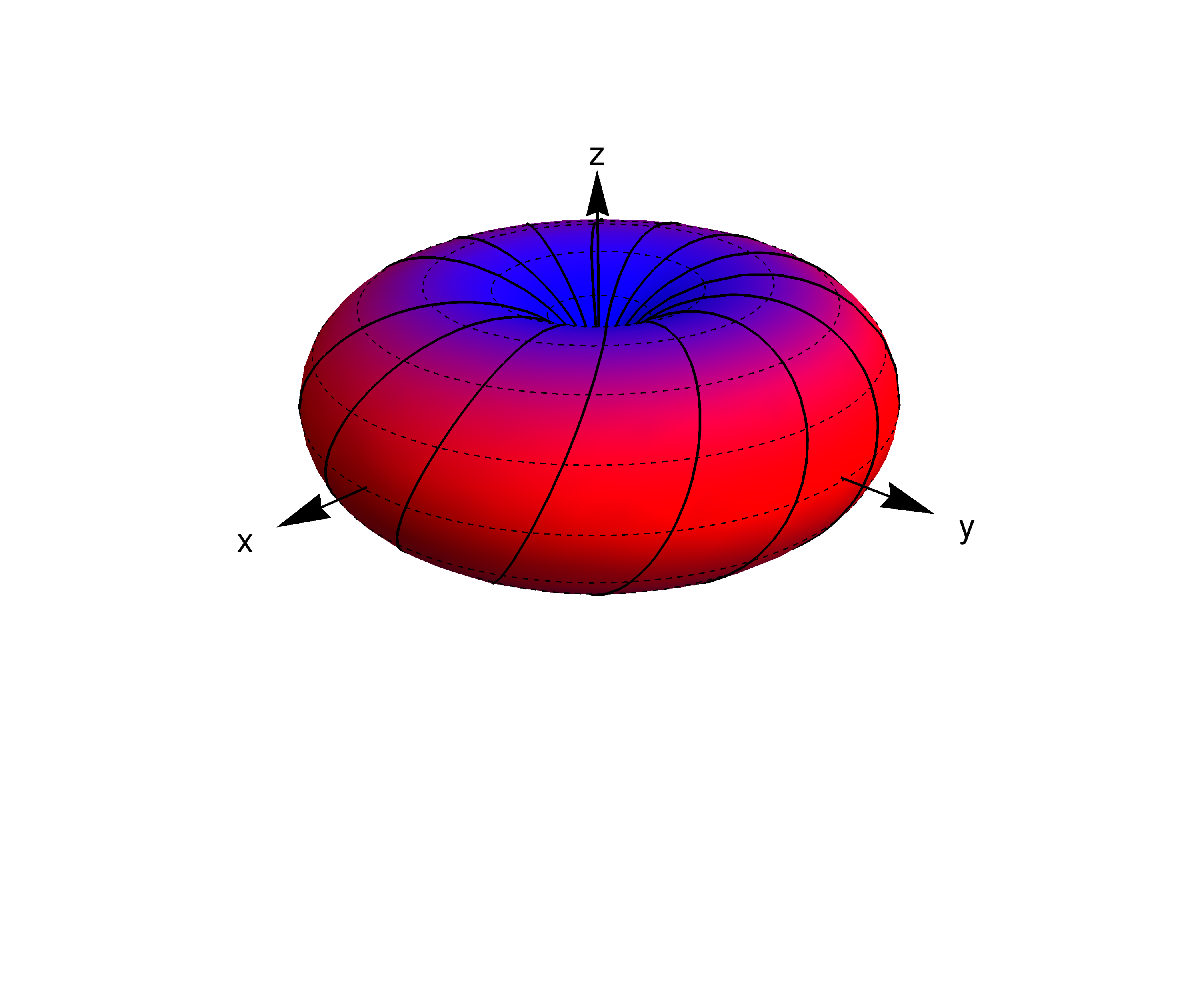}c)\includegraphics[width=1.5in]{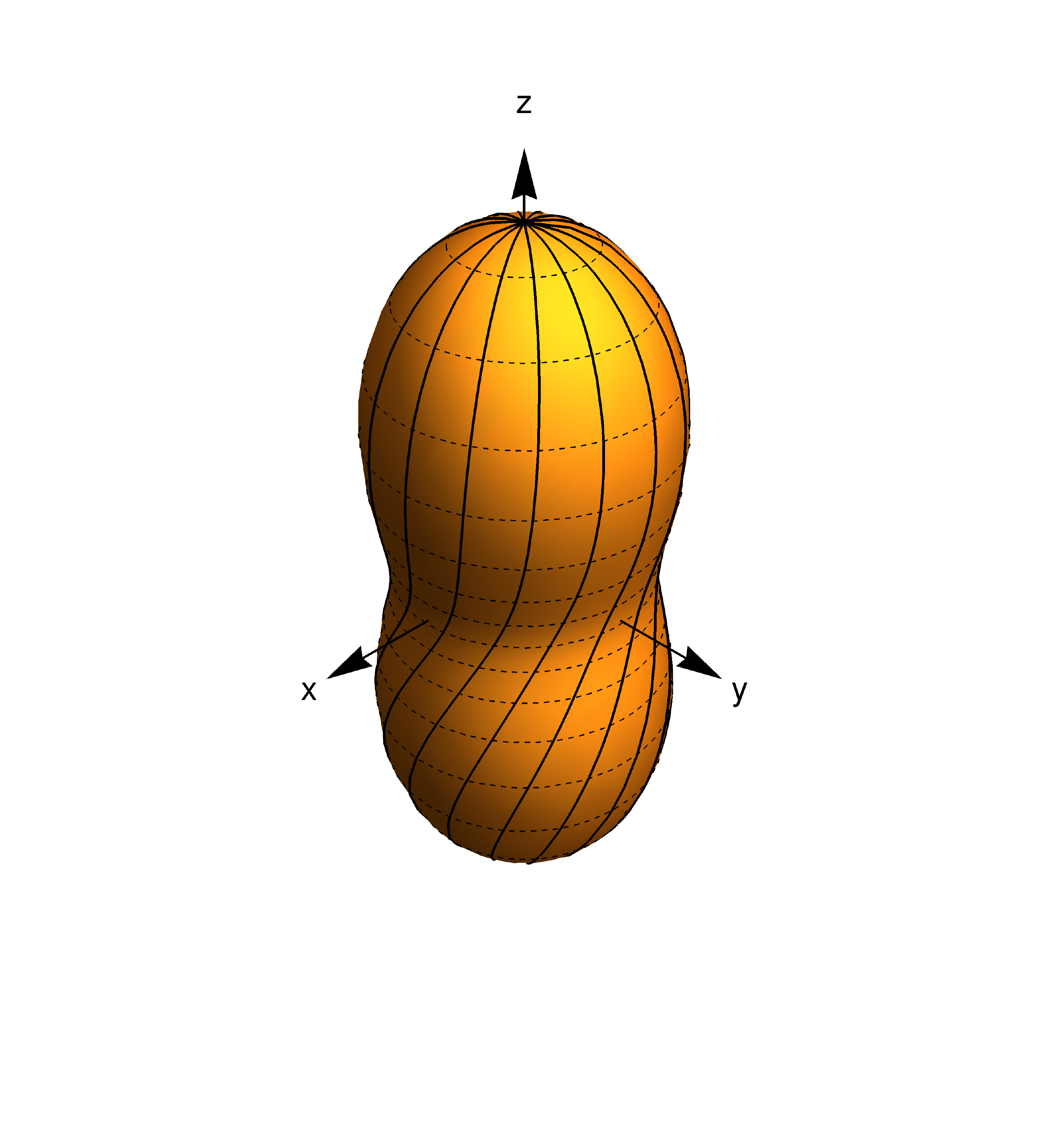}\caption{\label{fig:energyanddistribution}(a) Instantaneous state of a twisted coordinate system produced by a latitude dependent rate of rotation. (b) The hypothetical
energy surface of a system with rotational symmetry sharing the polar axis with the twisting coordinate system. The shape of this surface remains unchanged even though the coordinate system is twisted around it. (c) The corresponding Boltzmann distribution function for the same
magnetic system at a high temperature.}
\end{figure}
In a spherical coordinate system with $\mathbf{\hat m}_\mathrm{p}$ as the polar axis, $\mathbf{m}\cdot\mathbf{h}_{\mathrm{ST}}=\cos\Theta$, $\mathbf{m}\times\mathbf{\hat m}_\mathrm{p}=-\sin\Theta\hat{\Phi}$, and the second term in Eq.~(\ref{eq:crossLLGS}) becomes $$-\nu\left[\mathbf{m}\cdot\mathbf{h}_{\mathrm{ST}}\right]\mathbf{m}\times\mathbf{\hat m}_\mathrm{p}=\omega(\nu;\Theta)\sin\Theta\hat{\Phi}$$ 
with
$$\omega(\nu;\Theta)=\omega_{ST}\frac{\nu\cos\Theta}{\left(1+\nu\cos\Theta\right)}\, .$$ 
This would have the proper structure except that now the angular velocity of the rotating frame depends on $\Theta$. In that case, one could envision a twisted coordinate systems for which the meridians twist around $\mathbf{\hat m}_\mathrm{p}$ in a screw-like fashion, such as the one shown in Fig.~\ref{fig:energyanddistribution}a, with the degree of twist increasing with time. With the exception of the poles, the transformation between a fixed frame and a twisting reference frame is invertible.  Because none of of the energy terms depend on the azimuthal angle, the energy remains unchanged even as the meridians appears more twisted from the perspective of the stationary frame (Fig 1b). In consequence, the equilibrium distribution function is invariant even though the individual trajectories of the particle become more mixed. 

In summary, the introduction of a pseudo-energy associated with spin torque, the absorption of the extra precessional term into a change of reference frame, and the symmetry of the pseudo-energy with respect to the reference frame guarantee that at long times uniaxially symmetric systems reach thermal equilibrium in the presence of spin-torque even for cases where $\nu\ne0$.

The work of Pinna et al.~\cite{pinna_large_2016} treats the uniaxial problem using the Freidlin-Wentzell formalism. Their fluctuational trajectories are the macrospin version of the fluctuational trajectories proposed in  Sect.~\ref{sec:FW-connection}. They demonstrated that, provided toefere deviation from uniaxial symmetry is not too strong, the fluctuational paths will not cross and the optimal escape trajectory remains well-behaved (e.g., no singularities occur in the action). As shown in Sect.~\ref{sec:FW-connection}, the action along the escape fluctuational path remains $S=2\Delta \mathcal{E}_\mathrm{tot}$. 

Because of its symmetry, the uniaxial macrospin case provides a simpler situation which shows that these two distinct approaches provide consistent and complementary results. 

\section{\label{sec:Explicity-expresions-for}Explicit expresions for energy
and field in the rotating frame}

The pseudo-energy used in this paper is given by $\mathcal{E}_{\mathrm{tot}}=\mathcal{E}_{ex}+\mathcal{E}_{K}+\mathcal{E}_{Z}+\mathcal{E}_{\mathrm{ST}}$.
Although the pseudo-energy is invariant under changes of reference frame, the explicit form of all terms (except $\mathcal{E}_{ex}$) depends on the
orientation the coordinate axes. 
The fields in the rotating frame can be obtained using the general
expressions
\begin{equation}
h_{\Theta}=-\frac{1}{\mu_0 M_s^2}\frac{\delta\mathcal{E}}{\delta\Theta}\qquad h_{\Phi}=-\frac{1}{\mu_0 M_s^2}\frac{1}{\sin\Theta}\frac{\delta\mathcal{E}}{\delta\Phi}\, .
\end{equation}
Following~\cite{hoefer_theory_2010}, we will treat $\Phi$ and $\Theta$ perturbatively
with the expansions
\begin{equation}
\Theta=\Theta_{0}+\epsilon\Theta_{1}\quad\Phi=\Phi_{0}+\frac{\epsilon\Phi_{1}}{\sin\Theta_{0}}\, .
\end{equation}
Here $\Theta_{0}$ and $\Phi_{0}=0$ are solutions of Eqs.~(\ref{eq:thetamovingframe})
and~(\ref{eq:phimovingframe}) with $\epsilon$
set to zero; $\Theta_{1}$ and $\Phi_{1}$ are the perturbative solutions
to first order in $\epsilon$. The expansion incorporates
perturbations arising from small experimental misalignments $\theta_{z}$,
$\theta_{f}$ and the spin-torque asymmetry parameter~$\nu$. To denote the order of the perturbation we use a superscript, e.g. $h_{ex,\Phi}^{(1)}$ denotes the perturbation expansion to first order in the exchange-induced
field in the azimuthal direction.

To first order, the azimuthal time derivative is
\begin{equation}
\sin\Theta\dot{\Phi}\approx\left(\sin\Theta\dot{\Phi}\right)^{(0)}+\epsilon\Theta_{1}\cos\Theta_{0}\dot{\Phi_{0}}+\epsilon\dot{\Phi_{1}}\, .
\end{equation}
We now consider the contributions of each of the terms in the pseudo-energy.
\subsection{Exchange}

The exchange energy surface density is invariant with respect to changes in the coordinate
system 
\begin{equation}
\frac{\mathcal{E}_{ex}}{\mu_0M_s^2}=
\frac{l^2_{ex}}{2}\left(\left|\nabla\Theta\right|^{2}+\sin^{2}\Theta\left|\nabla\Phi\right|^{2}\right)\, .
\end{equation}
Similarly for the field expressions:
\begin{equation}
h_{ex,\Phi}=\sin\Theta\nabla^{2}\Phi+2\cos\Theta\nabla\Theta\cdot\nabla\Phi
\end{equation}
\begin{equation}
h_{ex,\Theta}=\nabla^{2}\Theta-\frac{1}{2}\sin2\Theta\left|\nabla\Phi\right|^{2}\, .
\end{equation}
Their perturbative expansions, to first order in $\epsilon$,
are
\begin{equation}
\begin{array}{ccc}
h_{ex,\Phi} & \approx & \epsilon\left[\nabla^{2}\Phi_{1}+\Phi_{1}\left(\nabla\Theta_{0}\right)^{2}-\Phi_{1}\frac{\cos\Theta_{0}}{\sin\Theta_{0}}\nabla^{2}\Theta_{0}\right]\end{array}
\end{equation}
and
\begin{equation}
h_{ex,\Theta}\approx h_{ex,\Theta}^{(0)}+\epsilon\left[\nabla^{2}\Theta_{1}\right]\, ,
\end{equation}
where $h_{ex,\Theta}^{(0)}=\nabla^{2}\Theta_{0}$. We note that the
term in square brackets with $\sin\Theta_{0}$ in the denominator
can be replaced by using the equation for $h_{\Theta}^{(0)}=0$ (Eq.~(\ref{eq:unperturbedthetafieldequation})). The first order term
is then
\begin{equation}
h_{ex,\Phi}^{(1)}\approx\left[\left[\left(\nabla\Theta_{0}\right)^{2}-\cos\Theta_{0}\left(\cos\Theta_{0}-\omega_{h}\right)\right]+\nabla^{2}\right]\Phi_{1}\, .
\end{equation}

\subsection{Anisotropy}

The anisotropy surface energy density
\begin{equation}
\frac{\mathcal{E}_{K}}{\mu_0M_s^2}=-\frac{Q-1}{2}\left(\mathbf{m}\cdot\mathbf{\hat{z}}_{L}\right)^{2}
\end{equation}
favors magnetization orientation along the anisotropy axis $\mathbf{\hat{z}}_{L}$,
which is oriented perpendicular to the sample plane. This axis rotates
around $\mathbf{\hat m}_\mathrm{p}$ in the rotating frame. These two vectors make
an angle $\theta_{f}$. The energy density for a given magnetization
orientation $\text{\ensuremath{\left(\Theta,\Phi\right)}}$ in the
rotating frame is time-dependent:
\begin{multline}
\frac{\mathcal{E}_{K}}{\mu_0M_s^2}=-\frac{\left[\sin\Theta\sin\theta_{f}\cos\left(\omega t+\Phi\right)+\cos\theta_{f}\cos\Theta\right]^{2}}{2/(Q-1)}\, .
\end{multline}
Consequently, the fields are
\begin{equation}
\begin{array}{ccc}
\frac{h_{K,\Theta}}{Q-1} & = & \frac{1}{2}\sin2\Theta\sin^{2}\theta_{f}\cos^{2}\left(\omega t+\Phi\right)\\
 &  & \frac{1}{4}\cos2\Theta\sin2\theta_{f}\cos\left(\omega t+\Phi\right)\\
 &  & -\frac{1}{2}\sin2\Theta\cos^{2}\theta_{f}
\end{array}
\end{equation}
and
\begin{equation}
\begin{array}{ccc}
\frac{h_{K,\Phi}}{Q-1} & = & -\frac{1}{2}\sin\Theta\sin^{2}\theta_{f}\sin\left[2\left(\omega t+\Phi\right)\right]\\
 &  & -\frac{1}{2}\cos\Theta\sin2\theta_{f}\sin\left(\omega t+\Phi\right)\, .
\end{array}
\end{equation}
We set $h_{K,\Theta}^{(0)}=\frac{Q-1}{2}\sin2\Theta_{0}$ for the
limit $\theta_{f}=0$. The perturbative expansions for small $\theta_{f}$
are
\begin{equation}
\begin{array}{ccc}
\frac{h_{K,\Theta}}{\left(Q-1\right)} & \approx & \frac{h_{K,\Theta}^{(0)}}{Q-1}\\
 &  & -\frac{\theta_{f}}{2}\cos2\Theta_{0}\cos\omega t+\epsilon\Theta_{1}\cos2\Theta_{0}.
\end{array}
\end{equation}
and
\begin{equation}
\begin{array}{ccc}
h_{K,\Phi} & \approx\frac{1}{2} & \left(Q-1\right)\theta_{f}\cos\Theta_{0}\sin\omega t.\end{array}
\end{equation}

\subsection{Zeeman}

To describe the orientation of the field, we use polar coordinates
in the laboratory reference frame
\begin{equation}
\mathbf{h_{Z}}=\left(\cos\phi_{Z}\sin\theta_{Z},\sin\phi_{Z}\sin\theta_{Z},\cos\theta_{Z}\right)
\end{equation}
where the field misalignment $\theta_{z}$ is presumed to be small.
The Zeeman energy density $
\frac{\mathcal{E}_{Z}}{\mu_0M_s^2}
=-\mathbf{m}\cdot\text{\ensuremath{\mathbf{h_{Z}}}}$
in the rotating frame is
\begin{equation}
\begin{array}{cc}
\frac{\mathcal{E}_{Z}}{-h_{Z}\mu_0M_s^2}= & \cos\phi_{Z}\sin\theta_{z}\left[\sin\Theta\cos\theta_{f}\cos\left(\omega t+\Phi\right)-\sin\theta_{f}\cos\Theta\right]\\
 & \sin\phi_{Z}\sin\theta_{z}\sin\Theta\sin\left(\omega t+\Phi\right)\\
 & \cos\theta_{z}\left[\sin\Theta\sin\theta_{f}\cos\left(\omega t+\Phi\right)+\cos\theta_{f}\cos\Theta\right].
\end{array}
\end{equation}
with corresponding fields
\begin{equation}
\begin{array}{cc}
\frac{h_{Z,\Theta}}{h_{Z}}= & \cos\phi_{Z}\sin\theta_{z}\left[\cos\Theta\cos\theta_{f}\cos\left(\omega t+\Phi\right)+\sin\theta_{f}\sin\Theta\right]\\
 & \sin\phi_{Z}\sin\theta_{z}\cos\Theta\sin\left(\omega t+\Phi\right)\\
 & \cos\theta_{z}\left[\cos\Theta\sin\theta_{f}\cos\left(\omega t+\Phi\right)-\cos\theta_{f}\sin\Theta\right]
\end{array}
\end{equation}
and
\begin{equation}
\begin{array}{cc}
\frac{h_{Z,\Phi}}{h_{Z}}= & -\cos\phi_{Z}\sin\theta_{z}\cos\theta_{f}\sin\left(\omega t+\Phi\right)\\
 & \sin\phi_{Z}\sin\theta_{z}\cos\left(\omega t+\Phi\right)\\
 & -\cos\theta_{z}\sin\theta_{f}\sin\left(\omega t+\Phi\right)\, .
\end{array}
\end{equation}

Using $h_{Z,\Theta}^{(0)}=-\sin\Theta_{0}$, the leading order perturbation
expansion for small $\theta_{z}$ and $\theta_{f}$ are
\begin{equation}
\begin{array}{cc}
\frac{h_{Z,\Theta}}{h_{Z}}= & h_{Z,\Theta}^{(0)}+\theta_{z}\cos\Theta_{0}\cos\left(\omega t-\phi_{Z}\right)\\
 & \theta_{f}\cos\Theta\cos\left(\omega t\right)-\epsilon\Theta_{1}\cos\Theta_{0}
\end{array}
\end{equation}
and
\begin{equation}
\begin{array}{cc}
\frac{h_{Z,\Phi}}{h_{Z}}= & -\theta_{z}\cos\phi_{Z}\sin\left(\omega t-\phi_{z}\right)\\
 & -\theta_{f}\sin\left(\omega t\right).
\end{array}
\end{equation}

\subsection{Spin torque}

The pseudo-potential energy density and field were introduced
in Eqs.~(\ref{eq:spintorquepseudopotential}) and~(\ref{eq:spintorquefield}).
The denominator of the spin-torque field term can be written as an expansion
in~$\nu$:
\begin{equation}
\frac{1}{1+\nu\mathbf{m}\cdot\mathbf{\hat m}_\mathrm{p}}=\sum_{n=0}^{\infty}\left(-\nu\mathbf{m\cdot p}\right)^{n}\, .\label{eq:nuexpansion}
\end{equation}
Notice that in the rotating frame $\mathbf{m}\cdot\mathbf{\hat m}_\mathrm{p}=\cos\Theta$,
$\mathbf{\hat m}_\mathrm{p}\cdot\hat{\Theta}=-\sin\Theta$, and $\mathbf{\hat m}_\mathrm{p}\cdot\hat{\Phi}=0$.
After rescaling, the field components are
\begin{equation}
h_{ST,\Theta}=+\text{\ensuremath{\frac{\sigma V}{\alpha}\sin\Theta}\ensuremath{\sum_{0}^{\infty}\left(-\nu\cos\Theta\right)^{n}}}
\end{equation}
and
\begin{equation}
h_{ST,\Phi}=0\, .
\end{equation}

Taking $\nu$ as a small parameter and $h_{ST,\Theta}^{(0)}=\frac{\sigma V}{\alpha}\sin\Theta_{0}$,
the perturbative expansion becomes

\begin{equation}
h_{ST,\Theta}=h_{ST,\Theta}^{(0)}+\ensuremath{\frac{\sigma V}{\alpha}}\ensuremath{\left[\epsilon\Theta_{1}\cos\Theta_{0}-\frac{\nu}{2}\sin2\Theta_{0}\right]}\, .
\end{equation}

To first order in $\nu$, the term in parenthesis in Eq.~(\ref{eq:phimovingframe}) is
\begin{equation}
\frac{\sigma' V}{\alpha'}\left(P_{\Theta}-\sin\Theta\right)\approx-\nu\frac{\sigma' V}{2\alpha'}\sin2\Theta_{0}\, .
\end{equation}

\bibliography{main}

\begin{thebibliography}{65}%
\makeatletter
\providecommand \@ifxundefined [1]{%
 \@ifx{#1\undefined}
}%
\providecommand \@ifnum [1]{%
 \ifnum #1\expandafter \@firstoftwo
 \else \expandafter \@secondoftwo
 \fi
}%
\providecommand \@ifx [1]{%
 \ifx #1\expandafter \@firstoftwo
 \else \expandafter \@secondoftwo
 \fi
}%
\providecommand \natexlab [1]{#1}%
\providecommand \enquote  [1]{``#1''}%
\providecommand \bibnamefont  [1]{#1}%
\providecommand \bibfnamefont [1]{#1}%
\providecommand \citenamefont [1]{#1}%
\providecommand \href@noop [0]{\@secondoftwo}%
\providecommand \href [0]{\begingroup \@sanitize@url \@href}%
\providecommand \@href[1]{\@@startlink{#1}\@@href}%
\providecommand \@@href[1]{\endgroup#1\@@endlink}%
\providecommand \@sanitize@url [0]{\catcode `\\12\catcode `\$12\catcode
  `\&12\catcode `\#12\catcode `\^12\catcode `\_12\catcode `\%12\relax}%
\providecommand \@@startlink[1]{}%
\providecommand \@@endlink[0]{}%
\providecommand \url  [0]{\begingroup\@sanitize@url \@url }%
\providecommand \@url [1]{\endgroup\@href {#1}{\urlprefix }}%
\providecommand \urlprefix  [0]{URL }%
\providecommand \Eprint [0]{\href }%
\providecommand \doibase [0]{https://doi.org/}%
\providecommand \selectlanguage [0]{\@gobble}%
\providecommand \bibinfo  [0]{\@secondoftwo}%
\providecommand \bibfield  [0]{\@secondoftwo}%
\providecommand \translation [1]{[#1]}%
\providecommand \BibitemOpen [0]{}%
\providecommand \bibitemStop [0]{}%
\providecommand \bibitemNoStop [0]{.\EOS\space}%
\providecommand \EOS [0]{\spacefactor3000\relax}%
\providecommand \BibitemShut  [1]{\csname bibitem#1\endcsname}%
\let\auto@bib@innerbib\@empty
\bibitem [{\citenamefont {Kosevich}\ \emph {et~al.}(1990)\citenamefont
  {Kosevich}, \citenamefont {Ivanov},\ and\ \citenamefont
  {Kovalev}}]{kosevich_magnetic_1990}%
  \BibitemOpen
  \bibfield  {author} {\bibinfo {author} {\bibfnamefont {A.~M.}\ \bibnamefont
  {Kosevich}}, \bibinfo {author} {\bibfnamefont {B.~A.}\ \bibnamefont
  {Ivanov}},\ and\ \bibinfo {author} {\bibfnamefont {A.~S.}\ \bibnamefont
  {Kovalev}},\ }\bibfield  {title} {\bibinfo {title} {Magnetic {Solitons}},\
  }\href@noop {} {\bibfield  {journal} {\bibinfo  {journal} {Phys. Rep.}\
  }\textbf {\bibinfo {volume} {194}},\ \bibinfo {pages} {117} (\bibinfo {year}
  {1990})}\BibitemShut {NoStop}%
\bibitem [{\citenamefont {Chen}\ \emph {et~al.}(2016)\citenamefont {Chen},
  \citenamefont {Dumas}, \citenamefont {Eklund}, \citenamefont {Muduli},
  \citenamefont {Houshang}, \citenamefont {Awad}, \citenamefont
  {D{\"u}rrenfeld}, \citenamefont {Malm}, \citenamefont {Rusu},\ and\
  \citenamefont {{\r A}kerman}}]{chen_spin-torque_2016}%
  \BibitemOpen
  \bibfield  {author} {\bibinfo {author} {\bibfnamefont {T.}~\bibnamefont
  {Chen}}, \bibinfo {author} {\bibfnamefont {R.~K.}\ \bibnamefont {Dumas}},
  \bibinfo {author} {\bibfnamefont {A.}~\bibnamefont {Eklund}}, \bibinfo
  {author} {\bibfnamefont {P.~K.}\ \bibnamefont {Muduli}}, \bibinfo {author}
  {\bibfnamefont {A.}~\bibnamefont {Houshang}}, \bibinfo {author}
  {\bibfnamefont {A.~A.}\ \bibnamefont {Awad}}, \bibinfo {author}
  {\bibfnamefont {P.}~\bibnamefont {D{\"u}rrenfeld}}, \bibinfo {author}
  {\bibfnamefont {B.~G.}\ \bibnamefont {Malm}}, \bibinfo {author}
  {\bibfnamefont {A.}~\bibnamefont {Rusu}},\ and\ \bibinfo {author}
  {\bibfnamefont {J.}~\bibnamefont {{\r A}kerman}},\ }\bibfield  {title}
  {\bibinfo {title} {Spin-{Torque} and {Spin}-{Hall} {Nano}-{Oscillators}},\
  }\href {https://doi.org/10.1109/JPROC.2016.2554518} {\bibfield  {journal}
  {\bibinfo  {journal} {Proc. IEEE}\ }\textbf {\bibinfo {volume} {104}},\
  \bibinfo {pages} {1919} (\bibinfo {year} {2016})}\BibitemShut {NoStop}%
\bibitem [{\citenamefont {Maci\`a}\ \emph {et~al.}(2014)\citenamefont
  {Maci\`a}, \citenamefont {Backes},\ and\ \citenamefont {Kent}}]{MBD14}%
  \BibitemOpen
  \bibfield  {author} {\bibinfo {author} {\bibfnamefont {F.}~\bibnamefont
  {Maci\`a}}, \bibinfo {author} {\bibfnamefont {D.}~\bibnamefont {Backes}},\
  and\ \bibinfo {author} {\bibfnamefont {A.~D.}\ \bibnamefont {Kent}},\
  }\bibfield  {title} {\bibinfo {title} {Stable magnetic droplet solitons in
  spin-transfer nanocontacts},\ }\href@noop {} {\bibfield  {journal} {\bibinfo
  {journal} {Nat. Nanotechnol.}\ }\textbf {\bibinfo {volume} {9}},\ \bibinfo
  {pages} {992 } (\bibinfo {year} {2014})}\BibitemShut {NoStop}%
\bibitem [{\citenamefont {Divinskiy}\ \emph {et~al.}(2017)\citenamefont
  {Divinskiy}, \citenamefont {Urazhdin}, \citenamefont {Demidov}, \citenamefont
  {Kozhanov}, \citenamefont {Nosov}, \citenamefont {Rinkevich},\ and\
  \citenamefont {Demokritov}}]{divinskiy_magnetic_2017}%
  \BibitemOpen
  \bibfield  {author} {\bibinfo {author} {\bibfnamefont {B.}~\bibnamefont
  {Divinskiy}}, \bibinfo {author} {\bibfnamefont {S.}~\bibnamefont {Urazhdin}},
  \bibinfo {author} {\bibfnamefont {V.~E.}\ \bibnamefont {Demidov}}, \bibinfo
  {author} {\bibfnamefont {A.}~\bibnamefont {Kozhanov}}, \bibinfo {author}
  {\bibfnamefont {A.~P.}\ \bibnamefont {Nosov}}, \bibinfo {author}
  {\bibfnamefont {A.~B.}\ \bibnamefont {Rinkevich}},\ and\ \bibinfo {author}
  {\bibfnamefont {S.~O.}\ \bibnamefont {Demokritov}},\ }\bibfield  {title}
  {\bibinfo {title} {Magnetic droplet solitons generated by pure spin
  currents},\ }\href {https://doi.org/10.1103/PhysRevB.96.224419} {\bibfield
  {journal} {\bibinfo  {journal} {Phys. Rev. B}\ }\textbf {\bibinfo {volume}
  {96}},\ \bibinfo {pages} {224419} (\bibinfo {year} {2017})}\BibitemShut
  {NoStop}%
\bibitem [{\citenamefont {Zhou}\ \emph {et~al.}(2015)\citenamefont {Zhou},
  \citenamefont {Iacocca}, \citenamefont {Awad}, \citenamefont {Dumas},
  \citenamefont {Zhang}, \citenamefont {Braun},\ and\ \citenamefont {{\r
  A}kerman}}]{zhou_dynamically_2015}%
  \BibitemOpen
  \bibfield  {author} {\bibinfo {author} {\bibfnamefont {Y.}~\bibnamefont
  {Zhou}}, \bibinfo {author} {\bibfnamefont {E.}~\bibnamefont {Iacocca}},
  \bibinfo {author} {\bibfnamefont {A.~A.}\ \bibnamefont {Awad}}, \bibinfo
  {author} {\bibfnamefont {R.~K.}\ \bibnamefont {Dumas}}, \bibinfo {author}
  {\bibfnamefont {F.~C.}\ \bibnamefont {Zhang}}, \bibinfo {author}
  {\bibfnamefont {H.~B.}\ \bibnamefont {Braun}},\ and\ \bibinfo {author}
  {\bibfnamefont {J.}~\bibnamefont {{\r A}kerman}},\ }\bibfield  {title}
  {\bibinfo {title} {Dynamically stabilized magnetic skyrmions},\ }\href
  {https://doi.org/10.1038/ncomms9193} {\bibfield  {journal} {\bibinfo
  {journal} {Nat. Commun.}\ }\textbf {\bibinfo {volume} {6}},\ \bibinfo {pages}
  {1} (\bibinfo {year} {2015})}\BibitemShut {NoStop}%
\bibitem [{\citenamefont {Ivanov}\ and\ \citenamefont
  {Kosevich}(1977)}]{ivanov_bound_1977}%
  \BibitemOpen
  \bibfield  {author} {\bibinfo {author} {\bibfnamefont {A.}~\bibnamefont
  {Ivanov}}\ and\ \bibinfo {author} {\bibfnamefont {A.~M.}\ \bibnamefont
  {Kosevich}},\ }\bibfield  {title} {\bibinfo {title} {Bound states of a large
  number of magnons in a ferromagnet with single ion anisotropy},\ }\href@noop
  {} {\bibfield  {journal} {\bibinfo  {journal} {Zh. Eksp. Teor. Fiz.}\
  }\textbf {\bibinfo {volume} {72}},\ \bibinfo {pages} {2000} (\bibinfo {year}
  {1977})}\BibitemShut {NoStop}%
\bibitem [{\citenamefont {Makhankov}(1978)}]{Makhankov78}%
  \BibitemOpen
  \bibfield  {author} {\bibinfo {author} {\bibfnamefont {V.~G.}\ \bibnamefont
  {Makhankov}},\ }\bibfield  {title} {\bibinfo {title} {Dynamics of classical
  solitons (in non-integrable systems)},\ }\href@noop {} {\bibfield  {journal}
  {\bibinfo  {journal} {Phys. Rep.}\ }\textbf {\bibinfo {volume} {35}},\
  \bibinfo {pages} {1} (\bibinfo {year} {1978})}\BibitemShut {NoStop}%
\bibitem [{\citenamefont {Mohseni}\ \emph {et~al.}(2013)\citenamefont
  {Mohseni}, \citenamefont {Sani}, \citenamefont {Persson}, \citenamefont
  {Nguyen}, \citenamefont {Chung}, \citenamefont {Pogoryelov}, \citenamefont
  {Muduli}, \citenamefont {Iacocca}, \citenamefont {Eklund}, \citenamefont
  {Dumas}, \citenamefont {Bonetti}, \citenamefont {Deac}, \citenamefont
  {Hoefer},\ and\ \citenamefont {{\r A}kerman}}]{mohseni_spin_2013}%
  \BibitemOpen
  \bibfield  {author} {\bibinfo {author} {\bibfnamefont {S.~M.}\ \bibnamefont
  {Mohseni}}, \bibinfo {author} {\bibfnamefont {S.~R.}\ \bibnamefont {Sani}},
  \bibinfo {author} {\bibfnamefont {J.}~\bibnamefont {Persson}}, \bibinfo
  {author} {\bibfnamefont {T.~N.~A.}\ \bibnamefont {Nguyen}}, \bibinfo {author}
  {\bibfnamefont {S.}~\bibnamefont {Chung}}, \bibinfo {author} {\bibfnamefont
  {Y.}~\bibnamefont {Pogoryelov}}, \bibinfo {author} {\bibfnamefont {P.~K.}\
  \bibnamefont {Muduli}}, \bibinfo {author} {\bibfnamefont {E.}~\bibnamefont
  {Iacocca}}, \bibinfo {author} {\bibfnamefont {A.}~\bibnamefont {Eklund}},
  \bibinfo {author} {\bibfnamefont {R.~K.}\ \bibnamefont {Dumas}}, \bibinfo
  {author} {\bibfnamefont {S.}~\bibnamefont {Bonetti}}, \bibinfo {author}
  {\bibfnamefont {A.}~\bibnamefont {Deac}}, \bibinfo {author} {\bibfnamefont
  {M.~A.}\ \bibnamefont {Hoefer}},\ and\ \bibinfo {author} {\bibfnamefont
  {J.}~\bibnamefont {{\r A}kerman}},\ }\bibfield  {title} {\bibinfo {title}
  {Spin {Torque}{\textendash}{Generated} {Magnetic} {Droplet} {Solitons}},\
  }\href {https://doi.org/10.1126/science.1230155} {\bibfield  {journal}
  {\bibinfo  {journal} {Science}\ }\textbf {\bibinfo {volume} {339}},\ \bibinfo
  {pages} {1295} (\bibinfo {year} {2013})}\BibitemShut {NoStop}%
\bibitem [{\citenamefont {Backes}\ \emph {et~al.}(2015)\citenamefont {Backes},
  \citenamefont {Maci\`a}, \citenamefont {Bonetti}, \citenamefont {Kukreja},
  \citenamefont {Ohldag},\ and\ \citenamefont {Kent}}]{Backes15}%
  \BibitemOpen
  \bibfield  {author} {\bibinfo {author} {\bibfnamefont {D.}~\bibnamefont
  {Backes}}, \bibinfo {author} {\bibfnamefont {F.}~\bibnamefont {Maci\`a}},
  \bibinfo {author} {\bibfnamefont {S.}~\bibnamefont {Bonetti}}, \bibinfo
  {author} {\bibfnamefont {R.}~\bibnamefont {Kukreja}}, \bibinfo {author}
  {\bibfnamefont {H.}~\bibnamefont {Ohldag}},\ and\ \bibinfo {author}
  {\bibfnamefont {A.~D.}\ \bibnamefont {Kent}},\ }\bibfield  {title} {\bibinfo
  {title} {Direct observation of a localized magnetic soliton in a
  spin-transfer nanocontact},\ }\href
  {https://doi.org/10.1103/PhysRevLett.115.127205} {\bibfield  {journal}
  {\bibinfo  {journal} {Phys. Rev. Lett.}\ }\textbf {\bibinfo {volume} {115}},\
  \bibinfo {pages} {127205} (\bibinfo {year} {2015})}\BibitemShut {NoStop}%
\bibitem [{\citenamefont {Hang}\ \emph {et~al.}(2018)\citenamefont {Hang},
  \citenamefont {Hahn}, \citenamefont {Statuto}, \citenamefont {Maci{\`a}},\
  and\ \citenamefont {Kent}}]{Hang2018}%
  \BibitemOpen
  \bibfield  {author} {\bibinfo {author} {\bibfnamefont {J.}~\bibnamefont
  {Hang}}, \bibinfo {author} {\bibfnamefont {C.}~\bibnamefont {Hahn}}, \bibinfo
  {author} {\bibfnamefont {N.}~\bibnamefont {Statuto}}, \bibinfo {author}
  {\bibfnamefont {F.}~\bibnamefont {Maci{\`a}}},\ and\ \bibinfo {author}
  {\bibfnamefont {A.~D.}\ \bibnamefont {Kent}},\ }\bibfield  {title} {\bibinfo
  {title} {Generation and annihilation time of magnetic droplet solitons},\
  }\href {https://doi.org/10.1038/s41598-018-25134-z} {\bibfield  {journal}
  {\bibinfo  {journal} {Sci. Rep.}\ }\textbf {\bibinfo {volume} {8}},\ \bibinfo
  {pages} {6847} (\bibinfo {year} {2018})}\BibitemShut {NoStop}%
\bibitem [{\citenamefont {Chung}\ \emph {et~al.}(2018)\citenamefont {Chung},
  \citenamefont {Le}, \citenamefont {Ahlberg}, \citenamefont {Awad},
  \citenamefont {Weigand}, \citenamefont {Bykova}, \citenamefont {Khymyn},
  \citenamefont {Dvornik}, \citenamefont {Mazraati}, \citenamefont {Houshang},
  \citenamefont {Jiang}, \citenamefont {Nguyen}, \citenamefont {Goering},
  \citenamefont {Sch\"utz}, \citenamefont {Gr\"afe},\ and\ \citenamefont
  {\AA{}kerman}}]{Chung18}%
  \BibitemOpen
  \bibfield  {author} {\bibinfo {author} {\bibfnamefont {S.}~\bibnamefont
  {Chung}}, \bibinfo {author} {\bibfnamefont {Q.~T.}\ \bibnamefont {Le}},
  \bibinfo {author} {\bibfnamefont {M.}~\bibnamefont {Ahlberg}}, \bibinfo
  {author} {\bibfnamefont {A.~A.}\ \bibnamefont {Awad}}, \bibinfo {author}
  {\bibfnamefont {M.}~\bibnamefont {Weigand}}, \bibinfo {author} {\bibfnamefont
  {I.}~\bibnamefont {Bykova}}, \bibinfo {author} {\bibfnamefont
  {R.}~\bibnamefont {Khymyn}}, \bibinfo {author} {\bibfnamefont
  {M.}~\bibnamefont {Dvornik}}, \bibinfo {author} {\bibfnamefont
  {H.}~\bibnamefont {Mazraati}}, \bibinfo {author} {\bibfnamefont
  {A.}~\bibnamefont {Houshang}}, \bibinfo {author} {\bibfnamefont
  {S.}~\bibnamefont {Jiang}}, \bibinfo {author} {\bibfnamefont {T.~N.~A.}\
  \bibnamefont {Nguyen}}, \bibinfo {author} {\bibfnamefont {E.}~\bibnamefont
  {Goering}}, \bibinfo {author} {\bibfnamefont {G.}~\bibnamefont {Sch\"utz}},
  \bibinfo {author} {\bibfnamefont {J.}~\bibnamefont {Gr\"afe}},\ and\ \bibinfo
  {author} {\bibfnamefont {J.}~\bibnamefont {\AA{}kerman}},\ }\bibfield
  {title} {\bibinfo {title} {Direct observation of {Zhang-Li} torque expansion
  of magnetic droplet solitons},\ }\href
  {https://doi.org/10.1103/PhysRevLett.120.217204} {\bibfield  {journal}
  {\bibinfo  {journal} {Phys. Rev. Lett.}\ }\textbf {\bibinfo {volume} {120}},\
  \bibinfo {pages} {217204} (\bibinfo {year} {2018})}\BibitemShut {NoStop}%
\bibitem [{\citenamefont {Sulymenko}\ \emph {et~al.}(2018)\citenamefont
  {Sulymenko}, \citenamefont {Prokopenko}, \citenamefont {Tyberkevych},
  \citenamefont {Slavin},\ and\ \citenamefont {Serga}}]{Sul18}%
  \BibitemOpen
  \bibfield  {author} {\bibinfo {author} {\bibfnamefont {O.~R.}\ \bibnamefont
  {Sulymenko}}, \bibinfo {author} {\bibfnamefont {O.~V.}\ \bibnamefont
  {Prokopenko}}, \bibinfo {author} {\bibfnamefont {V.~S.}\ \bibnamefont
  {Tyberkevych}}, \bibinfo {author} {\bibfnamefont {A.~N.}\ \bibnamefont
  {Slavin}},\ and\ \bibinfo {author} {\bibfnamefont {A.~A.}\ \bibnamefont
  {Serga}},\ }\bibfield  {title} {\bibinfo {title} {Bullets and droplets:
  Two-dimensional spin- wave solitons in modern magnonics},\ }\href@noop {}
  {\bibfield  {journal} {\bibinfo  {journal} {J. Low Temp. Phys.}\ }\textbf
  {\bibinfo {volume} {44}},\ \bibinfo {pages} {602} (\bibinfo {year}
  {2018})}\BibitemShut {NoStop}%
\bibitem [{\citenamefont {Iacocca}\ \emph {et~al.}(2014)\citenamefont
  {Iacocca}, \citenamefont {Dumas}, \citenamefont {Bookman}, \citenamefont
  {Mohseni}, \citenamefont {Chung}, \citenamefont {Hoefer},\ and\ \citenamefont
  {{\r A}kerman}}]{iacocca_confined_2014}%
  \BibitemOpen
  \bibfield  {author} {\bibinfo {author} {\bibfnamefont {E.}~\bibnamefont
  {Iacocca}}, \bibinfo {author} {\bibfnamefont {R.~K.}\ \bibnamefont {Dumas}},
  \bibinfo {author} {\bibfnamefont {L.}~\bibnamefont {Bookman}}, \bibinfo
  {author} {\bibfnamefont {M.}~\bibnamefont {Mohseni}}, \bibinfo {author}
  {\bibfnamefont {S.}~\bibnamefont {Chung}}, \bibinfo {author} {\bibfnamefont
  {M.~A.}\ \bibnamefont {Hoefer}},\ and\ \bibinfo {author} {\bibfnamefont
  {J.}~\bibnamefont {{\r A}kerman}},\ }\bibfield  {title} {\bibinfo {title}
  {Confined {Dissipative} {Droplet} {Solitons} in {Spin}-{Valve} {Nanowires}
  with {Perpendicular} {Magnetic} {Anisotropy}},\ }\href
  {https://doi.org/10.1103/PhysRevLett.112.047201} {\bibfield  {journal}
  {\bibinfo  {journal} {Phys. Rev. Lett.}\ }\textbf {\bibinfo {volume} {112}},\
  \bibinfo {pages} {047201} (\bibinfo {year} {2014})}\BibitemShut {NoStop}%
\bibitem [{\citenamefont {Xiao}\ \emph {et~al.}(2016)\citenamefont {Xiao},
  \citenamefont {Liu}, \citenamefont {Zhou}, \citenamefont {Mohseni},
  \citenamefont {Chung},\ and\ \citenamefont {{\r
  A}kerman}}]{xiao_merging_2016}%
  \BibitemOpen
  \bibfield  {author} {\bibinfo {author} {\bibfnamefont {D.}~\bibnamefont
  {Xiao}}, \bibinfo {author} {\bibfnamefont {Y.}~\bibnamefont {Liu}}, \bibinfo
  {author} {\bibfnamefont {Y.}~\bibnamefont {Zhou}}, \bibinfo {author}
  {\bibfnamefont {S.~M.}\ \bibnamefont {Mohseni}}, \bibinfo {author}
  {\bibfnamefont {S.}~\bibnamefont {Chung}},\ and\ \bibinfo {author}
  {\bibfnamefont {J.}~\bibnamefont {{\r A}kerman}},\ }\bibfield  {title}
  {\bibinfo {title} {Merging droplets in double nanocontact spin torque
  oscillators},\ }\href {https://doi.org/10.1103/PhysRevB.93.094431} {\bibfield
   {journal} {\bibinfo  {journal} {Phys. Rev. B}\ }\textbf {\bibinfo {volume}
  {93}},\ \bibinfo {pages} {094431} (\bibinfo {year} {2016})}\BibitemShut
  {NoStop}%
\bibitem [{\citenamefont {Xiao}\ \emph {et~al.}(2017)\citenamefont {Xiao},
  \citenamefont {Tiberkevich}, \citenamefont {Liu}, \citenamefont {Liu},
  \citenamefont {Mohseni}, \citenamefont {Chung}, \citenamefont {Ahlberg},
  \citenamefont {Slavin}, \citenamefont {{\r A}kerman},\ and\ \citenamefont
  {Zhou}}]{xiao_parametric_2017}%
  \BibitemOpen
  \bibfield  {author} {\bibinfo {author} {\bibfnamefont {D.}~\bibnamefont
  {Xiao}}, \bibinfo {author} {\bibfnamefont {V.}~\bibnamefont {Tiberkevich}},
  \bibinfo {author} {\bibfnamefont {Y.~H.}\ \bibnamefont {Liu}}, \bibinfo
  {author} {\bibfnamefont {Y.~W.}\ \bibnamefont {Liu}}, \bibinfo {author}
  {\bibfnamefont {S.~M.}\ \bibnamefont {Mohseni}}, \bibinfo {author}
  {\bibfnamefont {S.}~\bibnamefont {Chung}}, \bibinfo {author} {\bibfnamefont
  {M.}~\bibnamefont {Ahlberg}}, \bibinfo {author} {\bibfnamefont {A.~N.}\
  \bibnamefont {Slavin}}, \bibinfo {author} {\bibfnamefont {J.}~\bibnamefont
  {{\r A}kerman}},\ and\ \bibinfo {author} {\bibfnamefont {Y.}~\bibnamefont
  {Zhou}},\ }\bibfield  {title} {\bibinfo {title} {Parametric autoexcitation of
  magnetic droplet soliton perimeter modes},\ }\href
  {https://doi.org/10.1103/PhysRevB.95.024106} {\bibfield  {journal} {\bibinfo
  {journal} {Phys. Rev. B}\ }\textbf {\bibinfo {volume} {95}},\ \bibinfo
  {pages} {024106} (\bibinfo {year} {2017})}\BibitemShut {NoStop}%
\bibitem [{\citenamefont {Mohseni}\ \emph {et~al.}(2018)\citenamefont
  {Mohseni}, \citenamefont {Hamdi}, \citenamefont {Yazdi}, \citenamefont
  {Banuazizi}, \citenamefont {Chung}, \citenamefont {Sani}, \citenamefont {{\r
  A}kerman},\ and\ \citenamefont {Mohseni}}]{mohseni_magnetic_2018}%
  \BibitemOpen
  \bibfield  {author} {\bibinfo {author} {\bibfnamefont {M.}~\bibnamefont
  {Mohseni}}, \bibinfo {author} {\bibfnamefont {M.}~\bibnamefont {Hamdi}},
  \bibinfo {author} {\bibfnamefont {H.~F.}\ \bibnamefont {Yazdi}}, \bibinfo
  {author} {\bibfnamefont {S.~A.~H.}\ \bibnamefont {Banuazizi}}, \bibinfo
  {author} {\bibfnamefont {S.}~\bibnamefont {Chung}}, \bibinfo {author}
  {\bibfnamefont {S.~R.}\ \bibnamefont {Sani}}, \bibinfo {author}
  {\bibfnamefont {J.}~\bibnamefont {{\r A}kerman}},\ and\ \bibinfo {author}
  {\bibfnamefont {M.}~\bibnamefont {Mohseni}},\ }\bibfield  {title} {\bibinfo
  {title} {Magnetic droplet soliton nucleation in oblique fields},\ }\href
  {https://doi.org/10.1103/PhysRevB.97.184402} {\bibfield  {journal} {\bibinfo
  {journal} {Phys. Rev. B}\ }\textbf {\bibinfo {volume} {97}},\ \bibinfo
  {pages} {184402} (\bibinfo {year} {2018})}\BibitemShut {NoStop}%
\bibitem [{\citenamefont {Jiang}\ \emph
  {et~al.}(2018{\natexlab{a}})\citenamefont {Jiang}, \citenamefont {Chung},
  \citenamefont {Le}, \citenamefont {Mazraati}, \citenamefont {Houshang},\ and\
  \citenamefont {{\r A}kerman}}]{jiang_using_2018}%
  \BibitemOpen
  \bibfield  {author} {\bibinfo {author} {\bibfnamefont {S.}~\bibnamefont
  {Jiang}}, \bibinfo {author} {\bibfnamefont {S.}~\bibnamefont {Chung}},
  \bibinfo {author} {\bibfnamefont {Q.~T.}\ \bibnamefont {Le}}, \bibinfo
  {author} {\bibfnamefont {H.}~\bibnamefont {Mazraati}}, \bibinfo {author}
  {\bibfnamefont {A.}~\bibnamefont {Houshang}},\ and\ \bibinfo {author}
  {\bibfnamefont {J.}~\bibnamefont {{\r A}kerman}},\ }\bibfield  {title}
  {\bibinfo {title} {Using {Magnetic} {Droplet} {Nucleation} to {Determine} the
  {Spin} {Torque} {Efficiency} and {Asymmetry} in
  ${\mathrm{co}}_{x}{\mathrm{(ni,}\mathrm{fe)}}_{1\ensuremath{-}x}$ {Thin}
  {Films}},\ }\href {https://doi.org/10.1103/PhysRevApplied.10.054014}
  {\bibfield  {journal} {\bibinfo  {journal} {Phys. Rev. Applied}\ }\textbf
  {\bibinfo {volume} {10}},\ \bibinfo {pages} {054014} (\bibinfo {year}
  {2018}{\natexlab{a}})}\BibitemShut {NoStop}%
\bibitem [{\citenamefont {Jiang}\ \emph
  {et~al.}(2018{\natexlab{b}})\citenamefont {Jiang}, \citenamefont
  {Ruhollah~Etesami}, \citenamefont {Chung}, \citenamefont {Le}, \citenamefont
  {Houshang},\ and\ \citenamefont {{\r A}kerman}}]{jiang_impact_2018}%
  \BibitemOpen
  \bibfield  {author} {\bibinfo {author} {\bibfnamefont {S.}~\bibnamefont
  {Jiang}}, \bibinfo {author} {\bibfnamefont {S.}~\bibnamefont
  {Ruhollah~Etesami}}, \bibinfo {author} {\bibfnamefont {S.}~\bibnamefont
  {Chung}}, \bibinfo {author} {\bibfnamefont {Q.~T.}\ \bibnamefont {Le}},
  \bibinfo {author} {\bibfnamefont {A.}~\bibnamefont {Houshang}},\ and\
  \bibinfo {author} {\bibfnamefont {J.}~\bibnamefont {{\r A}kerman}},\
  }\bibfield  {title} {\bibinfo {title} {Impact of the {Oersted} {Field} on
  {Droplet} {Nucleation} {Boundaries}},\ }\href
  {https://doi.org/10.1109/LMAG.2018.2850007} {\bibfield  {journal} {\bibinfo
  {journal} {IEEE Magn. Lett.}\ }\textbf {\bibinfo {volume} {9}},\ \bibinfo
  {pages} {1} (\bibinfo {year} {2018}{\natexlab{b}})}\BibitemShut {NoStop}%
\bibitem [{\citenamefont {Statuto}\ \emph {et~al.}(2019)\citenamefont
  {Statuto}, \citenamefont {Hahn}, \citenamefont {Hern{\`a}ndez}, \citenamefont
  {Kent},\ and\ \citenamefont {Maci{\`a}}}]{statuto_multiple_2019}%
  \BibitemOpen
  \bibfield  {author} {\bibinfo {author} {\bibfnamefont {N.}~\bibnamefont
  {Statuto}}, \bibinfo {author} {\bibfnamefont {C.}~\bibnamefont {Hahn}},
  \bibinfo {author} {\bibfnamefont {J.~M.}\ \bibnamefont {Hern{\`a}ndez}},
  \bibinfo {author} {\bibfnamefont {A.~D.}\ \bibnamefont {Kent}},\ and\
  \bibinfo {author} {\bibfnamefont {F.}~\bibnamefont {Maci{\`a}}},\ }\bibfield
  {title} {\bibinfo {title} {Multiple magnetic droplet soliton modes},\ }\href
  {https://doi.org/10.1103/PhysRevB.99.174436} {\bibfield  {journal} {\bibinfo
  {journal} {Phys. Rev. B}\ }\textbf {\bibinfo {volume} {99}},\ \bibinfo
  {pages} {174436} (\bibinfo {year} {2019})}\BibitemShut {NoStop}%
\bibitem [{\citenamefont {Mohseni}\ \emph
  {et~al.}(2020{\natexlab{a}})\citenamefont {Mohseni}, \citenamefont {Wang},
  \citenamefont {Mohseni}, \citenamefont {Br{\"a}cher}, \citenamefont
  {Hillebrands},\ and\ \citenamefont {Pirro}}]{mohseni_propagating_2020}%
  \BibitemOpen
  \bibfield  {author} {\bibinfo {author} {\bibfnamefont {M.}~\bibnamefont
  {Mohseni}}, \bibinfo {author} {\bibfnamefont {Q.}~\bibnamefont {Wang}},
  \bibinfo {author} {\bibfnamefont {M.}~\bibnamefont {Mohseni}}, \bibinfo
  {author} {\bibfnamefont {T.}~\bibnamefont {Br{\"a}cher}}, \bibinfo {author}
  {\bibfnamefont {B.}~\bibnamefont {Hillebrands}},\ and\ \bibinfo {author}
  {\bibfnamefont {P.}~\bibnamefont {Pirro}},\ }\bibfield  {title} {\bibinfo
  {title} {Propagating {Magnetic} {Droplet} {Solitons} as {Moveable}
  {Nanoscale} {Spin}-{Wave} {Sources} with {Tunable} {Direction} of
  {Emission}},\ }\href {https://doi.org/10.1103/PhysRevApplied.13.024040}
  {\bibfield  {journal} {\bibinfo  {journal} {Phys. Rev. Applied}\ }\textbf
  {\bibinfo {volume} {13}},\ \bibinfo {pages} {024040} (\bibinfo {year}
  {2020}{\natexlab{a}})}\BibitemShut {NoStop}%
\bibitem [{\citenamefont {Mohseni}\ \emph
  {et~al.}(2020{\natexlab{b}})\citenamefont {Mohseni}, \citenamefont
  {Rodrigues}, \citenamefont {Saghafi}, \citenamefont {Chung}, \citenamefont
  {Ahlberg}, \citenamefont {Yazdi}, \citenamefont {Wang}, \citenamefont
  {Banuazizi}, \citenamefont {Pirro}, \citenamefont {{\r A}kerman},\ and\
  \citenamefont {Mohseni}}]{mohseni_chiral_2020}%
  \BibitemOpen
  \bibfield  {author} {\bibinfo {author} {\bibfnamefont {M.}~\bibnamefont
  {Mohseni}}, \bibinfo {author} {\bibfnamefont {D.~R.}\ \bibnamefont
  {Rodrigues}}, \bibinfo {author} {\bibfnamefont {M.}~\bibnamefont {Saghafi}},
  \bibinfo {author} {\bibfnamefont {S.}~\bibnamefont {Chung}}, \bibinfo
  {author} {\bibfnamefont {M.}~\bibnamefont {Ahlberg}}, \bibinfo {author}
  {\bibfnamefont {H.~F.}\ \bibnamefont {Yazdi}}, \bibinfo {author}
  {\bibfnamefont {Q.}~\bibnamefont {Wang}}, \bibinfo {author} {\bibfnamefont
  {S.~A.~H.}\ \bibnamefont {Banuazizi}}, \bibinfo {author} {\bibfnamefont
  {P.}~\bibnamefont {Pirro}}, \bibinfo {author} {\bibfnamefont
  {J.}~\bibnamefont {{\r A}kerman}},\ and\ \bibinfo {author} {\bibfnamefont
  {M.}~\bibnamefont {Mohseni}},\ }\bibfield  {title} {\bibinfo {title} {Chiral
  excitations of magnetic droplet solitons driven by their own inertia},\
  }\href {https://doi.org/10.1103/PhysRevB.101.020417} {\bibfield  {journal}
  {\bibinfo  {journal} {Phys. Rev. B}\ }\textbf {\bibinfo {volume} {101}},\
  \bibinfo {pages} {020417} (\bibinfo {year} {2020}{\natexlab{b}})}\BibitemShut
  {NoStop}%
\bibitem [{\citenamefont {Slonczewski}(1996)}]{Slonczewski96}%
  \BibitemOpen
  \bibfield  {author} {\bibinfo {author} {\bibfnamefont {J.~C.}\ \bibnamefont
  {Slonczewski}},\ }\bibfield  {title} {\bibinfo {title} {Current-driven
  excitation of magnetic multilayers},\ }\href@noop {} {\bibfield  {journal}
  {\bibinfo  {journal} {J. Magn. Magn. Mater.}\ }\textbf {\bibinfo {volume}
  {159}},\ \bibinfo {pages} {L1 } (\bibinfo {year} {1996})}\BibitemShut
  {NoStop}%
\bibitem [{\citenamefont {Berger}(1996)}]{Berger96}%
  \BibitemOpen
  \bibfield  {author} {\bibinfo {author} {\bibfnamefont {L.}~\bibnamefont
  {Berger}},\ }\bibfield  {title} {\bibinfo {title} {Emission of spin waves by
  a magnetic multilayer traversed by a current},\ }\href@noop {} {\bibfield
  {journal} {\bibinfo  {journal} {Phys. Rev. B}\ }\textbf {\bibinfo {volume}
  {54}},\ \bibinfo {pages} {9352} (\bibinfo {year} {1996})}\BibitemShut
  {NoStop}%
\bibitem [{\citenamefont {Bookman}\ and\ \citenamefont {Hoefer}(2013)}]{BH13}%
  \BibitemOpen
  \bibfield  {author} {\bibinfo {author} {\bibfnamefont {L.~D.}\ \bibnamefont
  {Bookman}}\ and\ \bibinfo {author} {\bibfnamefont {M.~A.}\ \bibnamefont
  {Hoefer}},\ }\bibfield  {title} {\bibinfo {title} {Analytic theory of
  modulated magnetic solitons},\ }\href@noop {} {\bibfield  {journal} {\bibinfo
   {journal} {Phys. Rev. B}\ }\textbf {\bibinfo {volume} {88}},\ \bibinfo
  {pages} {184401} (\bibinfo {year} {2013})}\BibitemShut {NoStop}%
\bibitem [{\citenamefont {Hoefer}\ \emph {et~al.}(2010)\citenamefont {Hoefer},
  \citenamefont {Silva},\ and\ \citenamefont {Keller}}]{hoefer_theory_2010}%
  \BibitemOpen
  \bibfield  {author} {\bibinfo {author} {\bibfnamefont {M.~A.}\ \bibnamefont
  {Hoefer}}, \bibinfo {author} {\bibfnamefont {T.~J.}\ \bibnamefont {Silva}},\
  and\ \bibinfo {author} {\bibfnamefont {M.~W.}\ \bibnamefont {Keller}},\
  }\bibfield  {title} {\bibinfo {title} {Theory for a dissipative droplet
  soliton excited by a spin torque nanocontact},\ }\href
  {https://doi.org/https://doi.org/10.1103/PhysRevB.82.054432} {\bibfield
  {journal} {\bibinfo  {journal} {Phys. Rev. B}\ }\textbf {\bibinfo {volume}
  {82}},\ \bibinfo {pages} {054432} (\bibinfo {year} {2010})}\BibitemShut
  {NoStop}%
\bibitem [{\citenamefont {Wills}\ \emph {et~al.}(2016)\citenamefont {Wills},
  \citenamefont {Iacocca},\ and\ \citenamefont {Hoefer}}]{WIH16}%
  \BibitemOpen
  \bibfield  {author} {\bibinfo {author} {\bibfnamefont {P.}~\bibnamefont
  {Wills}}, \bibinfo {author} {\bibfnamefont {E.}~\bibnamefont {Iacocca}},\
  and\ \bibinfo {author} {\bibfnamefont {M.~A.}\ \bibnamefont {Hoefer}},\
  }\bibfield  {title} {\bibinfo {title} {Deterministic drift instability and
  stochastic thermal perturbations of magnetic dissipative droplet solitons},\
  }\href@noop {} {\bibfield  {journal} {\bibinfo  {journal} {Phys. Rev. B}\
  }\textbf {\bibinfo {volume} {93}},\ \bibinfo {pages} {144408} (\bibinfo
  {year} {2016})}\BibitemShut {NoStop}%
\bibitem [{\citenamefont {Moore}\ and\ \citenamefont {Hoefer}(2019)}]{MH19}%
  \BibitemOpen
  \bibfield  {author} {\bibinfo {author} {\bibfnamefont {R.~O.}\ \bibnamefont
  {Moore}}\ and\ \bibinfo {author} {\bibfnamefont {M.~A.}\ \bibnamefont
  {Hoefer}},\ }\bibfield  {title} {\bibinfo {title} {Stochastic ejection of
  nanocontact droplet solitons via drift instability},\ }\href
  {https://doi.org/10.1103/PhysRevB.82.054432} {\bibfield  {journal} {\bibinfo
  {journal} {Phys. Rev. B}\ }\textbf {\bibinfo {volume} {100}},\ \bibinfo
  {pages} {014402} (\bibinfo {year} {2019})}\BibitemShut {NoStop}%
\bibitem [{\citenamefont {Lend{\'i}nez}\ \emph {et~al.}(2015)\citenamefont
  {Lend{\'i}nez}, \citenamefont {Statuto}, \citenamefont {Backes},
  \citenamefont {Kent},\ and\ \citenamefont
  {Maci{\`a}}}]{lendinez_observation_2015}%
  \BibitemOpen
  \bibfield  {author} {\bibinfo {author} {\bibfnamefont {S.}~\bibnamefont
  {Lend{\'i}nez}}, \bibinfo {author} {\bibfnamefont {N.}~\bibnamefont
  {Statuto}}, \bibinfo {author} {\bibfnamefont {D.}~\bibnamefont {Backes}},
  \bibinfo {author} {\bibfnamefont {A.~D.}\ \bibnamefont {Kent}},\ and\
  \bibinfo {author} {\bibfnamefont {F.}~\bibnamefont {Maci{\`a}}},\ }\bibfield
  {title} {\bibinfo {title} {Observation of droplet soliton drift resonances in
  a spin-transfer-torque nanocontact to a ferromagnetic thin film},\ }\href
  {https://doi.org/10.1103/PhysRevB.92.174426} {\bibfield  {journal} {\bibinfo
  {journal} {Phys. Rev. B}\ }\textbf {\bibinfo {volume} {92}},\ \bibinfo
  {pages} {174426} (\bibinfo {year} {2015})}\BibitemShut {NoStop}%
\bibitem [{\citenamefont {H\"anggi}\ \emph {et~al.}(1990)\citenamefont
  {H\"anggi}, \citenamefont {Talkner},\ and\ \citenamefont {Borkovec}}]{HTB90}%
  \BibitemOpen
  \bibfield  {author} {\bibinfo {author} {\bibfnamefont {P.}~\bibnamefont
  {H\"anggi}}, \bibinfo {author} {\bibfnamefont {P.}~\bibnamefont {Talkner}},\
  and\ \bibinfo {author} {\bibfnamefont {M.}~\bibnamefont {Borkovec}},\
  }\bibfield  {title} {\bibinfo {title} {Reaction-rate theory: {F}ifty years
  after {K}ramers},\ }\href@noop {} {\bibfield  {journal} {\bibinfo  {journal}
  {Rev. Mod. Phys.}\ }\textbf {\bibinfo {volume} {62}},\ \bibinfo {pages} {251}
  (\bibinfo {year} {1990})}\BibitemShut {NoStop}%
\bibitem [{\citenamefont {Freidlin}\ and\ \citenamefont
  {Wentzell}(2012)}]{FW12}%
  \BibitemOpen
  \bibfield  {author} {\bibinfo {author} {\bibfnamefont {M.~I.}\ \bibnamefont
  {Freidlin}}\ and\ \bibinfo {author} {\bibfnamefont {A.~D.}\ \bibnamefont
  {Wentzell}},\ }\href@noop {} {\emph {\bibinfo {title} {Random Perturbations
  of Dynamical Systems, Third Ed.}}}\ (\bibinfo  {publisher} {Springer-Verlag
  Berlin Heidelberg},\ \bibinfo {year} {2012})\BibitemShut {NoStop}%
\bibitem [{\citenamefont {Matkowsky}\ and\ \citenamefont
  {Schuss}(1983)}]{MS83}%
  \BibitemOpen
  \bibfield  {author} {\bibinfo {author} {\bibfnamefont {B.~J.}\ \bibnamefont
  {Matkowsky}}\ and\ \bibinfo {author} {\bibfnamefont {Z.}~\bibnamefont
  {Schuss}},\ }\bibfield  {title} {\bibinfo {title} {On the lifetime of a
  metastable state at low noise},\ }\href@noop {} {\bibfield  {journal}
  {\bibinfo  {journal} {Phys. lett. A}\ }\textbf {\bibinfo {volume} {95}},\
  \bibinfo {pages} {213} (\bibinfo {year} {1983})}\BibitemShut {NoStop}%
\bibitem [{\citenamefont {Maier}\ and\ \citenamefont {Stein}(1993)}]{MS93}%
  \BibitemOpen
  \bibfield  {author} {\bibinfo {author} {\bibfnamefont {R.~S.}\ \bibnamefont
  {Maier}}\ and\ \bibinfo {author} {\bibfnamefont {D.~L.}\ \bibnamefont
  {Stein}},\ }\bibfield  {title} {\bibinfo {title} {Escape problem for
  irreversible systems},\ }\href@noop {} {\bibfield  {journal} {\bibinfo
  {journal} {Phys. Rev. E}\ }\textbf {\bibinfo {volume} {48}},\ \bibinfo
  {pages} {931} (\bibinfo {year} {1993})}\BibitemShut {NoStop}%
\bibitem [{\citenamefont {Brown{,}~Jr.}(1963)}]{Brown63}%
  \BibitemOpen
  \bibfield  {author} {\bibinfo {author} {\bibfnamefont {W.~F.}\ \bibnamefont
  {Brown{,}~Jr.}},\ }\bibfield  {title} {\bibinfo {title} {Thermal fluctuations
  of a single-domain particle},\ }\href@noop {} {\bibfield  {journal} {\bibinfo
   {journal} {Phys. Rev.}\ }\textbf {\bibinfo {volume} {130}},\ \bibinfo
  {pages} {1677} (\bibinfo {year} {1963})}\BibitemShut {NoStop}%
\bibitem [{\citenamefont {Braun}(1993)}]{Braun93}%
  \BibitemOpen
  \bibfield  {author} {\bibinfo {author} {\bibfnamefont {H.-B.}\ \bibnamefont
  {Braun}},\ }\bibfield  {title} {\bibinfo {title} {Thermally activated
  magnetic reversal in elongated ferromagnetic particles},\ }\href@noop {}
  {\bibfield  {journal} {\bibinfo  {journal} {Phys. Rev. Lett.}\ }\textbf
  {\bibinfo {volume} {71}},\ \bibinfo {pages} {3557} (\bibinfo {year}
  {1993})}\BibitemShut {NoStop}%
\bibitem [{\citenamefont {E}\ \emph {et~al.}(2003)\citenamefont {E},
  \citenamefont {Ren},\ and\ \citenamefont {vanden Eijnden}}]{ERvE03}%
  \BibitemOpen
  \bibfield  {author} {\bibinfo {author} {\bibfnamefont {W.}~\bibnamefont {E}},
  \bibinfo {author} {\bibfnamefont {W.}~\bibnamefont {Ren}},\ and\ \bibinfo
  {author} {\bibfnamefont {E.}~\bibnamefont {vanden Eijnden}},\ }\bibfield
  {title} {\bibinfo {title} {String method for the study of rare events},\
  }\href@noop {} {\bibfield  {journal} {\bibinfo  {journal} {J. Appl. Phys.}\
  }\textbf {\bibinfo {volume} {93}},\ \bibinfo {pages} {2275} (\bibinfo {year}
  {2003})}\BibitemShut {NoStop}%
\bibitem [{\citenamefont {Martens}\ \emph {et~al.}(2006)\citenamefont
  {Martens}, \citenamefont {Stein},\ and\ \citenamefont {Kent}}]{MSK06}%
  \BibitemOpen
  \bibfield  {author} {\bibinfo {author} {\bibfnamefont {K.}~\bibnamefont
  {Martens}}, \bibinfo {author} {\bibfnamefont {D.~L.}\ \bibnamefont {Stein}},\
  and\ \bibinfo {author} {\bibfnamefont {A.~D.}\ \bibnamefont {Kent}},\
  }\bibfield  {title} {\bibinfo {title} {Magnetic reversal in nanoscopic
  ferromagnetic rings},\ }\href@noop {} {\bibfield  {journal} {\bibinfo
  {journal} {Phys. Rev. B}\ }\textbf {\bibinfo {volume} {73}},\ \bibinfo
  {pages} {054413} (\bibinfo {year} {2006})}\BibitemShut {NoStop}%
\bibitem [{\citenamefont {Pinna}\ \emph
  {et~al.}(2016{\natexlab{a}})\citenamefont {Pinna}, \citenamefont {Kent},\
  and\ \citenamefont {Stein}}]{PKS16}%
  \BibitemOpen
  \bibfield  {author} {\bibinfo {author} {\bibfnamefont {D.}~\bibnamefont
  {Pinna}}, \bibinfo {author} {\bibfnamefont {A.~D.}\ \bibnamefont {Kent}},\
  and\ \bibinfo {author} {\bibfnamefont {D.~L.}\ \bibnamefont {Stein}},\
  }\bibfield  {title} {\bibinfo {title} {Large fluctuations and singular
  behavior of nonequilibrium systems},\ }\href@noop {} {\bibfield  {journal}
  {\bibinfo  {journal} {Phys. Rev. E}\ }\textbf {\bibinfo {volume} {93}},\
  \bibinfo {pages} {012114} (\bibinfo {year} {2016}{\natexlab{a}})}\BibitemShut
  {NoStop}%
\bibitem [{\citenamefont {Graham}\ and\ \citenamefont {T\'el}(1986)}]{GT86}%
  \BibitemOpen
  \bibfield  {author} {\bibinfo {author} {\bibfnamefont {R.}~\bibnamefont
  {Graham}}\ and\ \bibinfo {author} {\bibfnamefont {E.}~\bibnamefont {T\'el}},\
  }\bibfield  {title} {\bibinfo {title} {Nonequilibrium potential for
  coexisting attractors},\ }\href@noop {} {\bibfield  {journal} {\bibinfo
  {journal} {Phys. Rev. A}\ }\textbf {\bibinfo {volume} {33}},\ \bibinfo
  {pages} {1322} (\bibinfo {year} {1986})}\BibitemShut {NoStop}%
\bibitem [{\citenamefont {{M. J. Donahue}}\ and\ \citenamefont {{D.G.
  Porter}}(1999)}]{m._j._donahue_oommf_1999}%
  \BibitemOpen
  \bibfield  {author} {\bibinfo {author} {\bibnamefont {{M. J. Donahue}}}\ and\
  \bibinfo {author} {\bibnamefont {{D.G. Porter}}},\ }\href@noop {} {\emph
  {\bibinfo {title} {{OOMMF} {User}'s {Guide}, {Version} 1.0}}},\ \bibinfo
  {type} {Interagency {Report}}\ \bibinfo {number} {NISTIR 6376}\ (\bibinfo
  {institution} {National Institute of Standards and Technology},\ \bibinfo
  {address} {Gaithersburg, MD},\ \bibinfo {year} {1999})\BibitemShut {NoStop}%
\bibitem [{\citenamefont {Garc{\'i}a-Palacios}\ and\ \citenamefont
  {L{\'a}zaro}(1998)}]{garcia-palacios_langevin-dynamics_1998}%
  \BibitemOpen
  \bibfield  {author} {\bibinfo {author} {\bibfnamefont {J.~L.}\ \bibnamefont
  {Garc{\'i}a-Palacios}}\ and\ \bibinfo {author} {\bibfnamefont {F.~J.}\
  \bibnamefont {L{\'a}zaro}},\ }\bibfield  {title} {\bibinfo {title}
  {Langevin-dynamics study of the dynamical properties of small magnetic
  particles},\ }\href {https://doi.org/10.1103/PhysRevB.58.14937} {\bibfield
  {journal} {\bibinfo  {journal} {Phys. Rev. B}\ }\textbf {\bibinfo {volume}
  {58}},\ \bibinfo {pages} {14937} (\bibinfo {year} {1998})}\BibitemShut
  {NoStop}%
\bibitem [{\citenamefont {Berkov}(2007)}]{berkov_magnetization_2007}%
  \BibitemOpen
  \bibfield  {author} {\bibinfo {author} {\bibfnamefont {D.~V.}\ \bibnamefont
  {Berkov}},\ }\bibfield  {title} {\bibinfo {title} {Magnetization {Dynamics}
  {Including} {Thermal} {Fluctuations}: {Basic} {Phenomenology}, {Fast}
  {Remagnetization} {Processes} and {Transitions} {Over} {High}-energy
  {Barriers}},\ }in\ \href {https://doi.org/10.1002/9780470022184.hmm204}
  {\emph {\bibinfo {booktitle} {Handbook of Magnetism and Advanced Magnetic
  Materials}}}\ (\bibinfo  {publisher} {American Cancer Society},\ \bibinfo
  {year} {2007})\BibitemShut {NoStop}%
\bibitem [{\citenamefont {Bertotti}\ \emph {et~al.}(2006)\citenamefont
  {Bertotti}, \citenamefont {Mayergoyz},\ and\ \citenamefont
  {Serpico}}]{bertotti_nonlinear_2006}%
  \BibitemOpen
  \bibfield  {author} {\bibinfo {author} {\bibfnamefont {G.}~\bibnamefont
  {Bertotti}}, \bibinfo {author} {\bibfnamefont {I.~D.}\ \bibnamefont
  {Mayergoyz}},\ and\ \bibinfo {author} {\bibfnamefont {C.}~\bibnamefont
  {Serpico}},\ }\bibfield  {title} {\bibinfo {title} {Nonlinear {Magnetization}
  {Dynamics}. {Switching} and {Relaxation} {Phenomena}},\ }in\ \href@noop {}
  {\emph {\bibinfo {booktitle} {The {Science} of {Hysteresis}}}},\
  Vol.~\bibinfo {volume} {2}\ (\bibinfo  {publisher} {Elsevier},\ \bibinfo
  {year} {2006})\ pp.\ \bibinfo {pages} {435--565}\BibitemShut {NoStop}%
\bibitem [{cha()}]{chaves-oflynn_sup}%
  \BibitemOpen
  \href@noop {} {}\bibinfo {note} {See {Supplemental} {Material} at URL for the
  source-code used in the different parts of this paper}\BibitemShut {NoStop}%
\bibitem [{\citenamefont {Serpico}\ \emph {et~al.}(2007)\citenamefont
  {Serpico}, \citenamefont {Bertotti}, \citenamefont {Mayergoyz},\ and\
  \citenamefont {d'Aquino}}]{serpico_nonlinear_2007}%
  \BibitemOpen
  \bibfield  {author} {\bibinfo {author} {\bibfnamefont {C.}~\bibnamefont
  {Serpico}}, \bibinfo {author} {\bibfnamefont {G.}~\bibnamefont {Bertotti}},
  \bibinfo {author} {\bibfnamefont {I.~D.}\ \bibnamefont {Mayergoyz}},\ and\
  \bibinfo {author} {\bibfnamefont {M.}~\bibnamefont {d'Aquino}},\ }\bibfield
  {title} {\bibinfo {title} {Nonlinear {Magnetization} {Dynamics} in
  {Nanomagnets}},\ }in\ \href {https://doi.org/10.1002/9780470022184.hmm206}
  {\emph {\bibinfo {booktitle} {Handbook of {Magnetism} and {Advanced}
  {Magnetic} {Materials}}}}\ (\bibinfo  {publisher} {American Cancer Society},\
  \bibinfo {year} {2007})\BibitemShut {NoStop}%
\bibitem [{\citenamefont {Wedemann}\ \emph {et~al.}(2016)\citenamefont
  {Wedemann}, \citenamefont {Plastino},\ and\ \citenamefont
  {Tsallis}}]{wedemann_curl_2016}%
  \BibitemOpen
  \bibfield  {author} {\bibinfo {author} {\bibfnamefont {R.~S.}\ \bibnamefont
  {Wedemann}}, \bibinfo {author} {\bibfnamefont {A.~R.}\ \bibnamefont
  {Plastino}},\ and\ \bibinfo {author} {\bibfnamefont {C.}~\bibnamefont
  {Tsallis}},\ }\bibfield  {title} {\bibinfo {title} {Curl forces and the
  nonlinear {Fokker}-{Planck} equation},\ }\href
  {https://doi.org/10.1103/PhysRevE.94.062105} {\bibfield  {journal} {\bibinfo
  {journal} {Phys. Rev. E}\ }\textbf {\bibinfo {volume} {94}},\ \bibinfo
  {pages} {062105} (\bibinfo {year} {2016})}\BibitemShut {NoStop}%
\bibitem [{\citenamefont {Berry}\ and\ \citenamefont
  {Shukla}(2016)}]{berry_curl_2016}%
  \BibitemOpen
  \bibfield  {author} {\bibinfo {author} {\bibfnamefont {M.~V.}\ \bibnamefont
  {Berry}}\ and\ \bibinfo {author} {\bibfnamefont {P.}~\bibnamefont {Shukla}},\
  }\bibfield  {title} {\bibinfo {title} {Curl force dynamics: symmetries, chaos
  and constants of motion},\ }\href
  {https://doi.org/10.1088/1367-2630/18/6/063018} {\bibfield  {journal}
  {\bibinfo  {journal} {New Journal of Physics}\ }\textbf {\bibinfo {volume}
  {18}},\ \bibinfo {pages} {063018} (\bibinfo {year} {2016})}\BibitemShut
  {NoStop}%
\bibitem [{\citenamefont {{Berry M. V.}}\ and\ \citenamefont {{Shukla
  Pragya}}(2015)}]{berry_m._v._hamiltonian_2015}%
  \BibitemOpen
  \bibfield  {author} {\bibinfo {author} {\bibnamefont {{Berry M. V.}}}\ and\
  \bibinfo {author} {\bibnamefont {{Shukla Pragya}}},\ }\bibfield  {title}
  {\bibinfo {title} {Hamiltonian curl forces},\ }\href
  {https://doi.org/10.1098/rspa.2015.0002} {\bibfield  {journal} {\bibinfo
  {journal} {Proc. R. Soc. A.}\ }\textbf {\bibinfo {volume} {471}},\ \bibinfo
  {pages} {20150002} (\bibinfo {year} {2015})}\BibitemShut {NoStop}%
\bibitem [{\citenamefont {Newhall}\ and\ \citenamefont
  {Vanden-Eijnden}(2013)}]{newhall_averaged_2013}%
  \BibitemOpen
  \bibfield  {author} {\bibinfo {author} {\bibfnamefont {K.~A.}\ \bibnamefont
  {Newhall}}\ and\ \bibinfo {author} {\bibfnamefont {E.}~\bibnamefont
  {Vanden-Eijnden}},\ }\bibfield  {title} {\bibinfo {title} {Averaged equation
  for energy diffusion on a graph reveals bifurcation diagram and thermally
  assisted reversal times in spin-torque driven nanomagnets},\ }\href
  {https://doi.org/10.1063/1.4804070} {\bibfield  {journal} {\bibinfo
  {journal} {J. Appl. Phys}\ }\textbf {\bibinfo {volume} {113}},\ \bibinfo
  {pages} {184105} (\bibinfo {year} {2013})}\BibitemShut {NoStop}%
\bibitem [{\citenamefont {Pinna}\ \emph {et~al.}(2013)\citenamefont {Pinna},
  \citenamefont {Kent},\ and\ \citenamefont {Stein}}]{pinna_thermally_2013}%
  \BibitemOpen
  \bibfield  {author} {\bibinfo {author} {\bibfnamefont {D.}~\bibnamefont
  {Pinna}}, \bibinfo {author} {\bibfnamefont {A.~D.}\ \bibnamefont {Kent}},\
  and\ \bibinfo {author} {\bibfnamefont {D.~L.}\ \bibnamefont {Stein}},\
  }\bibfield  {title} {\bibinfo {title} {Thermally assisted spin-transfer
  torque dynamics in energy space},\ }\href
  {https://doi.org/10.1103/PhysRevB.88.104405} {\bibfield  {journal} {\bibinfo
  {journal} {Phys. Rev. B}\ }\textbf {\bibinfo {volume} {88}},\ \bibinfo
  {pages} {104405} (\bibinfo {year} {2013})}\BibitemShut {NoStop}%
\bibitem [{\citenamefont {Rippard}\ \emph {et~al.}(2010)\citenamefont
  {Rippard}, \citenamefont {Deac}, \citenamefont {Pufall}, \citenamefont
  {Shaw}, \citenamefont {Keller}, \citenamefont {Russek}, \citenamefont
  {Bauer},\ and\ \citenamefont {Serpico}}]{rippard_spin-transfer_2010}%
  \BibitemOpen
  \bibfield  {author} {\bibinfo {author} {\bibfnamefont {W.~H.}\ \bibnamefont
  {Rippard}}, \bibinfo {author} {\bibfnamefont {A.~M.}\ \bibnamefont {Deac}},
  \bibinfo {author} {\bibfnamefont {M.~R.}\ \bibnamefont {Pufall}}, \bibinfo
  {author} {\bibfnamefont {J.~M.}\ \bibnamefont {Shaw}}, \bibinfo {author}
  {\bibfnamefont {M.~W.}\ \bibnamefont {Keller}}, \bibinfo {author}
  {\bibfnamefont {S.~E.}\ \bibnamefont {Russek}}, \bibinfo {author}
  {\bibfnamefont {G.~E.~W.}\ \bibnamefont {Bauer}},\ and\ \bibinfo {author}
  {\bibfnamefont {C.}~\bibnamefont {Serpico}},\ }\bibfield  {title} {\bibinfo
  {title} {Spin-transfer dynamics in spin valves with out-of-plane magnetized
  {CoNi} free layers},\ }\href {https://doi.org/10.1103/PhysRevB.81.014426}
  {\bibfield  {journal} {\bibinfo  {journal} {Phys. Rev. B}\ }\textbf {\bibinfo
  {volume} {81}},\ \bibinfo {pages} {014426} (\bibinfo {year}
  {2010})}\BibitemShut {NoStop}%
\bibitem [{\citenamefont {Mohseni}\ \emph {et~al.}(2011)\citenamefont
  {Mohseni}, \citenamefont {Sani}, \citenamefont {Persson}, \citenamefont
  {Anh~Nguyen}, \citenamefont {Chung}, \citenamefont {Pogoryelov},\ and\
  \citenamefont {Akerman}}]{Mohseni11}%
  \BibitemOpen
  \bibfield  {author} {\bibinfo {author} {\bibfnamefont {S.~M.}\ \bibnamefont
  {Mohseni}}, \bibinfo {author} {\bibfnamefont {S.~R.}\ \bibnamefont {Sani}},
  \bibinfo {author} {\bibfnamefont {J.}~\bibnamefont {Persson}}, \bibinfo
  {author} {\bibfnamefont {T.~N.}\ \bibnamefont {Anh~Nguyen}}, \bibinfo
  {author} {\bibfnamefont {S.}~\bibnamefont {Chung}}, \bibinfo {author}
  {\bibfnamefont {Y.}~\bibnamefont {Pogoryelov}},\ and\ \bibinfo {author}
  {\bibfnamefont {J.}~\bibnamefont {Akerman}},\ }\bibfield  {title} {\bibinfo
  {title} {High frequency operation of a spin-torque oscillator at low field},\
  }\href@noop {} {\bibfield  {journal} {\bibinfo  {journal}
  {Phys.~Stat.~Sol.~RRL}\ }\textbf {\bibinfo {volume} {5}},\ \bibinfo {pages}
  {432} (\bibinfo {year} {2011})}\BibitemShut {NoStop}%
\bibitem [{\citenamefont {Xiao}\ \emph {et~al.}(2004)\citenamefont {Xiao},
  \citenamefont {Zangwill},\ and\ \citenamefont
  {Stiles}}]{xiao_boltzmann_2004}%
  \BibitemOpen
  \bibfield  {author} {\bibinfo {author} {\bibfnamefont {J.}~\bibnamefont
  {Xiao}}, \bibinfo {author} {\bibfnamefont {A.}~\bibnamefont {Zangwill}},\
  and\ \bibinfo {author} {\bibfnamefont {M.~D.}\ \bibnamefont {Stiles}},\
  }\bibfield  {title} {\bibinfo {title} {Boltzmann test of {Slonczewski}'s
  theory of spin-transfer torque},\ }\href
  {https://doi.org/10.1103/PhysRevB.70.172405} {\bibfield  {journal} {\bibinfo
  {journal} {Phys. Rev. B}\ }\textbf {\bibinfo {volume} {70}},\ \bibinfo
  {pages} {172405} (\bibinfo {year} {2004})}\BibitemShut {NoStop}%
\bibitem [{\citenamefont {Hoefer}\ \emph {et~al.}(2008)\citenamefont {Hoefer},
  \citenamefont {Silva},\ and\ \citenamefont {Stiles}}]{Hoefer_2008}%
  \BibitemOpen
  \bibfield  {author} {\bibinfo {author} {\bibfnamefont {M.~A.}\ \bibnamefont
  {Hoefer}}, \bibinfo {author} {\bibfnamefont {T.~J.}\ \bibnamefont {Silva}},\
  and\ \bibinfo {author} {\bibfnamefont {M.~D.}\ \bibnamefont {Stiles}},\
  }\bibfield  {title} {\bibinfo {title} {Model for a collimated spin-wave beam
  generated by a single-layer spin torque nanocontact},\ }\href
  {https://doi.org/10.1103/PhysRevB.77.144401} {\bibfield  {journal} {\bibinfo
  {journal} {Phys. Rev. B}\ }\textbf {\bibinfo {volume} {77}},\ \bibinfo
  {pages} {144401} (\bibinfo {year} {2008})}\BibitemShut {NoStop}%
\bibitem [{\citenamefont {Bookman}\ and\ \citenamefont {Hoefer}(2015)}]{BH15}%
  \BibitemOpen
  \bibfield  {author} {\bibinfo {author} {\bibfnamefont {L.~D.}\ \bibnamefont
  {Bookman}}\ and\ \bibinfo {author} {\bibfnamefont {M.~A.}\ \bibnamefont
  {Hoefer}},\ }\bibfield  {title} {\bibinfo {title} {Perturbation theory for
  propagating magnetic droplet solitons},\ }\href
  {https://doi.org/10.1098/rspa.2015.0042} {\bibfield  {journal} {\bibinfo
  {journal} {Proc. R. Soc. A.}\ }\textbf {\bibinfo {volume} {471}},\ \bibinfo
  {pages} {20150042} (\bibinfo {year} {2015})}\BibitemShut {NoStop}%
\bibitem [{\citenamefont {Chung}\ \emph {et~al.}(2016)\citenamefont {Chung},
  \citenamefont {Eklund}, \citenamefont {Iacocca}, \citenamefont {Mohseni},
  \citenamefont {Sani}, \citenamefont {Bookman}, \citenamefont {Hoefer},
  \citenamefont {Dumas},\ and\ \citenamefont {{\r
  A}kerman}}]{chung_magnetic_2016}%
  \BibitemOpen
  \bibfield  {author} {\bibinfo {author} {\bibfnamefont {S.}~\bibnamefont
  {Chung}}, \bibinfo {author} {\bibfnamefont {A.}~\bibnamefont {Eklund}},
  \bibinfo {author} {\bibfnamefont {E.}~\bibnamefont {Iacocca}}, \bibinfo
  {author} {\bibfnamefont {S.~M.}\ \bibnamefont {Mohseni}}, \bibinfo {author}
  {\bibfnamefont {S.~R.}\ \bibnamefont {Sani}}, \bibinfo {author}
  {\bibfnamefont {L.}~\bibnamefont {Bookman}}, \bibinfo {author} {\bibfnamefont
  {M.~A.}\ \bibnamefont {Hoefer}}, \bibinfo {author} {\bibfnamefont {R.~K.}\
  \bibnamefont {Dumas}},\ and\ \bibinfo {author} {\bibfnamefont
  {J.}~\bibnamefont {{\r A}kerman}},\ }\bibfield  {title} {\bibinfo {title}
  {Magnetic droplet nucleation boundary in orthogonal spin-torque
  nano-oscillators},\ }\href {https://doi.org/10.1038/ncomms11209} {\bibfield
  {journal} {\bibinfo  {journal} {Nat. Commun.}\ }\textbf {\bibinfo {volume}
  {7}},\ \bibinfo {pages} {1} (\bibinfo {year} {2016})}\BibitemShut {NoStop}%
\bibitem [{\citenamefont {Lend{\'i}nez}\ \emph {et~al.}(2017)\citenamefont
  {Lend{\'i}nez}, \citenamefont {Hang}, \citenamefont {V{\'e}lez},
  \citenamefont {Hern{\'a}ndez}, \citenamefont {Backes}, \citenamefont {Kent},\
  and\ \citenamefont {Maci{\`a}}}]{lendinez_effect_2017}%
  \BibitemOpen
  \bibfield  {author} {\bibinfo {author} {\bibfnamefont {S.}~\bibnamefont
  {Lend{\'i}nez}}, \bibinfo {author} {\bibfnamefont {J.}~\bibnamefont {Hang}},
  \bibinfo {author} {\bibfnamefont {S.}~\bibnamefont {V{\'e}lez}}, \bibinfo
  {author} {\bibfnamefont {J.~M.}\ \bibnamefont {Hern{\'a}ndez}}, \bibinfo
  {author} {\bibfnamefont {D.}~\bibnamefont {Backes}}, \bibinfo {author}
  {\bibfnamefont {A.~D.}\ \bibnamefont {Kent}},\ and\ \bibinfo {author}
  {\bibfnamefont {F.}~\bibnamefont {Maci{\`a}}},\ }\bibfield  {title} {\bibinfo
  {title} {Effect of {Temperature} on {Magnetic} {Solitons} {Induced} by
  {Spin}-{Transfer} {Torque}},\ }\href
  {https://doi.org/10.1103/PhysRevApplied.7.054027} {\bibfield  {journal}
  {\bibinfo  {journal} {Phys. Rev. Applied}\ }\textbf {\bibinfo {volume} {7}},\
  \bibinfo {pages} {054027} (\bibinfo {year} {2017})}\BibitemShut {NoStop}%
\bibitem [{\citenamefont {Kohn}\ and\ \citenamefont {Slastikov}(2005)}]{KS05}%
  \BibitemOpen
  \bibfield  {author} {\bibinfo {author} {\bibfnamefont {R.~V.}\ \bibnamefont
  {Kohn}}\ and\ \bibinfo {author} {\bibfnamefont {V.~V.}\ \bibnamefont
  {Slastikov}},\ }\bibfield  {title} {\bibinfo {title} {Another thin-film limit
  of micromagnetics},\ }\href@noop {} {\bibfield  {journal} {\bibinfo
  {journal} {Arch. Rat. Mech. Anal.}\ }\textbf {\bibinfo {volume} {178}},\
  \bibinfo {pages} {227} (\bibinfo {year} {2005})}\BibitemShut {NoStop}%
\bibitem [{\citenamefont {Miltat}\ and\ \citenamefont
  {Donahue}(2007)}]{miltat_numerical_2007}%
  \BibitemOpen
  \bibfield  {author} {\bibinfo {author} {\bibfnamefont {J.~E.}\ \bibnamefont
  {Miltat}}\ and\ \bibinfo {author} {\bibfnamefont {M.~J.}\ \bibnamefont
  {Donahue}},\ }\bibfield  {title} {\bibinfo {title} {Numerical
  {Micromagnetics}: {Finite} {Difference} {Methods}},\ }in\ \href
  {https://doi.org/10.1002/9780470022184.hmm202} {\emph {\bibinfo {booktitle}
  {Handbook of {Magnetism} and {Advanced} {Magnetic} {Materials}}}}\ (\bibinfo
  {publisher} {American Cancer Society},\ \bibinfo {year} {2007})\BibitemShut
  {NoStop}%
\bibitem [{\citenamefont {Statuto}\ \emph {et~al.}(2018)\citenamefont
  {Statuto}, \citenamefont {Hern{\`a}ndez}, \citenamefont {Kent},\ and\
  \citenamefont {Maci{\`a}}}]{statuto_generation_2018}%
  \BibitemOpen
  \bibfield  {author} {\bibinfo {author} {\bibfnamefont {N.}~\bibnamefont
  {Statuto}}, \bibinfo {author} {\bibfnamefont {J.~M.}\ \bibnamefont
  {Hern{\`a}ndez}}, \bibinfo {author} {\bibfnamefont {A.~D.}\ \bibnamefont
  {Kent}},\ and\ \bibinfo {author} {\bibfnamefont {F.}~\bibnamefont
  {Maci{\`a}}},\ }\bibfield  {title} {\bibinfo {title} {Generation and
  stability of dynamical skyrmions and droplet solitons},\ }\href
  {https://doi.org/10.1088/1361-6528/aac411} {\bibfield  {journal} {\bibinfo
  {journal} {Nanotechnology}\ }\textbf {\bibinfo {volume} {29}},\ \bibinfo
  {pages} {325302} (\bibinfo {year} {2018})}\BibitemShut {NoStop}%
\bibitem [{\citenamefont {Ralph}\ and\ \citenamefont
  {Stiles}(2008)}]{ralph_spin_2008}%
  \BibitemOpen
  \bibfield  {author} {\bibinfo {author} {\bibfnamefont {D.~C.}\ \bibnamefont
  {Ralph}}\ and\ \bibinfo {author} {\bibfnamefont {M.~D.}\ \bibnamefont
  {Stiles}},\ }\bibfield  {title} {\bibinfo {title} {Spin transfer torques},\
  }\href {https://doi.org/10.1016/j.jmmm.2007.12.019} {\bibfield  {journal}
  {\bibinfo  {journal} {J. Magn. Magn. Mater.}\ }\textbf {\bibinfo {volume}
  {320}},\ \bibinfo {pages} {1190} (\bibinfo {year} {2008})}\BibitemShut
  {NoStop}%
\bibitem [{\citenamefont {Vansteenkiste}\ \emph {et~al.}(2014)\citenamefont
  {Vansteenkiste}, \citenamefont {Leliaert}, \citenamefont {Dvornik},
  \citenamefont {Helsen}, \citenamefont {Garcia-Sanchez},\ and\ \citenamefont
  {Van~Waeyenberge}}]{vansteenkiste_design_2014}%
  \BibitemOpen
  \bibfield  {author} {\bibinfo {author} {\bibfnamefont {A.}~\bibnamefont
  {Vansteenkiste}}, \bibinfo {author} {\bibfnamefont {J.}~\bibnamefont
  {Leliaert}}, \bibinfo {author} {\bibfnamefont {M.}~\bibnamefont {Dvornik}},
  \bibinfo {author} {\bibfnamefont {M.}~\bibnamefont {Helsen}}, \bibinfo
  {author} {\bibfnamefont {F.}~\bibnamefont {Garcia-Sanchez}},\ and\ \bibinfo
  {author} {\bibfnamefont {B.}~\bibnamefont {Van~Waeyenberge}},\ }\bibfield
  {title} {\bibinfo {title} {The design and verification of {MuMax}3},\ }\href
  {https://doi.org/10.1063/1.4899186} {\bibfield  {journal} {\bibinfo
  {journal} {AIP Advances}\ }\textbf {\bibinfo {volume} {4}},\ \bibinfo {pages}
  {107133} (\bibinfo {year} {2014})}\BibitemShut {NoStop}%
\bibitem [{\citenamefont {Butler}\ \emph {et~al.}(2012)\citenamefont {Butler},
  \citenamefont {Mewes}, \citenamefont {Mewes}, \citenamefont {Visscher},
  \citenamefont {Rippard}, \citenamefont {Russek},\ and\ \citenamefont
  {Heindl}}]{butler_switching_2012}%
  \BibitemOpen
  \bibfield  {author} {\bibinfo {author} {\bibfnamefont {W.~H.}\ \bibnamefont
  {Butler}}, \bibinfo {author} {\bibfnamefont {T.}~\bibnamefont {Mewes}},
  \bibinfo {author} {\bibfnamefont {C.~K.~A.}\ \bibnamefont {Mewes}}, \bibinfo
  {author} {\bibfnamefont {P.~B.}\ \bibnamefont {Visscher}}, \bibinfo {author}
  {\bibfnamefont {W.~H.}\ \bibnamefont {Rippard}}, \bibinfo {author}
  {\bibfnamefont {S.~E.}\ \bibnamefont {Russek}},\ and\ \bibinfo {author}
  {\bibfnamefont {R.}~\bibnamefont {Heindl}},\ }\bibfield  {title} {\bibinfo
  {title} {Switching {Distributions} for {Perpendicular} {Spin}-{Torque}
  {Devices} {Within} the {Macrospin} {Approximation}},\ }\href
  {https://doi.org/10.1109/TMAG.2012.2209122} {\bibfield  {journal} {\bibinfo
  {journal} {IEEE Trans. Magn}\ }\textbf {\bibinfo {volume} {48}},\ \bibinfo
  {pages} {4684} (\bibinfo {year} {2012})}\BibitemShut {NoStop}%
\bibitem [{\citenamefont {Taniguchi}\ and\ \citenamefont
  {Imamura}(2011)}]{taniguchi_thermally_2011}%
  \BibitemOpen
  \bibfield  {author} {\bibinfo {author} {\bibfnamefont {T.}~\bibnamefont
  {Taniguchi}}\ and\ \bibinfo {author} {\bibfnamefont {H.}~\bibnamefont
  {Imamura}},\ }\bibfield  {title} {\bibinfo {title} {Thermally assisted spin
  transfer torque switching in synthetic free layers},\ }\href
  {https://doi.org/10.1103/PhysRevB.83.054432} {\bibfield  {journal} {\bibinfo
  {journal} {Phys. Rev. B}\ }\textbf {\bibinfo {volume} {83}},\ \bibinfo
  {pages} {054432} (\bibinfo {year} {2011})}\BibitemShut {NoStop}%
\bibitem [{\citenamefont {He}\ \emph {et~al.}(2007)\citenamefont {He},
  \citenamefont {Sun},\ and\ \citenamefont {Zhang}}]{he_switching_2007}%
  \BibitemOpen
  \bibfield  {author} {\bibinfo {author} {\bibfnamefont {J.}~\bibnamefont
  {He}}, \bibinfo {author} {\bibfnamefont {J.~Z.}\ \bibnamefont {Sun}},\ and\
  \bibinfo {author} {\bibfnamefont {S.}~\bibnamefont {Zhang}},\ }\bibfield
  {title} {\bibinfo {title} {Switching speed distribution of
  spin-torque-induced magnetic reversal},\ }\href
  {https://doi.org/10.1063/1.2668365} {\bibfield  {journal} {\bibinfo
  {journal} {J. Appl. Phys}\ }\textbf {\bibinfo {volume} {101}},\ \bibinfo
  {pages} {09A501} (\bibinfo {year} {2007})}\BibitemShut {NoStop}%
\bibitem [{\citenamefont {Pinna}\ \emph
  {et~al.}(2016{\natexlab{b}})\citenamefont {Pinna}, \citenamefont {Kent},\
  and\ \citenamefont {Stein}}]{pinna_large_2016}%
  \BibitemOpen
  \bibfield  {author} {\bibinfo {author} {\bibfnamefont {D.}~\bibnamefont
  {Pinna}}, \bibinfo {author} {\bibfnamefont {A.~D.}\ \bibnamefont {Kent}},\
  and\ \bibinfo {author} {\bibfnamefont {D.~L.}\ \bibnamefont {Stein}},\
  }\bibfield  {title} {\bibinfo {title} {Large fluctuations and singular
  behavior of nonequilibrium systems},\ }\href
  {https://doi.org/10.1103/PhysRevE.93.012114} {\bibfield  {journal} {\bibinfo
  {journal} {Phys. Rev. E}\ }\textbf {\bibinfo {volume} {93}},\ \bibinfo
  {pages} {012114} (\bibinfo {year} {2016}{\natexlab{b}})}\BibitemShut
  {NoStop}%
\end{thebibliography}%

\end{document}